**Nalin Asanka Gamagedara Arachchilage**

**Serious Games for Cyber Security Education**

**Nalin Asanka Gamagedara Arachchilage**

# Serious Games for Cyber Security Education

# Abstract

The research reported in this book focuses on developing a framework for game design to protect computer users against phishing attacks. A comprehensive literature review was conducted to understand the research domain, support the proposed research work and identify the research gap to fulfil the contribution to knowledge. Two studies and one theoretical design were carried out to achieve the aim of this research reported in this book. A quantitative approach was used in the first study while engaging both quantitative and qualitative approaches in the second study. The first study reported in this book was focused to investigate the key elements that should be addressed in the game design framework to avoid phishing attacks. The proposed game design framework was aimed to enhance the user avoidance behaviour through motivation to thwart phishing attack. The results of this study revealed that perceived threat, safeguard effectiveness, safeguard cost, self-efficacy, perceived severity and perceived susceptibility elements should be incorporated into the game design framework for computer users to avoid phishing attacks through their motivation. The theoretical design approach was focused on designing a mobile game to educate computer users against phishing attacks. The elements of the framework were addressed in the mobile game design context. The main objective of the proposed mobile game design was to teach users how to identify phishing website addresses (URLs), which is one of many ways of identifying a phishing attack. The mobile game prototype was developed using MIT App inventor emulator. In the second study, the formulated game design framework was evaluated through the deployed mobile game prototype on a HTC One X touch screen smart phone. Then a discussion is reported in this book investigating the effectiveness of the developed mobile game prototype compared to traditional online learning to thwart phishing threats. Finally, the research reported in this book found that the mobile game is somewhat effective in enhancing the user's phishing awareness. It also revealed that the participants who played the mobile game were better able to identify fraudulent websites compared to the participants who read the website without any training. Therefore, the research reported in this book determined that perceived threat, safeguard effectiveness, safeguard cost, self-efficacy, perceived threat and perceived susceptibility elements have a significant impact on avoidance behaviour through motivation to thwart phishing attacks as addressed in the game design framework.

## Acknowledgements

I would like to express my deepest gratitude to many people for their continuous support and encouragement, without which this achievement would not have been possible. In particular, I am extremely grateful to my supervisors: Dr Steve Love and Dr Mark Perry of the Department of Information Systems and Computing at Brunel University, London. Dr Steve Love whose insight changed the way I view the research, without whose support there would not be an appearance of this book. I would like to thank him for his enthusiastic guidance, invaluable advice, immense patience and also encouragement when times were tough throughout my PhD journey and making my stay at Brunel University memorable and wonderful.

Thanks to Dr Melissa Cole who helped me to open my eyes to see the world as a researcher. Thanks also go to Dr Tim French who pushed and continuously encouraged me to start a PhD journey in the first place. For my ability to survive the research office environment I owe a lot to Professor Carsten Maple and Professor Yong Yue at the University of Bedfordshire who offered relentless cheer and support. I also would like to thank my academic staff mates in the Department of Computer Science and Technology at University of Bedfordshire.

Thanks also to Jeremy Baxter and Gareth Hirst from the technical support unit in the Department of Information Systems and Computing at Brunel University, who generously provided experimental equipment and software licenses, which significantly enhanced the quality of the experiments conducted. A big thank to you all of the students, colleagues and friends who participated in the studies and experiments conducted for this book.

I would like to acknowledge the academic staff, colleagues and friends in the Department of Information Systems and Computing at Brunel University for their invaluable assistance and advice contributed to this research.

Special thanks must go to my dearest parents Nanda and Somapala, who have always encouraged me in my academic pursuits, given me the self-belief to take on new challenges and supported me wholeheartedly which act like butterfly wings in my future.

*This book is dedicated to my lovely Mother and Father*

# Table of Contents

**Chapter 3**

**Chapter 4**

**Chapter 5**

**Chapter 6**

# List of Figures

# List of Tables

## Publications

The following peer reviewed papers have been published as a result of the research conducted for this book.

# Chapter 1

# Introduction

## 1.1 Overview

This book examines the security awareness issues surrounding computer use with particular attention to a malicious IT threat avoidance perspective. Security exploits can include malicious IT threats such as computer programs which can disturb the normal behaviour of computer systems (viruses), malicious software (malware), unsolicited e-mail (spam), monitoring software (spyware), attempting to make computer resources unavailable to its intended users (Distributed Denial-of-Service or DDoS attack), the art of human hacking (social engineering) and online identity theft (phishing). These attacks are prepared to target either financial or social gain (Ng, Kankanhalli and Xu, 2009; Workman, Bommer and Straub, 2008; Woon, Tan and Low, 2005). For example, a DDoS attack could target a bank in order to break down their email server and the attacker can exhort a lump sum of money to give the email server back to the bank.

On the other hand, perhaps some people are on a mission of fun and accomplishment rather than financial or social gain. For example, teenager scan hack 'his' friend's Facebook account to have fun or show off their capabilities. The BBC has reported that one in four young Britons attempts to access Facebook accounts of their friends just for fun as claimed by a survey (BBC News, 2010). This situation could be severe with the rapid growth of information and communication technology. The same broadcasting corporation has recently reported that the police have arrested a 17-year-old boy alleged to be the spokesman for a notorious hacking group (BBC News, 2012). This poses a danger not only to individuals but also to the entire society as pervasiveness of Internet technology today provides a paradise to hackers for their 'hactivism'.

Why is it that security breaches and human hacking increase regularly? This is mainly because people tempt to disclose their personal information to public via online. It can be through Facebook, Twitter, Hi5, Orkut, Skype and even more professional social networking sites like LinkedIn. Previous research has revealed that the human factor is still infancy and a major threat in the field of information security (CNN. com, 2005 and Pike, 2011). Therefore, this research comprehends the need of which the human aspect of performing security can be influenced to avoid malicious IT threats in the context of computer use. Current work in the emerging field of "usable security" is to design and develop a tool to educate computer users to protect themselves against malicious IT threats.

Computer users play a significant role in helping to make cyberspace a safer place for everyone due to the Internet technology growth. Internet technology is so pervasive today that it provides the backbone for modern living enabling ordinary people to shop, socialize and be entertained all through the use of their home computers. As people's reliance on Internet technology grows, so the possibility of hacking and other security breaches increases (Liang and Xue, 2010). Therefore, the message "security is vital" should be spread to all computer users (Ngand Rahim, 2005).

In addition, as organizations have become increasingly 'virtual' there has been a technological shift from work to the domestic environment (O'Brien, et al., 1999). Employees are free to work at home or bring unfinished work home due to the pervasiveness of Internet technology. This increases the opportunity for individual users to open themselves to vulnerable IT threats. Unlike employees in organisations, these computer users at home are unlikely to have a sufficient IT infrastructure to protect themselves from malicious IT attacks, or may not have proper standard or strict IT security policies in place. For example, most computer users are not IT professionals and lack a high degree of computer literacy to set up a secure home computing system (Doswell, 2008). In addition, computer users tend to display unsafe computer behaviour which is particularly vulnerable to malicious IT threats. Further examples are of browsing unsafe websites, downloading suspicious software, sharing passwords among family and peers and using unprotected home wireless networks (Liang and Xue, 2010).

## 1.2 Outline of Research Problem

One such IT threat that is particularly dangerous to computer users is phishing (Dhamija, Tygar and Hearst, 2006). This is a type of *semantic attack* (Schneier, 2000), in which victims get invited by scam emails to visit fraudulent websites. The attacker creates a fraudulent website which has the look-and-feel of the legitimate website. Users are invited by sending scam emails to access the fraudulent website and steal their money. Phishing attacks get more sophisticated day by day as and when attackers learn new techniques and change their strategies accordingly (Kumaraguru, et al., 2007). The most popular is email (James, 2005; Richmond, 2006). Phishing emails occupy a variety of tactics to trick people to disclose their confidential information such as usernames, passwords, national insurance numbers and credit/debit card numbers. For example, asking people to take part in a survey or urging people to verify their bank account information, in which, they must provide their bank details to be compensated. The increasing sophistication of these techniques makes it a challenge to protect individual users against phishing attacks (Drake, Oliver and Koontz, 2006).

A number of automated software tools have been developed to help users to identify fraudulent emails and websites (Kumaraguru, et al., 2007; Arachchilage and Love, 2014). For example, Ye and Sean (2002) and Dhamija and Tygar (2005) have developed a prototype called "trusted paths" for the Mozilla web browser that is designed to help users verify that their browser has made a secure connection to a trusted website. However, these systems are not totally reliable in detecting phishing attacks (Sheng, et al., 2007; Abbasi, Zahedi and Chen, 2012). In addition, previous research has revealed that the available anti-phishing tools such as CallingID Toolbar, Cloudmark Anti-Fraud Toolbar, EarthLink Toolbar, Firefox 2, eBay Toolbar and Netcraft Anti-Phishing Toolbar are insufficient to combat phishing threats (Robila and Ragucci, 2006). On the one hand, software application designers and developers with the help of computer security experts will continue to improve phishing and spam detection. However, the human factor risk is still high and human is the weakest link in information security (Purkait, 2012; CNN. com, 2005). On the other hand, human factor risks can be mitigated by educating users against phishing threats (Brody, Mulig and Kimball, 2007; Robila and Ragucci, 2006; Arachchilage and Cole, 2011; Abbasi, Zahedi and Chen, 2012; Purkait, 2012).

Computer users are susceptible for phishing attacks (Dhamija, Tygar and Hearst, 2006; Kumaraguru, et al., 2007; Purkait, 2012; Arachchilage and Love, 2014; Arachchilage, 2013). This is because users can have a lack of security awareness and sensitive trust decisions that they make during online activities such as online banking transactions or bill payments. To protect individual users against phishing threats, phishing education needs to be considered (Kumaraguru, et al., 2007; Abbasi, Zahedi and Chen, 2012; Purkait, 2012). Previous studies have reported end-user education as a frequently recommended approach to countering phishing attacks (Hiner, 2002; Wu, Miller and Garfinkel, 2005; Allen, 2006; Kumaraguru, et al., 2008; Purkait, 2012). Therefore, this research work effort is to investigate how to better educate computer users to protect themselves against phishing attacks.

## 1.3 Theoretical Background

Previous research has noticed that technology alone is insufficient to ensure critical IT security issues. So far, there has been little work on the end user behaviour of performing security and preventing users from attacks which are imperative to cope up with malicious IT threats such as phishing attacks (Anderson and Agarwal, 2006; Aytes and Terry, 2004; Liang and Xue, 2009; Ng and Rahim, 2005; Susan, Catherine and Ritu, 2006; Woon, Tan and Low, 2005; Workman, Bommer and Straub, 2008; Purkait, 2012). Many discussions have terminated with the conclusion of "if we could only remove the user from the system, we would be able to make it secure" (Gorling, 2006). Where it is not possible to completely eliminate the user, for example in home use, the best possible approach for computer security is to educate the user in security prevention (Mitnick and Simon, 2002; Schneier, 2000). Previous research has revealed how well designed user security education can be effective (Kumaraguru, et al., 2007; Sheng, et al., 2007). This could be web-based training materials, contextual training and embedded training to improve users' ability to avoid phishing attacks.

Therefore, the research work reported in this book attempts to find an effective way of educating people to identify and prevent phishing websites.

So, how does one can educate the computer user to protect against phishing attacks? To address the issue, the research work reported in this book focuses on developing a game design framework for computer users to protect themselves from phishing attacks. This concept is based on the notion that not only can a computer game provide education (Raybourn and Waern, 2004), but also games potentially provide a better learning environment, because they motivate the user and keep attention by providing immediate feedback (Amory and Seagram, 2003; Prensky, 2001). Still, the game based education indicates its potential to improve learning. The game at least moves users into a different state where they are mentally ready for learning by creating the right environment and where they also already experience some social oriented learning (Parsons, Ryu and Cranshaw, 2006).

Sheng, et al. (2010) have conducted a role-play survey with 1001 online survey respondents to study who falls for phishing attacks. The study discovered that women are more susceptible than men to phishing attacks and participants between the ages of 18 and 25 are more susceptible to phishing than other age groups. The studies reported in this book were targeted towards to examining participants' phishing threat avoidance behaviour by using game based anti-phishing education. Therefore, participants included ages ranged from 18 to 25 were students from Brunel University and University of Bedfordshire (including people who were concerned about computer security).

## 1.4 Aims and Objectives

The aim of the research reported in this book is to investigate how one can educate computer users against phishing threats. The proposed research work accomplishes by developing a game design framework as a tool to educate computer users to thwart phishing attacks.

The research objectives are as follows:

- To investigate the level of security challenges for computer users with respect to phishing threats.
- To explore the knowledge gap of the research with the published literature.
- To investigate the key elements that should be addressed in a game design framework to thwart phishing attacks.
- To develop a mobile game prototype based on the proposed game design framework.
- To evaluate the game design framework through the mobile game prototype.
- To investigate the effectiveness of the mobile game prototype compared to traditional online learning.
- To formulate a game design framework to thwart phishing attacks.

## 1.5 The Research Approach

A comprehensive literature review is conducted to understand the research domain, support the work of the proposed research and identify the research gap to fulfil the contribution to knowledge. A cognitive activity is also used to illustrate some of the difficulties with understanding challenges for computer users with respect to phishing threat.

Two studies and one theoretical design were used for the research work reported in this book. There are several research methodology approaches existing in IS (Oates, 2006). For example, qualitative approach, quantitative approach and mixed method approach. A quantitative research approach was engaged in the first study while using both quantitative and qualitative approaches in the second study. Furthermore, the data were collected using a questionnaire, a usability study and a think-aloud study.

The first study reported in this book focused on developing a game design framework for computer users to thwart phishing attacks. The study asked what key elements that should be addressed in the game design framework. The game design framework enhances the user's avoidance behaviour though motivation to thwart phishing attacks. Therefore, this is a positivist approach (Oates, 2006). A questionnaire was designed and used in this study to evaluate the proposed game design framework for computer users to thwart phishing attacks.

The second study reported in this book attempted to empirically evaluate the proposed game design framework though the mobile game prototype. This sounds more interpretive because the study focuses on determining the user's impact on the proposed game design framework after playing the mobile game prototype. A questionnaire was designed and used in the second study to assess the user subjective satisfaction of a mobile game prototype interface. Then a think-aloud study was employed along with a pre- and post-test to assess the proposed game design framework through the mobile game prototype.

## 1.6 Structure of Book

The reminder of this book is structured in the following manner. Using the literature, survey studies and experiments, Chapter 2 explores how to protect computer users against phishing threat with a particular focus on security education. Therefore, game based security education is explored. In addition, the computer user's security behaviour is observed in order to structure the literature into a meaningful story. The chapter contains a number of sections, which are: computer users' security behaviour; Cyber security threats; Phishing threats; Educational games. This can be used to identify similarities and disparities between research and the current knowledge gap.

A detailed description of the research approach employed is offered in Chapter 3. The chapter is divided into five main sections, which are: Overview of the research question; Overview of the research approach used in this book; Setting up laboratory; Data collection; Data analysis.

Chapter 4 describes the first study. The aim of this study was to develop a game design framework for computer users to protect themselves from phishing attacks. The chapter reports the first study of this book in detail. Aims and objectives, pilot study, main study, model testing and game design framework are then explained. Finally, the results are reported and discussed and a chapter summary is presented.

Chapter 5 discusses a theoretical mobile game design approach for computer users to thwart phishing attacks. The chapter is divided into nine main sections which are: Aims and objectives; Game design issues; Game description; Game based learning; Learning science principles; Mobile game design; Storyboard design; Mobile game prototype.

Chapter 6 describes the second study. In this chapter the game design framework (presented in Chapter 4) was assessed through a mobile game prototype (presented in Chapter 5). This chapter reports the second study of the book in detail. The research approach is presented, including the sampling technique, data collection instruments, procedure and data analysis. The results are then presented, followed by a discussion section, concluding with a chapter summary.

Chapter 7 is a discussion chapter that focuses on investigating the effectiveness of the developed mobile game prototype (presented in Chapter 5) compared to traditional online learning to thwart phishing threats. Furthermore, it discusses the implications for the research work reported in this book. Therefore, an evaluation study is conducted to examine the effectiveness of the developed mobile game prototype compared to traditional online learning to thwart phishing threats. To accomplish this study, a website developed by Anti-Phishing Work Group (APWG) for the purpose of public anti-phishing education initiative was employed as a traditional web based learning source. This chapter reports the evaluation study of the book in detail. The research approach is presented, including the sampling technique, data collection instruments, procedure and data analysis. The results are then presented, followed by implications for research, concluding with a chapter summary.

Finally, the book ends up with presenting a discussion and conclusion in Chapter 8. The chapter highlights a summary of the research findings of this book and the contribution to knowledge of the research work reported in this book. The chapter also identifies potential limitations of this research and opens up opportunities for future work that may extend the current research findings.

## 1.7 Summary

This introduction chapter presented a brief explanation of the research work reported in this book. A background to the research was explained and then the research aim and objectives were presented. The methodological approach that undertakes to achieve the research aim of this book was then introduced. Finally, the structure of the book was presented.

The next chapter is a detailed background of relevant research. This provides a background discussion for the aim and objectives of this research.

# Chapter 2

# Literature Review

## 2.1 Introduction

The research reported in this book focuses on how to educate computer users against phishing threat with particular focus on game based learning. The research related to computer users' phishing threats avoidance behaviour and game based learning is reviewed in order to organise the literature into a meaningful story, which can be used to identify similarities and disparities between research and gap in the current knowledge. First, computer users' security behaviour is examined to give the reader a background of importance of the research work reported in this book. More recent work related to cyber threats in computer use is addressed in detail. The dangerous of phishing threat is then stressed highlighting the current state of security education to thwart phishing threats. After that, the importance of game based learning in the educational context is discussed. Moreover, mobile game based learning approaches were discussed in order to provide a better education for computer users to thwart phishing threats. Finally, a summary is given to highlight the main points of interest relating to the problem at hand.

## 2.2 Computer Users' Security Behaviour

The computer user plays a vital role in protecting the cyberspace; however the security of computers is left to the initiative of the user (Ng and Rahim, 2005). This is mainly because sensitive trust decisions are made during their online activities; such as online banking transactions. The pervasiveness of Internet technology provides media enabling them to engage with online activities such as online shopping, banking and entertaining much easily. Unfortunately, most computer users are unaware of this technique and will open a back door for hackers. This is one of the main reasons to increase a possibility of hacking and other security breaches. Therefore, it is worth looking into human aspects of security within their regular computer use.

Protection Motivation Theory (PMT) describes that motivation to protect oneself from threats is related to the individual's cognitive beliefs that one is personally vulnerable to the threat (Tu and Yuan, 2012). However, there is relatively little research on the human aspect of performing computer security and how it can be influenced to practice security (Ng, Kankanhalli and Xu, 2009).

Why is the human aspect security so important? Of course, it is challenging knowing the complexity of human behaviour and lack of computer literacy in the computer use. This enables researchers and security practitioners to reveal possible ways of protecting them from malicious IT threats. Personal computers are a popular target by intruders (Doswell, 2008; Rogers, 2006), because they are interested in what users' store in their computers. For example, confidential personal information such as usernames, passwords and bank account details. On the other hand it can argue that individual computer users can be easily fooled due to the lack of security awareness in their computer use. Therefore, understanding the user's security behaviour in the context of computer use is important to protect them from malicious IT attacks.

Furthermore, it is not just only money-related information that they are after, they also would be interested in the computer user's resources such as hard disk space, Internet connection and fast processors. They might use these resources of the computer user to attack someone else on the Internet. For example, hackers can access into someone's laptop illegally to compromise his office computers. The worst situation is this kind of attack cannot be traced in most of the cases.

Why do hackers pay attention for personal computer users specifically? Personal computer environment is typically insecure and also easy to break into individual computer systems compared to office environment (Rogers, 2006). This is problematic, because many personal computer users may not realise that they are exposed to vulnerabilities and therefore need to secure their personal computer systems.

The complexity of individual computer systems is ever increasing due to the rapid development of Information and Communication Technology (Hui, 2007). For example, this could include broadband Internet access, audio and video systems, file sharing and network printing services. However, on the one hand, this could be a positive remark to assume that the trend to embrace individual technologies continuing growth. On the other hand, combating against malicious IT threats in the computer use would be a great challenge.

The environment of personal computer systems operates considerably different from a typical organisational environment (Art and Glenn, 2005). A centralised decision making power that exercises operational control over the aspect of computer systems management by the organisational environment. This enables to manage organisational environment occupying sufficient IT infrastructure and strict IT security policies and procedures developed by professional security experts. Therefore, employees adhere to best practices within the organisational context. For example, this includes predetermined installation of software applications such anti- virus and anti-spyware, proper firewall configurations, locking the systems down to prevent unauthorized software installation and use strong passwords policies and backup procedures. Furthermore, in the managed organisation setup may occupy system administrators to keep computers up-to-date and properly protect against malicious IT threats. This might be because organisations want to typically manage their IT infrastructure relieving the end user from his responsibility.

However, personal computer environment does not typically have this sort of managed set up with the advantages of resource intensive central management authority (Art and Glenn, 2005). In addition, systems maintenance and configuration functions are left to the individual user. The reliance upon a system administrator is a less optimal solution due to the lack of education, awareness, training and professionalism in the unmanaged personal computer environment. Despite this, still social networking websites are wide spread among individual computer users, friends, peers and families. This unmanaged personal computer environment along with their engagement of social networking websites enable to disclose a rich set of confidential information to hackers for them to succeed with the attacks. Therefore, the scale of the threat may dramatically increase due to the pervasiveness of unmanaged individual computers connected to high-speed broadband services.

Human aspect of security in the computer use could be concerned as the most exciting and challenging area in Information Security because general users are not equipped with a certain degree of computer literacy (Liang and Xue, 2009a; 2010b). However, in many occasions, teenagers are given the responsibilities to look after home computer systems as concerned their better exposure to latest technology (Hui, 2007). The worst situation could be most teenagers are interested to contribute as full-time unpaid workers at home providing technical support, administration and troubleshooting services. Perhaps this can be a false assumption as long as concern is with their computer security knowledge.

It is sometimes much easier and less risky to manipulate humans who have access to sensitive information rather than try and break into systems straight away (Pike, 2011). For example, spear phishing attack, which is an e-mail spoofing fraud, attempting to target a specific user or an organisation.

Unfortunately most computer users have a lack of security awareness due to the deficiency of education, professionalism and training (Hui, 2007). Security awareness, attitude towards trust and privacy protection depend on cultural background and society norms (Hui, 2007). Hofstede describes a number of cultural values indices to measure cultural differences between societies (Hofstede, 1991; Hofstede, 2003). As Hofstede defines *Collectivism*, represents a preference for a tightly-knit framework in the society in which individuals can expect their relatives, peers or members of a particular in-group to look after them in exchange for unquestioning loyalty. Therefore, the culture may play a significant role in shaping peoples' attitudes towards privacy (Kumaraguru and Cranor, 2005). For example, collectivism in some societies has created a better trust attitude among each other and therefore they may not hesitate to share online identities such as usernames, passwords and credit or debit card details among peers and family. Similarly, they may not attempt to have a proper authentication mechanism to access their wireless network or share home wireless networks authentication such as usernames and passwords among their neighbours. This poor security behaviour in computer use exposes computer users to vulnerable malicious IT attacks or security breaches.

Computer users should realise that they are responsible for their own personal computer's security as the same way that they are responsible for having insurance when they buy a house or vehicle. However, knowledge about the human aspect of performing computer security is yet far from complete (Anderson and Agarwal, 2006; Aytes and Terry, 2004; Liang and Xue, 2009; Ng and Rahim, 2005; Susan, Catherine and Ritu, 2006; Woon, Tan and Low, 2005; Workman, Bommer and Straub, 2008; Purkait, 2012). The purpose of the research reported in this book is to investigate how one can educate computer users to combat phishing threats. Therefore, the introduced game design framework to educate computer users against phishing attacks is coming from an IS perspective. The study attempted to empirically investigate what key elements that should be incorporated into the game design framework for computer users to avoid phishing attacks through motivation. The elements of a theoretical model derived from Technology Threat Avoidance Theory (TTAT) were used to address in the game design framework (Liang and Xue, 2010).

The model investigates how individuals avoid malicious IT threats by using a given safeguarding measure. The safeguarding measure does not necessarily have to be an IT source such as anti-phishing tools; rather it could be behaviour such as game based anti-phishing education (Liang and Xue, 2010). Liang and Xue (2010) revealed that computer users' IT threat avoidance behaviour is predicted by avoidance motivation, which, in turn, is determined by perceived threat, safeguard effectiveness, safeguard cost and self-efficacy. Furthermore, they stressed that computer users develop a threat perception when they believe that the malicious IT is likely to attack them (perceived susceptibility) and the negative consequences will be severe if they are attacked (perceived severity). When threatened, computer users are more motivated to avoid the threat if they believe that the safeguarding measure is effective (safeguard effectiveness), inexpensive (safeguard cost) and they have confidence in using it (self-efficacy). Moreover, they found that perceived threat and safeguard effectiveness have a negative interaction effect on avoidance motivation.

Liang and Xue (2010) developed the measurements based on their theoretical meaning and relevant literature. Perceived threat was measured by a number of items created on the basis of its substantive meaning (Rosenstock, 1974). The items assessed participants' perception of the potential harms, peril or danger that a malicious IT threat (in this case spyware) imposes. They developed the scales for perceived susceptibility based on the research related to health behaviour (Saleeby, 2000); which assess the likelihood and possibility of a malicious IT threat occurrence.

TTAT forms that computer users' IT-related well-being includes two dimensions: information privacy and computer performance where a malicious IT attack could damage both dimensions (Liang and Xue, 2009). Therefore, computer users' severity perceptions should relate to the two dimensions. To develop a list of items to measure perceived severity, they referred to the privacy literature in IS (Smith, et al., 1996) and practitioner journals that point out the negative impacts of a malicious IT threat (Schultz, 2003; Shaw, 2003; Sipior, et al., 2005; Stafford and Urbaczewski, 2004). The items measured based on computer users' concerns about both the loss of personal information and degraded computer performance related to processing speed, Internet connection and software applications.

Liang and Xue (2010) developed the items of safeguard effectiveness based on relevant health behaviour research (Champion and Scott, 1997; Saleeby, 2000). They used the items for safeguard cost from Milne, et al. (2002) and Saleeby (2000). They assessed self-efficacy with the items developed by Compeau and Higgins (1995), making minor adjustments to adapt it to the malicious IT threat context (in this case spyware context). The items developed to measure avoidance motivation were based on behavioural intention measures from technology adoption research (Davis, 1989; Davis, et al., 1983), with a focus on malicious IT threat avoidance rather than IT adoption. Finally, Liang and Xue (2010) measured IT threat avoidance behaviour with two self-developed items.

## 2.3 Cyber Security Threats

Cyber security threats are critical issues continuously faced by people in their computer use (Bishop, 2003). This is due to more and more individual users experiencing the immense benefit of Internet technology. However, Internet technology brings new and dangerous malicious IT threats to the society regularly such as viruses, malicious software (malware), unsolicited e-mails (spam), monitoring software (spyware), the art of human hacking (social engineering) and online identity theft (phishing). Attackers find easy ways to break into individual computer systems due to the lack of security countermeasures. For example, most of the time, individual users' Internet connection is always turned on when using high-speed broadband connection. They may not have secure authentication techniques such as strong usernames and password. Therefore, attackers can easily find a way and attack this kind of unmanaged computer system.

How do hackers break into individuals computer systems? (Rogers, 2006) There is no particular standard way of breaking into a system. In some cases, they send emails with attached viruses. Opening the attachment happens to activate the virus. Then hackers may use this as an entry point to break into the computer system. Once hackers are compromised the computer system, then they can easily accomplish their tasks by installing new malicious programs. For example, hackers can collect the information about users without their knowledge by installing spyware. The presence of spyware is usually hidden from the user and can be difficult to detect. Therefore, the art of manipulating human into performing actions (Social engineering) causes to succeed cyber-attacks.

Social engineering is one of the main strategies that attackers use in the first place. It is a method (Allen, 1993; Hiner, 2002; Pike, 2011; Timko, 2008) that gains information from people without directly asking. Social engineering is a constant cyber threat as people give away too much information in social media (Allen, 1993; Hiner, 2002; Timko, 2008). This could be through Internet enable services such as Facebook, Twitter, Hi5, Orkut, Skype and even more professional social networking website like LinkedIn. Security practitioners argue that this happens due to the lack of security awareness, professionalism and training, the poor computer user opens the back door for hackers using social networking websites (Allen, 1993; Hiner, 2002; Pike, 2011).

In addition, hackers may attempt to find out the user's IT security measures in a preparation for hacking attacks. In this case, victims could be contacted and asked to complete a telephone or email survey, which is most of about general IT. But there would be a small section at the end asking what firewall, anti-virus products and other security tools in individual uses. To motivate participants who engaged in the survey, there may be a price draw such as an Apple iPad if they complete the survey successfully. Therefore, social engineering is hard to stop, yet a fancy tool in which hackers urge innocent people to give away sensitive information.

Virus programs are another severe threat faced by computer users (Foreman, 2004). Viruses are usually a set of programs, which can disturb the normal behaviour of the computer system. For example, once viruses are on the computer system, they often occupy whole hard disk by replicating files and ultimately exploit it. Therefore, computer users must be careful when installing third party software applications from unknown vendors. This would be the same way as someone ringing the doorbell and wanted to come into the house to sell something. At this stage, the house owner needs to make a decision whether or not to let him in. The problem is people who visit home to sell phony products or services. Once inside, they may try to steal valuable items or try to harm in an armed robbery. Therefore, it is important to watch them carefully while they were in the house. Similarly, viruses can also infect personal computer systems in many ways such as through flash drives, floppy disks, CD-ROMs, websites, emails and downloaded files (Rogers, 2006). Therefore, it is advisable to scan for viruses before each time use them. For example, when receiving an email with an attachment or inserting a flash drive to the computer system, need to scan them for viruses before using it. The problem is how many computer users are practicing this?

What can one do if a house appliance broke? The response would be try to repair or by calling a repairperson get the work done, or replace it with new one. Software programs work much the same way. For example, what happens when the operating system itself malfunctioned due to a cyber-attack? To restoring the function that it provides is important. Therefore, the most appropriate solution would be installing patches that fix the bugs provided by a trusted vendor before intruders have a chance to exploit by the malicious attack. For example, the Microsoft provides patches for all windows operating systems on their website (Microsoft, 2012). The BBC reported that Microsoft has recommended for all their customers to install patches as soon as possible or update to the latest version of the web browser for "improved security" after they found out Internet Explorer is the weakest link in a sophisticated and targeted cyber-attack on Google in January, 2010 (BBC News, 2010). However, it is important to consider how many computer users did realise the worth of installing patches in order to protect their web browser.

We have heard stories about receiving a letter or parcel through postal services that in some way caused harm. For example, Anthrax-laden letters caused the death of 5 people while infecting 17 others seriously in 2001 (Centre for Counter Proliferation Research, 2002). In the same way computer users may receive lots of emails each day, most of them which are unsolicited. Those unsolicited emails can use social engineering techniques to lure people by entering the contest or the prize that you may have won. The attacker may attempt to encourage opening and reading the email by the receiver. There is not much harmful in only opening and reading those emails, but on clicking links and downloading attachments in emails is extremely dangerous. Emails attached to viruses and worms are very common (Rogers, 2006), which may cause much harm for the computer system. For example, once an email-born virus is opened, it may spread to all the contacts in the email address book. This may even be more harmful as the attack spread among the people. Therefore, the computer user should create awareness of avoiding this kind of cyber threats.

The job of a security guard is to assess everybody who wants to enter or leave the premises. The guard decides if he or she should continue to enter or be stopped. In other words, the guard keeps unwanted people out and only permits authorised people to enter to the premises by checking their identities. Firewall also acts similar to same way as the security guard. It observes the traffic when sending or receiving data packets from another computer via the Internet (Rogers, 2006). Finally, it decides whether or not to send the packet to the receiver. If the data packet is suspicious, then he may not send to the receiver. Therefore, proper configuration of a firewall is very important in individual computer use. Otherwise, there may be a threat of receiving suspicious data packets, which can cause harm to personal computer system. Clearly, users should be aware of best security practices to protect themselves from cyber security threats in their computer use (Anderson and Agarwal, 2006; Ng and Rahim, 2005; Susan, Catherine and Ritu, 2006).

Every house has doors, windows and sometimes keys to lock and protect householders from strangers. People should be discouraged of sharing the key with strangers, sometimes even close friends or neighbours. Using passwords for emails or social networking accounts work much the same way. Therefore, it is worth noticing to use strong passwords in individual computer use. Why should use strong passwords? This is one of the major sources of leaking users' information to hackers. For example, computer users can attempt to proceed with online transactions in their day-to-day life. Before proceeding with the online transaction they need to register with the website providing personal details such as username, password, date of birth and address. The hacker can attack by trying every possible combination of the password (searching likely possibilities). If the password is weak, the attack could be successful. This is a trial-and-error process where the hacker tries until the right password is found. This type of attack is well known amongst computer users and so called dictionary attack (Rogers, 2006). The attack method would be effective although it takes a long time to success.

Among the other malicious IT threats that exist in the cyberspace such as viruses, malware, spam and spyware, online identity theft which is well known as phishing is particularly dangerous to computer users (Arachchilage and Cole, 2011). This is because phishing attacks are easy to success and hard to stop.

## 2.4 Phishing Threats

"*Sometimes a 'friendly' email message tempts recipients to reveal more online than they otherwise would, playing right into the sender's hand"* (Jagatic, et al., 2007). Phishing is one of the rapidly growing online crimes, in which, the attacker attempts to fraudulently gain sensitive information from the victim by impersonating a trustworthy entity (Anderson, 2001; Iacovos and Sasse, 2012; Purkait, 2012). Victims are lured by an email asking to log onto a website that appears legitimate, but it's actually created to steal their confidential information such as usernames, passwords and banking details. Though phishing websites are shut down once identified, new ones spring up daily. However, it is often difficult to shutdown phishing websites quickly due to cross-border jurisdictional problems (Kumaraguru, et al., 2010). As reported in Anti-Phishing Work Group (APWG) phishing websites stay online average 4-5 days (APWG, 2006; Kumaraguru, et al., 2010).

Online crimes like phishing attacks are easy to do, but hard to stop, because these attacks involve social and psychological techniques along with technical tools. For example, the attacker can send an email insisting the victim logs on to his bank website for identity verification. Actually, the bank website is a fake site which has a look and feel of the legitimate website. Hackers may use their technical skills to develop fake websites whilst using social engineering techniques attempting to reveal victims more online. Phishing is a type of semantic attack, which synbookes social engineering techniques along with technical skills to make the attack successful (Downs, Holbrook and Cranor, 2007). Social engineering attacks are increasingly being used to bypass robust secure systems due to the weakness of human tendency to trust (Kumaraguru and Cranor, 2009).

Phishing threat has been problematic for online shoppers in the past 15 years (Iacovos and Sasse, 2012). The probability of an online consumer coming through a phishing website is considerably high due to the growth of online transactions. The UK police revealed in a recent investigation, seven out of top 10 Google search results for a popular brand of boots were advertised in fraudulent websites which have the same look and feel of legitimate websites (Iacovos and Sasse, 2012). Furthermore, it has been reported that one in 12 online customers of buying event tickets have been caught by fake ticket websites, with the average loss of each victim being £80.

Stajano and Wilson (2011) have described the 'need and greed principle', which attackers exploit successfully. Once the attacker knows what users want, they can easily manipulate. In fact, most online shoppers are behind good deals. So they tempt to use web search engines to look for what they want. Once they are presented very tempting offers with links to websites, they do not give up the opportunity to save money that they do not have to spend. This grand opportunity is taken by hackers to commit online fraud activities. Perhaps, online shoppers may be too late in realising what has actually happened. As a consequence, they may lose all their money on credit or debit cards.

Over the last decade phishing attacks have increased dramatically in the World Wide Web (Purkait, 2012; Arachchilage and Love, 2013). This is because phishing has become the most common channel  for thieves to steal personal information to aid them in identity theft (Mercuri, 2006;  Brody, et al., 2007; Anderson, et al., 2008; Eisenstein, 2008). Therefore, phishers are  now  aiming  computer  users  and organisations all over the world (Sullins, 2006).  Previous studies have shown a steady increase in phishing activities as well as the  related cost. Phishing activity trend report published by APWG (2013) in April 2013  reported 76,123 unique phishing sites detected from October 2012 through December  2012 as shown in Table 2.1. Furthermore, the report emphasised that 142,862 unique  phishing e-mail reports (campaigns) received by APWG from its consumers in the same period. It is worth to notice that usually 56.04 percent of phishing attacks contain some form of target name in URL. Therefore, the end user phishing URL  awareness need should be considered to combat phishing threats.

**Table 2.1: The APWG Phishing Activity Trends Report, 4[th] Quarter 2012 (APWG, 2013)**

| Statistical Highlights for 4th Quarter 2012 | | | |
|---|---|---|---|
| | October | November | December |
| Number of unique phishing e-mail reports (campaigns) received by APWG from consumers | 23,365 | 24,563 | 28,195 |
| Number of unique phishing websites detected | 51,232 | 46,002 | 45,628 |
| Number of brands targeted by phishing campaigns | 401 | 430 | 418 |
| Country hosting the most phishing websites | USA | USA | USA |
| Contain some form of target name in URL | 60.31% | 54.23% | 53.59% |
| No hostname; just IP address | 1.63% | 1.87% | 1.93% |
| Percentage of sites not using port 80 | 0.30% | 0.24% | 1.04% |

It has been seen that phishing attacks against online game players have a massive increase (Figure 2.1 and 2.2). This increase is from 2.7 percent of all phishing attacks in Q3 (3[rd] Quarter 2012) to 14.7 percent in Q4 (4[th] Quarter 2012). The financial service industry continued to be the most-targeted sector in the fourth quarter of 2012 while payment services being close behind. Furthermore, as stated in the phishing activity trend reports, the most targeted victims for the attacks are personal computer users (APWG, 2013). This is mainly because most online gamers are personal computer users.

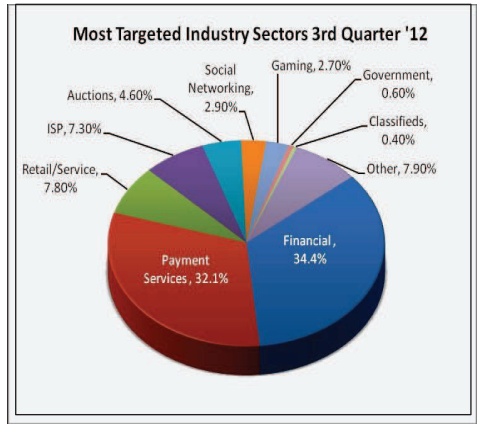

**Figure 2.1: Phishing attacks: most targeted industry sectors, 3$^{rd}$ Quarter 2012 (APWG, 2013)**

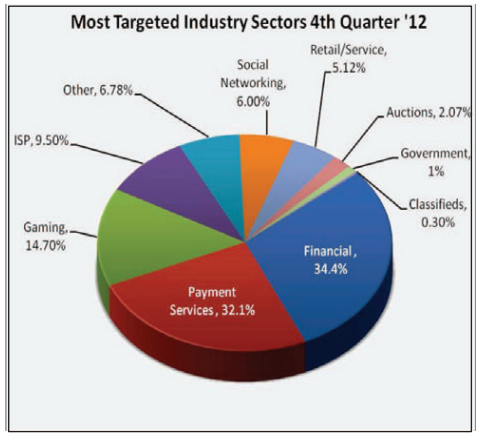

**Figure 2.2: Phishing attacks: most targeted industry sectors, 4$^{th}$ Quarter 2012 (APWG, 2013)**

PhishTank, the online website which collects information on web sites engaged in phishing activities received 25,021 valid submissions of phishing websites only in the month of January 2013 (PhishTank, 2013). Furthermore, the website stated 860 (3 percent of valid phishes that month) phishes used an IP address (i.e.

http://12.34.56.78) and 24,169 (or 97 percent) used a domain name (i.e. http://example.com) in that month. Therefore, the criminals use a variety of tactics to enticing people into visiting fake websites and persuading them to disclose their personal information.

Disclosing customer confidential information such as personal information or banking details to fake websites, can lead not only to immediate financial losses but also to become a victim of identity theft or its consequences (Brody, Mulig and Kimball, 2007; Iacovos and Sasse, 2012; Kumaraguru, et al., 2007). For example, this can lead to damage a person's credit rating or being linked to illegal or criminal activities. Furthermore, this sort of situation could lead to an overall loss of trustworthiness of online consumers shopping and discourage them from engaging online financial transactions. Therefore, phishing is considered as a dangerous threat that exists in cyberspace. However, academic researchers and security practitioners are constantly in a battle against phishing threats (Brody, Mulig and Kimball, 2007; Dahamija, Tygar and Hearst, 2006; Downs, Holbrook and Cranor, 2007; Garera, et al., 2007; Iacovos and Sasse, 2012; Kumaraguru, et al., 2007a; 2007b; Robila and Ragucci, 2006; Sheng, et al., 2007; Wu, Miller and Garfinkel, 2006).

Previous research has reported on phishing falls into three categories: research related to understand why people fall for phishing attacks, what tools to protect people against phishing attacks and methods for educating people not to fall for phishing attacks (Arachchilage and Cole, 2011; Iacovos and Sasse, 2012; Kumaraguru, et al., 2007a; 2008b; Sheng, et al., 2007). Furthermore, it is important to dig deeper into the above categories in order to provide a better solution for phishing threats.

## 2.4.1 Why Do People Fall for Phishing Attacks?

Dhamija, Tygar and Hearst (2006) conducted a laboratory-based experiment showing twenty-two participants to twenty websites, asking them to determine which ones were legitimate. They found out that participants made mistakes on the test 40 percent of the time. Furthermore, they noted that 23 percent of their participants ignored all cues in the web browser address bar and status bar as well as all security indicators. A considerable amount of literature work has been reported that this is one of major reasons why people fall for phishing attacks (Downs, Holbrook and Cranor, 2007; Kumaraguru, et al., 2007a; 2007b; Sheng, et al., 2007; Wu, Miller and Garfinkel, 2006).

It is important to note that users' perception of such phishing threats may encourage them to avoid potential vulnerabilities. Downs, Holbrook and Cranor (2006) have conducted a role playing study aimed at understanding why people fall for phishing emails and what cues they look for to avoid such attacks. Their research finding revealed two key things. First, even those people that are aware of phishing attacks, they do not think their potential vulnerabilities or strategies to trace phishing attacks. Second, while people can secure themselves from known risks, they have difficulties of generalising their known to unfamiliar risks. Wu, Miller and Garfinkel (2006) stated that many users do not understand phishing attacks or realise how sophisticated such attacks can be. A possible reason for this would be the user's lack of phishing threat perception.

Wu, Miller and Garfinkel (2006) evaluated three simulated anti-phishing toolbars to determine how they were effective at protecting users from visiting fraudulent websites. They revealed that many participants ignored passive toolbar security indicators and instead used the site's content to decide whether or not it is a fraudulent website. In some cases participants did not notice warning signals and in other cases they assumed warnings were invalid though they noticed them. Perhaps the possible reason would be users struggle to interact with toolbars due to a lack of usability. Therefore, it is advisable to conduct usability studies to determine how users interact with security toolbars.

In a follow-up study, the authors tested anti-phishing toolbars to generate pop-up warning messages that blocked access to fraudulent web sites until overridden by the user. These pop-up warning messages reduced the rate at which users fell for fraudulent websites. However, it did not completely eliminate all users from falling for phishing websites. The authors concluded in their research that Internet users are not very good at interpreting security warnings and familiar with common phishing attacks techniques and strategies. Furthermore, they recommended user education to phishing prevention and online safety practices.

Reuven Harrison, co-founder of Tufin Technologies which commissioned the survey, told the BBC that young people need better education in order to protect themselves from hacking and other security breaches (BBC News, 2010). Sheng, et al. (2010) have conducted a role-play survey instrument administered to 1001 online survey respondents to study who falls for phish. The study revealed that women are more susceptible than men to phishing; further, participants between the ages of 18 and 25 are more susceptible to phishing than other age groups. Participants were selected from a diverse group of staff and students, including people who were concerned about computer security.

Kumaraguru, et al. (2009) conducted a real world evaluation of anti-phishing training with 515 participants, which extended their previous work "PhishGuru". In this real world study, they focused on long-term anti-phishing knowledge retention and the reaction to two received training emails. It has been reported that 17.5 percent of participants still entered their personal details into simulated phishing websites after 28 days from the first email message. This was a considerable improvement from the 40.1 percent measured in a control group before the study. However, the study revealed that one in five users is vulnerable for phishing attacks. They also found that participants between the age group of 18 to 25 years were consistently more vulnerable to phishing attacks than older participants.

## 2.4.2 Automated Software Tools for Phishing Detection

Anti-phishing tools are now provided by Internet service providers (Sheng, et al., 2007; Iacovos and Sasse, 2012; Sanchez and Duan, 2012), for instance, available as web browser toolbars. There are currently various anti-phishing tools available for free downloading (Zhang, et al., 2007). CallingID Toolbar, Cloudmark Anti-Fraud Toolbar, EarthLink Toolbar, Firefox 2, eBay Toolbar and Netcraft Anti-Phishing Toolbar are a few examples. However, these tools are not entirely reliable in detecting phishing attacks (Aggarwaly, Rajadesingan and Kumaraguru, 2012; Sanchez and Duan, 2012; Sheng, et al., 2007; Wu, Miller and Garfinkel, 2006). Even the best anti-phishing tools missed over 20 percent of phishing websites (Zhang, et al., 2007).

Zhang, et al. (2007) have developed an automated test bed for testing available anti-phishing tools. Their automated anti-phishing test bed was implemented in C# and was comprised of 2000 lines of code. Figure 2.1 shows the high-level system architecture used by Zhang, et al. (2007).

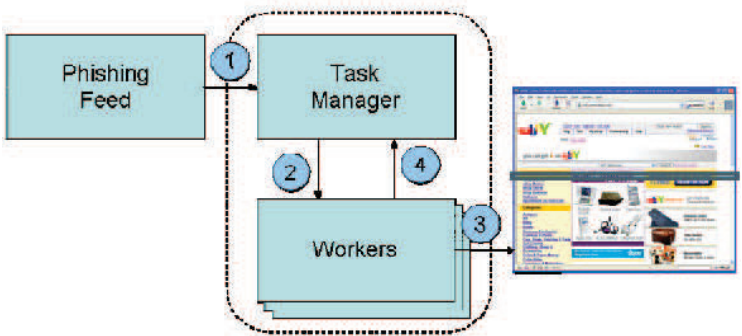

**Figure 2.3: High-level system architecture for the anti-phishing evaluation test bed (Zhang, et al., 2007)**

The task manager gets an updated list of Unified Resource Locators (URLs) from a phishing feed and then sends those URLs to a set of workers. Each worker retrieves a web page, checks whether the web page was labelled as a phishing scam or not, and sends the result back to the task manager to aggregate all of the results. The task manager and workers are grouped together because they can be run on the same machine or on separate machines. The experiment was conducted through automating this process by extracting URLs from a feed of validated phishing URLs and/or by using own heuristics to select phishing URLs from a feed of invalidated phishing email messages.

They used 200 verified phishing URLs and 516 legitimate URLs to test the effectiveness of ten popular anti-phishing tools. Their findings revealed that only one tool, namely 'SpoofGuard', was able to consistently identify more than 90 percent of phishing URLs correctly. Nevertheless, it also incorrectly identified 42 percent of legitimate URLs as phish. Other tools varied considerably depending on the source of phishing URLs as shown in Table 2.2 as provided by Zhang, et al. (2007). Wrongly identified positive (False positives) results are summarised in Table 2.3 as provided by Zhang, et al. (2007). Of these remaining tools, only one correctly identified over 60 percent of phishing URLs. However, it is interesting to note that attackers and tool developers are continuously on an arms race.

**Table 2.2: Number of phishing sites correctly identified by anti-phishing tools. Note, SpoofGuard's catch rate is estimated after time 0 (Zhang, et al., 2007)**

| Time since URL extraction | PhishTank | | | | | APWG | | | | |
|---|---|---|---|---|---|---|---|---|---|---|
| | 0 hours | 1 hour | 2 hours | 12 hours | 24 hours | 0 hours | 1 hour | 2 hours | 12 hours | 24 hours |
| CallingID | NA | NA | NA | NA | NA | 23 (23%) | 26 (26%) | 25 (27%) | 30 (38%) | 24 (36%) |
| Cloudmark | 68 (68%) | 68 (68%) | 68 (69%) | 64 (67%) | 47 (67%) | 22 (22%) | 24 (24%) | 21 (22%) | 25 (31%) | 25 (37%) |
| EarthLink | 83 (83%) | 83 (83%) | 81 (82%) | 78 (84%) | 59 (84%) | 54 (54%) | 53 (54%) | 51 (54%) | 51 (64%) | 47 (70%) |
| eBay | 28 (28%) | 28 (28%) | 26 (27%) | 24 (26%) | 18 (26%) | 52 (52%) | 52 (53%) | 51 (54%) | 43 (54%) | 35 (52%) |
| IE7 | 68 (68%) | 68 (68%) | 67 (68%) | 62 (67%) | 47 (67%) | 75 (75%) | 74 (75%) | 72 (77%) | 67 (84%) | 58 (87%) |
| Firefox | NA | NA | NA | NA | NA | 28 (28%) | 50 (50%) | 51 (54%) | 47 (59%) | 44 (66%) |
| Firefox/ Google | 70 (70%) | 70 (70%) | 70 (71%) | 71 (76%) | 59 (84%) | 53 (53%) | 54 (55%) | 56 (60%) | 56 (70%) | 49 (73%) |
| Netcraft | 77 (77%) | 77 (77%) | 73 (74%) | 69 (74%) | 56 (80%) | 60 (60%) | 59 (60%) | 57 (61%) | 62 (78%) | 49 (73%) |
| Netscape | 8 (8%) | 10 (10%) | 10 (10%) | 9 (10%) | 15 (21%) | 31 (31%) | 31 (31%) | 32 (34%) | 37 (46%) | 30 (45%) |
| SpoofGuard | 91 (91%) | 91(91%) | 89(91%) | 85(91%) | 64(91%) | 96 (96%) | 95(96%) | 90(96%) | 78(98%) | 65(97%) |
| TrustWatch | 49 (49%) | 49 (49%) | 48 (49%) | 45 (48%) | 36 (51%) | 44 (44%) | 43 (43%) | 44 (47%) | 45 (56%) | 45 (67%) |
| ActiveURLs | 100 | 100 | 98 | 93 | 70 | 100 | 99 | 94 | 80 | 67 |

**Table 2.3 Number of legitimate sites (out of 516 tested) falsely identified as phishing sites by anti-phishing tools (Zhang, et al., 2007)**

|  | Falsely identified as phishing | | Unsure | |
|---|---|---|---|---|
| CallingID | 10 | (2%) | 177 | (34%) |
| Cloudmark | 5 | (1%) | 497 | (96%) |
| EarthLink | 5 | (1%) | 493 | (96%) |
| eBay | 0 | (0%) | 0 | (0%) |
| Firefox | 0 | (0%) | 0 | (0%) |
| Firefox/Google | 0 | (0%) | 0 | (0%) |
| IE7 | 0 | (0%) | 0 | (0%) |
| Netcraft | 0 | (0%) | 0 | (0%) |
| Netscape | 0 | (0%) | 0 | (0%) |
| SpoofGuard | 218 | (42%) | 256 | (50%) |
| TrustWatch | 0 | (0%) | 256 | (50%) |

Dhamija, Tygar and Hearst (2006) showed that anti-phishing indicators developed in automated software tools are ineffective due to a significant percentage of users' ignorance towards passive indicators (Iacovos and Sasse, 2012). Even though users noticed indicators, they may not understand what they signify. In addition, inconsistent placing on heterogeneous web browsers makes the task of identifying a phishing site difficult.

Schechter (2007) reported that 53 percent of their research study participants still tried to access a web site after their tasks were interrupted by a security-warning message. They also revealed in the same study that deletion of Hyper Text Transfer Protocol Security (HTTPS) did not make any effect on participants disclosing their personal details on a site. Furthermore, 97 percent of participants were tempted to enter their personal details even after website authentication images were removed.

Abbasi, Zahediand Chen (2012) conducted an experiment involving over 400 participants to evaluate the impact of anti-phishing tools' accuracy on users' ability to avoid phishing attacks. Each participant was given a high accuracy (90 percent) or low accuracy (60 percent) tool and asked to make a decision on several legitimate and phishing websites. Their experimental results revealed that participants using the high accuracy anti-phishing tool performed much better than those using the less accuracy tool in terms of avoiding phishing attacks. However, even users of the high accuracy tool often ignored its correct recommendations. Therefore, the authors suggested that while the accuracy of anti-phishing tools is a critical factor, reducing the success rates of phishing attacks require other considerations such as enhancing the user's awareness of phishing.

For the mentioned reasons, the current study concludes effective anti-phishing education must complement any technical measures to enhance users' ability to detect phishing attacks (Kumaraguru, et al., 2007a, 2007b; 2008; Robila and Ragucci, 2006).

### 2.4.3 Anti-Phishing Education

It has been shown that both academic institutions and government organisations have made a significant effort to provide end user education to enable public understanding of security (Iacovos and Sasse, 2012; Arachchilage, Tarhini, Love, 2015; Arachchilage, 2015). The Anti-Phishing Work Group (APWG) is a non-profit organisation working to provide anti-phishing education to enhance the public understanding of security. They cover various areas: 1) What is a phishing threat? 2) How can it be severe? 3) What is the usefulness of having a safeguarding measure? 4) Where to report a suspected phishing email or website? and 5) Phishing education to thwart phishing attacks. The US Computer Emergency Readiness Team also offers free advice on its website about common security breaches for computer users who have a lack of computer literacy.

While a great deal of effort has been dedicated to resolving the phishing threat problem by prevention and detection of phishing emails, URLs and web sites, little research has been done in the area of educating users to protect themselves from phishing attacks (Iacovos and Sasse, 2012). Therefore, more research needs to focus on anti-phishing education to protect users from phishing threats.

However, is anti-phishing training or education effective? Even though there are usability experts who claim that user education and training does not work (Sheng, et al., 2007; Evers, 2007), other researchers have revealed that well designed end-user education could be a recommended approach to combating phishing attacks (Allen, 1993; Hiner, 2002; Iacovos and Sasse, 2012; Kumaraguru, et al., 2008; Schechter, 2007; Sheng, et al., 2007;Timko, 2008). In line with Herley (2009), also Iacovos and Sasse (2012) and other researchers argue that current security education on phishing threats offers little protection to end users, who access potentially malicious websites (Kumaraguru et al., 2009; Sheng, et al., 2007; Arachchilage and Love, 2014).

Another reason for ineffectiveness of current security education for phishing prevention is because security education providers assume that users are keen to avoid risks and thus likely to adopt behaviours that might protect them. Iacovos and Sasse (2012) claimed that security education should consider the drivers of end user behaviour rather than warning users of dangers. Therefore, well-designed security education should develop threat perception where users are aware that such a threat is present in the cyberspace. It should also encourage users to enhance avoidance behaviour through motivation to protect them from malicious IT threats.

Kumaraguru, et al. (2007) designed and evaluated an embedded training email system that teaches people to protect themselves from phishing attacks during their normal use of email. The authors conducted lab experiments contrasting the effectiveness of standard security notices about phishing with two embedded training designs they developed. They found that embedded training works better than the current practice of sending security notices.

Robila and Ragucci (2006) evaluated the impact of end user education in differentiating phishing emails from legitimate ones. They provided an overview of phishing education, targeting on context aware attacks and introduced a new strategy for end user education by combining phishing IQ tests and class room discussions. The technique involved displaying both legitimate and fraudulent emails to users and asking them to identify the phishing attempts from the authentic emails. The study concluded that users identified phishing emails correctly after having the phishing IQ test and classroom discussions. Users also acknowledged the usefulness of the IQ test and classroom discussion. Researchers at Indiana University also conducted a similar study of 1700 students in which they collected websites frequently visited by students and either sent them phishing messages or spoofed their email addresses (Jagatic, et al., 2007).

Sheng, et al. (2007) conducted a study to evaluate participants' ability to identify fraudulent websites before and after spending 15 minutes engaged in one of their three anti-phishing training activities: playing the game; reading an anti-phishing tutorial created based on the game; or reading existing online training materials. They found that participants who played the game were better able to identify phishing websites after engaging 15 minutes of training compared to participants in other conditions. They reported the false-positive rate (phishing website identified as real) was reduced from 30 percent to 14 percent, and the false-negative rate (legitimate website identified as fake) was reduced from 34 percent to 17 percent. However, they also reported 31 percent of users could still not differentiate between good websites and bad ones. Iacovos and Sasse (2012) argued that the probability of an online buyer coming across a phishing website is relatively high, because nowadays many fake websites appear in popular web search engines results such as Google and Yahoo. Therefore, teaching people how to use search engines to find legitimate websites is yet crucial and problematic.

Tseng, et al. (2011) also developed a game to teach users about phishing based on the content of the website. The authors proposed the phishing attack frame hierarchy to describe stereotype features of phishing attack techniques. The inheritance and instantiation properties of the frame model allowed them to extend the original phishing pages to increase game contents. Finally, they developed an anti-phishing educational game to evaluate the effectiveness of proposed frame hierarchy. The evaluation results showed that most of the lecturers and experts were satisfied with this proposed system.

To date, end user education on phishing threat has tried to encourage users to check URLs, but resulted in limited success (Iacovos and Sasse, 2012). Kumaraguru, et al. (2007) conducted a research on teaching "Johnny not to fall for phish," which focused on educating users about phishing and helping them to make better trust decisions. They came across a number of challenges in the anti-phishing education in particular: users are not motivated to learn security; security is secondary task for most users; it is difficult to teach people how to identify threats without also increasing their tendency to misjudge non-threats as threats.

Kumaraguru, et al. (2007) also designed educational materials using learning science principles to teach users not to fall for phishing attacks. Their focus was to motivate users to pay attention to the training materials. This is due to the fact that in most of the situations users ignore security educational materials.

The possible phishing attacks can be identified in several ways, such as carefully looking at the website address, so called Universal Resource Locator (URL), signs and content of the web page, the lock icons and jargons of the webpage, the context of the email message, and the general warning messages displayed in the website (Downs, Holbrook and Cranor, 2007; Wu, Miller and Garfinkel, 2006). Previous research has identified that existing anti-phishing techniques based on URLs are not robust enough for phishing detection (Garera, et al., 2007; Sheng, et al., 2007; Iacovos and Sasse, 2012). Iacovos and Sasse (2012) stated that two main approaches have been used to protect users against phishing: anti-phishing indicators and end-user education. However, Dhamija, Tygar and Hearst (2006) emphasised that the first approach is still ineffective because a significant percentage of users ignore passive indicators. They conducted a laboratory-based experiment showing 22 participants to 20 websites and asking them to determine which ones were legitimate. The results revealed that 23 percent of the participants ignored all cues in the web browser address bar, and status bar as well as all security indicators. This is due to the users' lack of phishing URL awareness (Dhamija, Tygar and Hearst, 2006; Liang and Xue, 2009; 2010). Garera, et al. (2007) emphasised that it is often possible to tell whether or not a particular URL belongs to a phishing website without having any knowledge of the corresponding page information. In their study, they have identified several fine-grained heuristics that can be used to distinguish between a phishing URL and a benign URL. The classification technique based on these heuristics achieved an accuracy of 97.3 percent demonstrating that URL analysis alone can reach a high degree of accuracy in phishing detection.

The research work reported in Chapter 5 of this book attempts to design a mobile game for computer users to protect themselves from phishing attacks. The objective of mobile game design is to teach users how to avoid phishing URLs. However, the mobile game design should create the awareness of identifying the features of potential phishing URLs. For example, legitimate websites usually do not have numbers at the beginning of their URLs such as http://81.153.192.106/.www.hsbc.co.uk.

Anandpara, et al. (2007) revealed that users who read existing online training materials are overly cautious and identify many legitimate sites as fraudulent sites. However, they do not learn useful techniques for identifying phishing attacks because many of the online training resources do not teach specific cues and strategies (Kumaraguru, et al., 2007). Even most existing training materials were not designed based on learning science principles (Kumaraguru, et al., 2007). Therefore, it has been suggested that researchers and system developers should attempt to apply e-Learning science principles to maximise the effectiveness of training materials. Learning by doing, immediate feedback, contiguity principle, personalisation principles and story-based-agent environment principles could be used. Furthermore, e-Learning theories can be used to enhance the motivation of players in the game design.

The lack of motivation is one of the main barriers when educating the end user in phishing detection and prevention (Tseng, et al., 2011). Therefore, the game-based learning approach utilises interactive activities and interesting multimedia as a suitable solution for motivating end-users for phishing education (Sheng, et al., 2007).

## 2.5 Educational Games

Educational games and simulations have become increasingly acknowledged as an enormous and powerful teaching tool that may result in an "instructional revolution" (Walls, 2012; Cone, et al., 2007; Foreman, 2004). The main reason is that game based education allows users to learn through experience and the use of virtual environment while leading them to approach problem solving through critical thinking.

Educational games have become an important part of the social and cultural environment (Walls, 2012; Papastergiou, 2009; Oblinger, 2004). Those games are particularly interesting and pervasive among children and adolescents and are ranked as the most popular activity in computer use (Downes, 1999; Harris, 1999; Mumtaz, 2001). McFarlane, Sparrowhawk and Heald (2002) conducted a study on 7-16 year old students in the UK and revealed most of them were regular game players in home computer use.

Zyda (2005) describes Serious Games as "a mental contest, played with a computer in accordance with specific rules, which uses entertainment to further government or corporate training, education, health, public policy and strategic communication objectives". Michael and Chen (2005) proposed the following thematic classification based upon Zyda's definition of Serious Games; Military Games, Government Games, Educational Games, Corporate Games, Healthcare Games, Political and Religious Games. However, educational games are quite useful as an effective teaching medium (Froschauer, et al., 2010) because it enables users to learn in an interactive and attractive manner.

There is a considerable amount of published studies in the literature describing the role of games in the educational context. Bellotti, et al. (2009) designed a Massive Multiplayer Online Game (MMOG) for high-school called 'SeaGame' to promote best practices in sea-related behaviours, such as sailing or beach surveillance. The main focus of the 'SeaGame' was to embed the educational content into a gaming context in a meaningful, homogeneous and compelling whole, where the player can enjoy learning while having fun. Therefore, this type of games helps people to improve their best practices in behaviour.

In addition, game based education is further useful in motivating players to change their behaviour. Gustafsson and Bang (2008) designed a pervasive mobile-based game called 'Power Agent' to educate teenagers and their families to reduce energy consumption in their homes. They attempted to transform the home environment and its devices into a learning arena for hands-on experiences with electricity usage. The results suggested that the game concept was more efficient in engaging teenagers and their families to reduce their energy consumption during the game sessions.

Furthermore, game based learning can be effective not only for changing people's behaviour, but also for developing their logical thinking to solve mathematical problems. Eagle (2009) designed and empirically investigated 'Wu's Castle', which is a 2D (2-Dimensional) role-playing game where students can develop C++ code to solve in-game problems. The results showed that 'Wu's Castle' is more effective than a traditional programming assignment for learning how to solve problems on arrays and loops.

People nowadays travel a lot due to a mobile and global life style and affordable transportation costs. Generally travelers and tourists are interested to know about the culture and traditions before they arrive to the place. Reading books is often too-time consuming as many of them have a very busy life, juggling work, chores and family and friends. Therefore, it is worth for them if the information is quickly accessible. 'Travel in Europe' (Bellotti, et al., 2008) and 'Second China Island' (Fishwick, et al., 2008; Henderson, et al., 2008), are two projects developed concentrating on cultural heritage and tourism respectively. 'Travel in Europe' (TiE) focused to implement an innovative means to promote and reveal cultural heritage. It promoted the user's interaction within online virtual representations of European cities and encouraged the user to accomplish missions in a treasure hunt manner. 'The Second China Island' is a 3D (3-Dimensional) space developed in the 'Virtual World of Second Life' in order to experience the content of China and its culture virtually by the user.

The literature revealed that well-designed games focusing on education could be helpful for learning even when used without assistance. The 'Anti-phishing Phil' game developed by Sheng, et al. (2007) reported results confirm that games educate people about phishing and other security attacks in a more effective way than other educational approaches such as reading anti-phishing tutorial or reading existing online training materials. Anido, et al. (2011) designed a framework, which was developed as an open source software system that supports personalised learning with focus on serious games and simulations. They claimed that the most successful achievement of the project was easy-to-use, interoperable and powerful expressive, educational and adaptive system that enabled teachers with no technical expertise to design, develop and execute e-Games through a comprehensive platform and in a number of devices and environments.

Video games education is used more and more within schools to teach children various subjects such as English, Mathematics and Geography (Wang, Fsdahl and Morch-Storstein, 2008; Walls, 2012). Previous studies have reported video games education is to be effective in terms of children's learning (Gee, 2003; Jenkins, 2002; Papert, 1998; Wang, Fsdahl and Morch-Storstein, 2008; Walls, 2012). School children are able to practice certain skills using video games in a computer lab supervised by a teacher. Previous research showed that the use of dynamic video games within the classroom could be beneficial to improve children's academic achievements and motivation (Rosas, et al., 2003). Game based education is not only attractive to school children but also to students in universities (Sharples, 2000). Several studies have shown how video games can be used in higher education (Baker, Navarro and Hoek, 2003; Natvig, Line and Djupdal, 2004; Navarro and Hoek, 2004; Wang, Fsdahl and Morch-Storstein, 2008; Walls, 2012). Nevertheless, Wang, Fsdahl and Morch-Storstein (2008) strongly believed that this area needs to be explored more extensively.

A number of educational games have been designed and developed to protect computer users and to assert security issues in the cyberspace (Cone, et al., 2007). For example, some educational games teach information assurance concepts, whereas others teach pure entertainment with no basis in information assurance principles or reality. However, there is little research on engagement in the virtual world that also combines the human aspect of security (Sheng, et al., 2007). Therefore, it is worth to investigate further on game based learning in order to protect computer users from malicious IT threats such as phishing attacks.

In the past, educational games have received attention not only from the academic and research community, but also from the UK government (Oblinger, 2004). The BBC reported that the UK government has planned to introduce corporation tax relief from April 2013 for the video games, animation and high-end television industries (BBC News, 2012). Furthermore, they mentioned that the government wanted to promote the UK as the technological centre of Europe. This is mainly because multiple improvements of the learning experience are attributed to educational games and game based educational learning, as they have increasing potential to influence and enhance students' motivation. It is believed that games technologies have potential to improve the learning process in multiple ways (Torrente, et al., 2010).

One of the most well-established facts in the educational game research is the ability to enhance players' motivation towards learning as they are able to retain their attention and keep them engaged and immersed with games (Gee, 2003; Ghazvini, Earnshaw, Robison and Excell, 2009a; 2009b; Gunter, Kennyand Vick, 2008; Malone, 1981; Quinn, 2005). Other interesting facts of educational games are:

- The provision of immersive gaming environments that can be freely explored by players and promote self-directed learning (Oblinger, 2004; Squire, 2003)
- The immediate feedback process with perception of progress (Freitas and Oliver, 2006; Ghazvini, et al., 2009a; 2009b; Gunter, Kenny and Vick, 2008; Quinn, 2005)
- Their relation to constructivist theories and support of learning in augmented environment (Ghazvini, et al., 2009a; 2009b; Prensky, 2001)

It has been shown that games are being extensively spread among users due to the pervasiveness of mobile devices and technologies. The BBC reported 'World of Warcraft's' developers have already confirmed they are looking at bringing the hit video game to mobile devices (BBC News, 2012). Today, mobile devices can be used to play most of the games due to the fact that the software scale of these games is becoming smaller. This is an enormous challenge to academic researchers and cyber security experts who try to find solutions to protect people from malicious IT threats. Researchers have strongly argued that the mobile game based education has potential and worth researching for cyber security awareness (Cone, et al., 2007; Ghazvini, et al., 2009a; 2009b; Shih, Hou and Wu, 2011).

### 2.5.1 Mobile Game Based Learning

The pervasiveness of cellular phones and related technologies among the youth has made them an ideal platform to provide educational content for learning informally anywhere and anytime (Molnar and Martinez, 2011). Mobile devices have become a significantly important part of people's lives due to a pervasive platform that they carry out almost all the time. Unlike desktop computers, mobile phone penetration is significantly higher in emerging economies, leapfrogging landlines and broadband access (ITU ICT-Eye Free Statistics, 2012). This worldwide ubiquity has contributed to enhance game based education through mobile devices. A large collection of existing educational mobile games provides the evidence for the increase of mobile game based education (Molnar and Martinez, 2011; Daher, 2010).

Mobile game based learning is a method where users are given the opportunity to perform learning activities using a game or a series of games through mobile devices such as cellular phones and smart phones (Denk, Weber and Belfin, 2007; Ghazvini, et al., 2009a; 2009b; Klopfer, 2008; Kurkovsky, 2009). Those digital games based on mobile devices can offer fidelity of simulations and problem solving tasks, which are challenging, entertaining and interacting for students and thus enhance their learning motivation. Therefore, learning security awareness concepts using mini games based on handheld devices will become one of the key challenges and future trends in mobile game based education (Cone, et al., 2007; Shih, Hou and Wu, 2011). Although, mobile gaming is one of the most important application areas in educational technology, the literature revealed that relatively few educational mobile games have been designed effectively than other types in existence (Hosein and Beharry, 2010).

Mobile gaming started with the 'snake' game on a Nokia handset in 1997 and has progressed to the 'Angry Birds' game (Chasey, 2010; Shih, Hosein and Beharry, 2010; Hou and Wu, 2011). Domestic mobile online games began to increase since 2004 due to the rapid growth of handheld devices and related wireless technologies such as cellular phones, pagers, PDA, Bluetooth, Wi-Fi, and 3G (Garner, et al., 2006; Zhu, et al., 2009). Furthermore, emerging wireless technologies and open Internet standards allow communication across heterogeneous mobile devices via the network (Gee, 2003). Portable heterogeneous devices are central to mobile game based learning, since they provide access to the learning content in a game based environment.

Educational researchers and software developers have recognised those devices and technologies as powerful resources to support game based learning (Andrews, et al., 2003; Facer, et al., 2004; Gee, 2003; Roschelle and Pea, 2002; Wegerif, 2003). The mobile game industry professionals have analysed that today's online game is tomorrow's mobile platform; they also predict a great potential growth in domestic mobile phone games (Zhu, et al., 2009). It has been shown that major mobile communication operators, network operators, mobile device manufactures and software developers attempt to develop games for mobile platforms to increase income. The BBC reported Google Android shipments have increased by 886 percent while Apple showing the second largest growth in the smart phone sector with 61 percent growth in the year 2010 (BBC News, 2010). It has been emphasised that the overall smart phone sector grew by 64 percent in the same year. The same broadcasting corporation reported in 2012 that Apple has almost doubled its profits in the first three months of the year and 35 million Apple iPhones were sold in that quarter, which was almost double the level to the year ago (BBC News, 2012). Therefore, academic researchers and educational specialists can benefit from this growth of devices and related technologies to provide better learning in the educational context.

Some innovators strongly argue that desktop computers will disappear from the society while new handheld devices and their interfaces will turn into ubiquitous, pervasive, invisible and embedded technologies or devices in the souring environment (Shneiderman, 1987). It is also believed that those devices will be focused on context-awareness, attentive and perceptive, sensing users' desires and providing feedback through ambient displays that glow, hum, change shape or blow air. Furthermore, some researchers and technology experts predict advanced mobile devices that are wearable, or even implemented under the human skin. For example, through adapting technology like individual wireless sensors that can be used to track users entering premises.

Mobile games and related technologies can also enrich family and peer relationships (Shneiderman, 1987). For example, mobile multiplayer games aid to develop a strong relationship among users, which is useful to share their knowledge, expertise and experience likewise, emotions and love among each other. It has been seen that some of those relationships perhaps end up with future marriage.

Currently, there is an enormous trend in games technology targeting on dedicated handheld programming devices such as Nintendo DS and PlayStation Portable or PDAs and mobile phones (Denk, Weber and Belfin, 2007; Klopfer, 2008; Kurkovsky, 2009). For example, touch-based interfaces introduced on Nintendo DS or iPhone/iPod changed computer based educational games to the emerging mobile- based platform. Those touch-based interfaces enable the player to interact with digital objects within the game environment more easily than navigating through the keyboard. As a consequence of those emerging mobile technologies, iPhone and Android devices and low-cost apps stores are awash with games.

It has been proved that mobile games can improve recruitment and retention by capturing players' enthusiasm (Kurkovsky, 2009). The mobility is a significant feature that supports the use of mobile games (Parsons, Ryu and Cranshaw, 2006). It can be conceptualised in many ways such as mobility of the user, mobility of the service and mobility of the device. For example, the mobility feature behind the mobile gaming environment enables players to use the game while they are away from the communication space. Most users play casual games in short breaks (Kurkovsky, 2009); during class breaks, travelling on a bus or train, and while waiting for transportation or waiting in a queue.

Although previous research on educational games has struggled to prove that mobile game based learning can seriously contribute to the education arena, other research in this area has revealed that signs of change have been detected (Molnar and Martinez, 2011). Banerjee, et al. (2007) discovered that mobile game based educational can significantly enhance the learning process when used to study. Kam, et al. (2009) have also shown that the use of mobile game based education can improve children literacy (Tian, 2010). Furthermore, previous research has developed meta-level reflections on learning strategies through mobile game based education (Facer, et al., 2004).

There have been many attempts using mobile gaming for teaching; widely popular in game based education are, for example, puzzle games (Hosein and Beharry, 2010). Previous research has in many attempts identified and defended mobile gaming as an educational tool. Costabile, et al. (2008) conducted an experimental study to explore how a mobile phone can be used to teach archaeology through game based learning. Yordanova (2007) has explained the overall importance of mobile learning integrated with technology in education. Furthermore, basic characteristics, advantages and existing challenges related to mobile learning are addressed in a study by Kurkovsky (2009). Kurkovsky's study shows how mobile game based learning can be used to engage students in Computer Science courses from an early stage on (Yordanova, 2007). The study concluded that the mobile games helped students to relate to their course materials and to establish strong connections to real world applications.

Also, the mobile gaming environment has become a mammoth in the entertainment sector (Wang, Fsdahl and Morch-Storstein, 2008). The reasons are convenience, low cost, enhancement of creativity and community of the mobile games. The popularity of mobile games is due to the ability to take them anywhere once the game has been downloaded to the mobile device. Creative mobile games can be designed and developed to a much lower cost than console games. While mobile phones were originally created for communication, the core values presented for users in mobile games are a distinct advantage.

Mobile devices and related technologies such as mobile phones, PDAs, pagers, Wi-Fi, Bluetooth and 3G are increasingly seen as fertile ground for the development of resources to support mobile game base learning. There is an increasing interest of young people for a digital culture outside of school and towards the interaction with mobile and games technologies, as opposed to desktop PC applications (Facer, et al., 2004; Holloway and Valentine, 2003; Facer, et al., 2003). Educational researchers have identified these tools and technologies as powerful resources in supporting learning experience and encouraging the development of cognitive skills (Andrews, et al., 2003; Facer, et al., 2004; Gee, 2003; Roschelle and Pea, 2002; Wegerif, 2003).

For example, mobile phone camera, Internet, Wi-Fi or Bluetooth are available in almost all smart phones. It is also known that more and more games use these features (Baillie, et al., 2010; Elina and Koivisto, 2010). According to the Office for National Statistics (2011), almost half of the UK Internet users are connected to the Internet using a mobile phone while away from the home or office. The figure was up 22 percent from the year 2009 to 2011 (Office for National Statistics, 2011) game developers can utilise this knowledge and existing tools and technologies to design and develop effective educational mobile games.

Mobile games and related technologies have become more and more complex and pervasive over the last years; currently they look like console and PC games (Korhonen and Koivisto, 2006). The volume of mobile phone ownership is certainly greater than desktop PC ownership in the UK (Office for National Statistics, 2011). Widespread ownership of mobile phones may ensure high market penetration for game developers. The mobile phone ownership in the UK has been dramatically increased by 45 percent during the last ten years (Office for National Statistics, 2011). Therefore, designing and developing games for pervasive platforms enables and enhances the user's accessibility.

## 2.6 Summary and Conclusion

This chapter discussed the existing research in the field of online usable security and mobile gaming in order to provide the necessary background for this book and in order to support the research question. Previous studies have reported that technology alone is insufficient to ensure critical IT security issues. There has not been much research work on the human behaviour of performing security or the protection of user against attacks, which are imperative to cope with malicious IT threats such as phishing attacks. However, well-designed security education can be effective to protect the end-user against these malicious IT attacks.

One objective of the background work reported in this book was to find effective ways to educate people to identify and prevent phishing attacks. The literature showed the importance of educating people to identify and prevent themselves from phishing attacks as the best possible approach. Furthermore, it discussed the notion that not only can a computer game provide education but also that games can potentially provide a better learning environment. Moreover, it indicated mobile games to improve the potential learning experience.

The next chapter will describe the research methods, which were employed to undertake the studies reported in this book.

# Chapter 3

# Research Methodology

## 3.1 Introduction

The following chapter contains four constituent parts. The first part describes the research approach used in this book. Then, the setup of the laboratory for the experiments is discussed. After that, the data collection techniques employed in this research work are addressed. Finally, the analytical procedures applied to the collected data are presented.

## 3.2 Overview of the Research Approach Used in this Book

Two studies and one theoretical design were incorporated for the research work reported in this book. Several research methodology approaches exist in the Information Systems literature (Creswell, 2003; Oates, 2006), for example, the qualitative approach, the quantitative approach and the mixed method approach.

Quantitative research refers to studies, which produce results that are concluded by statistical analyses and summary. Quantitative data analysis generally means producing data or evidence based on numbers. Researchers who follow this approach attempt to gather data by employing different strategies such as questionnaire surveys and experiments. Quantitative data is primarily used and analysed by positivist researchers (Oates, 2006), but critical and interpretive researchers sometimes also generate such data.

On the other hand, qualitative research approach refers to studies with findings that are not concluded by statistical analysis and summary. The data obtained from qualitative research are usually gathered from case studies, interviews and observations. Qualitative data includes non-numeric data such as words, sounds and images. Furthermore, qualitative data can be used to describe social behaviour and individual groups. Quantitative data is the main kind of data used and analysed by interpretive and critical researchers (Oates, 2006), but can be generated by positivist researchers too.

In the mixed method approach, researchers gather both quantitative and qualitative data to provide the best answer to the research question. This approach starts by gathering statistical data using a quantitative method (such as a survey) and then gathers qualitative data from interviews using a qualitative method. This mixed method approach can be useful in that it can make best use of both the quantitative and qualitative approaches.

The first study used a quantitative approach and both quantitative and qualitative approaches were used in in the second. The first study attempted to develop a game design framework for computer users to thwart phishing attacks. As a research methodology, although a qualitative approach could have been used in this research study, a quantitative approach was chosen because it offered the flexibility to represent the general population of computer users; quantitative research approach is also highly penetrated approach in IS (Oates, 2006). Additionally, this study focused on investigating the key elements that should be addressed in the game design framework, which enhances the user avoidance behaviour through motivation to protect themselves from phishing attacks. Quantitative research methods are mostly associated with the philosophical paradigm of positivism (Oates, 2006). Therefore, this sounds more like a positivist approach.

The other philosophical paradigm is the interpretive approach; which does not prove or disprove a hypobook, as compared to positivist research (Oates, 2006). If the research is based on surveys or experiments, it is more likely to be based on positivism. Therefore, the first study reported in this book takes a positivist approach since it is survey-based and tests hypobook derived from a research model. Positivism is more appropriate for this research study, because it enhances the user's phishing threat avoidance behaviour.

The second study was focused on evaluating the proposed game design framework through the mobile game prototype developed in MIT App Inventor Emulator. The study has employed a quantitative approach as the first step to measure the user's subjective satisfaction of the mobile game prototype interface. Then the study followed by a qualitative approach as the second step in order to determine users' impact on the framework after playing the mobile game prototype. Quantitative research methods are mostly associated with the philosophical paradigm of interpretivism (Oates, 2006). Therefore, this research study looks more like an interpretive approach, because it focuses on determining the user's impact on the proposed game design framework after playing the mobile game prototype.

Hypotheses are formal statements whose predictions are derived from the evidence of past research work and theory or simply the result of a guess (Breakwell, Hammond and Fife-Schaw, 1995). For example, researchers start developing theories about their topic of the research territory. This leads to a statement based on the theory that can be tested through an empirical investigation. Hypotheses are tested by manipulating one or more variables (Preece, Rogers and Sharp, 2002). An experiment is then designed to prove or disprove the defined hypobook (Oates, 2006). Robson stated that an experiment encompasses: assigning participants to different conditions; manipulating one or more independent variables; measuring the effects of this manipulation on one or more of the dependent variables; and controlling all other variables (Robson, 2002).

There are two types of experiments: those performed in the laboratory setting and those conducted in the field, the so-called "real world" (Oates, 2006). When an experiment is laboratory-based, the participant would normally use the system in a controlled environment. For example, an experiment is performed in the Microsoft Usability Laboratory. An advantage of the laboratory-based experiment is that it allows the control of variables in order to correctly measure cause and effect of the system (Coolican, 2004). For example, all participants can utilise the same mobile device throughout the experiment. This allows measuring participants' behaviour more precisely and keeping conditions identical for them.

Coolican (2004) has identified some weaknesses of the field experiment: lack of control brings problem of extraneous variables; difficult to replicate the experiment and record data accurately and most commonly ethical problems. The ethical aspect of the experiment would be more concerned, especially when it comes to investigate participants' computer security related behaviour involving sensitive information such as usernames, passwords or online banking details. There might be a risk to disclose their personal information to public in the real world experiment if they are unaware that it is an experiment. Therefore, the experiment is more robust and reliable when it is performed under a controlled environment.

In addition, the laboratory-based experiment can be occupied with technologies and equipment, which enables for data recordings while offering the participant an environment free from day-to-day distractions. Janssen and Ostrom (2006) have stated that the laboratory-based experiments provide a very abstract and controlled environment, in which, researchers test very precise hypotheses. On the other hand, this environment allows the control of variables where all participants can utilise the same equipment. The main reason to conduct this experiment in a laboratory is because all participants should be accessible for pre- and post-tests based on a computer connected to the Internet. Additionally, the researcher should be able to record the data accurately during the experiment. Therefore, it would be ideal to set up a lab before the experiment begins with all required equipment rather than arranging a fieldwork.

Previous research has reported on investigating the human aspect of security behaviour using laboratory-based experiments. Dhamija, Tygar and Hearst (2006) conducted a laboratory experiment with a group of twenty-two participants to categorize why people fall for phishing attacks. Egelman, Cranor and Hong (2008) conducted a laboratory-based experiment asking participants to buy a product on eBay and sent them a related eBay fake phishing email. Downs, Holbrook and Cranor (2006) also performed in a two-part laboratory study with 20 participants to analyses the reasons why people fall for phishing attacks. Therefore, it is worth mentioning that the laboratory-based experiments are very useful to researchers who wish to investigate specifically the human aspect of security behaviour in a given situation.

For the mentioned reasons, a survey study based on the quantitative approach was reported in Chapter 4, whilst a laboratory study based on the qualitative approach was reported in Chapter 6. The first study introduced in Chapter 4 was related to develop a game design framework for computer users to thwart phishing attacks. The second study conducted in Chapter 6 was focused to evaluate the game design framework using a mobile game prototype.

### 3.3 3 Setting up the Laboratory

It is important to define the research environment and conditions including the physical characteristics in order to carry out a laboratory based experiment (Rates, 2004). Furthermore, it requires a quiet and closed experimental room where only the participant and the experimenter sit together at some point. The experimental study room is equipped with necessary hardware devices such as a computer, Internet access, network cards and Internet cable. Additionally, required software needs to be pre-prepared for the experimental study. Therefore, the laboratory study was reported in Chapter 6 arranged as follows (Figure 3.1).

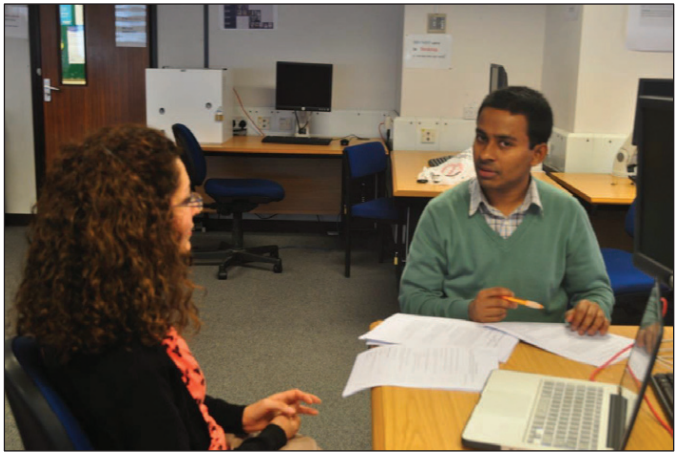

**Figure 3.1: Experiment room in the second study**

## 3.4 Data Collection

A number of instruments were employed to gather necessary data for the research work reported in this book. The data were collected using a questionnaire, a usability study and a think-aloud study. Each of these instruments is described in the following section.

### 3.4.1 Questionnaire

A questionnaire is a flexible and basic research technique for collecting structured information from individuals (Coolican, 2004). Zaharias and Poylymenakou (2009) argued that a questionnaire is one of the useful methods of gathering data in scientific research. It is widely used to recognise participants' opinions, preference, judgments and the patterns of behaviour in a particular structure. A questionnaire drives participants directly to research topics helping them to clearly see the focus. Furthermore, the same questionnaire can be delivered to all participants in order to maintain the consistency of the data collection process. In addition, an anonymous response style encourages participants to response truthfully, especially when they are talking about controversial issues (Walliman, 2001). Moreover, questionnaires can be delivered electronically (Root and Draper, 1983), which is also cheaper and takes less time compared to other methods (Zaharias and Poylymenakou, 2009).

There are some basic principles that the researcher should bear in mind when designing a questionnaire (Brooke, 1996; Coolican, 2004; Kirakowski and Corbett, 1988). First, requesting the minimum of information needed for the research purpose; because most respondents do not wish to spend a long time answering questions or their answers may depend on their mood. In addition, the questionnaire should not address questions that are no longer used or that can be obtained from elsewhere. Second, the researcher has to make sure that the participant can answer questions. For example, a question like "*how many hours have you used the Internet this year?*" is difficult to answer precisely for most people. Third, the researcher has to make sure that the participant should answer questions honestly and accurately. If the question is difficult, it is unlikely to be answered honestly and accurately. Although the question was being answered, it would be more based on well-known public opinion rather than the individual's real belief. Therefore, the researcher should avoid including difficult questions in designing a questionnaire. Finally, the researcher has to make sure all questions will be answered without any rejections, because participants may refuse to answer sensitive questions.

Thus, the questionnaire is considered to be an appropriate technique to collect data related to users' perception of usability (Brooke, 1996; Kirakowski and Corbett, 1988). The use of a questionnaire and attitude scale provides simplicity and speed for the evaluation method being used. For example, in most cases participants have to travel to the laboratory to go through the evaluation exercises lasting between 20 minutes and an hour. As can be imagined, participants might be discouraged if they come across any problems since no assistance was given. If they were presented with a short and simple questionnaire, which encompasses to assess subjective reactions to the system usability, it is very likely that they would not reject to complete it.

A questionnaire was designed and employed for each study in this research work. Details about each questionnaire will be presented in the related study later in chapters 4 and 6 of this book. However, a questionnaire was used in the first study, evaluated a game design framework for computer users to thwart phishing attacks. Another questionnaire was designed and employed in the second study to measure the user subjective satisfaction of a mobile game prototype interface. Falthzik and Carroll (1971) conducted email questionnaire survey of industrial organizations about the rate of return for closed versus open-ended questions. Of 100 firms receiving an "open-ended" question, 27 percent returned them while 78 percent of 100 receiving a "closed" question returned them. Furthermore, they stated that "closed" questions are easier and cheaper to analyse, ensuring that the form in which the data gathered correctly for the type of analysis planned and also provides a more meaningful basis for comparison. In addition, the alternatives presented with the "closed" question may help the participant to understand the question clearly. On the other hand, "open- ended" questions are more appropriate when the issue is complex, the relevant dimensions and critical factors are unknown and most importantly exploration is the main goal of the research work (Falthzik and Carroll, 1971). Therefore, "closed" questions were used in both the questionnaires that have many forms such as:

- Yes or No types of questions: e.g. "Do you use a mobile phone?"
- One true answer: e.g. "Please select your gender?"
- Choosing from different available answers; e.g. "Experience of using mobile devices"
    1. Mobile Phone
    2. Smart Phone
    3. Personal Digital Assistant (PDA)
    4. Tablet PC

A 5-Point Likert scale was used in the questionnaires of both the studies. Likert scales are used to assess attitudes, beliefs and opinions and also have been extensively used for evaluating the subjective user satisfaction with products. The researcher should design each scale to be a unitary measuring instrument rather than an opinion questionnaire. There are numerous types of popular scales that are mentioned by Coolican (2004): equal appearing intervals; the semantic differential; and summated rating.

Thurstone (1931) introduced *"equal appearing intervals"* on the scale a score equivalent to the strength of each statement that a participant agrees with is given. The researcher has to go through the following steps in order to structure this scale:

1. Present a number of both positive and negative statements towards the attitude object.
2. Ask a group of judges to rate the statements ranging from 1 (highly negative) to 11 (highly positive).
3. Find scale values by obtaining the mean value of all the ratings for each statement.
4. Reject statements which judges rated in a different way.
5. The overall attitude score is the total of all scale values on items participants agreed with.

However, this scale is still problematic. The judges cannot be utterly natural. It is also quite difficult to select most discriminating statements from the items which have the same scale value.

Osgood, Suci and Tannenbaum (1957) introduced *"the semantic differential scale"*, which can be used to measure the connotative meaning of an object for an individual. For example, a participant is asked to mark a scale between bi-polar adjectives according to their feeling where the object holds on the scale as given below:

"Doctor"

good _ _ _ _ _ _ _ bad

weak _ _ _ _ _ _ strong

active _ _ _ _ _ _ passive

Furthermore, they state that all bi-polar pairs can be attached to the next three general meaning factors: 'active' (along with 'slow-fast', 'hot-cold') is an example of the activity factor; 'strong' (along with 'rugged-delicate', 'thick-thin') is an example of potency factor; and 'good' (along with 'clean-dirty', 'pleasant-unpleasant') is an example of the estimative factor.

The semantic differential scale produces good reliability values and correlates well with other attitude scales. However, this is also again problematic when participants have a tendency towards a 'position response bias' where they usually rate at the extreme end of the scale or not using the extreme at all without considering possible weaker or stronger responses.

Likert (1932) introduced *"the summated rating"* that needs to following steps in order to structure in a scale:

1. This is similar to the proposed Thurstone's (1931) scale; present a set of favourable and unfavourable statements about an attitude object.

2. Then ask participants to provide their response to each statement using a scale ranging between strongly disagree to strongly agree. For example, "A mobile game is a good provider of learning" can be scaled as follows:

| 1 | 2 | 3 | 4 | 5 |
|---|---|---|---|---|
| Strongly disagree | Disagree | Neutral | Agree | Strongly agree |

Scales typically range from 1 to 3 points, to a maximum of 1 to 9 points, but it is usually agreed that obtaining the middle ground, by using scales of 1 to 5, or 1 to 7, is the most effective method (Dix, et al., 2004). For the work reported in this book, it was therefore decided to use a scale of 1 to 5 as the above example shows. Each value of the scale can be used as a score for the individual item of each respondent. For example, '5' will be the score for strongly agree with the satisfactory item while it will be "1" for strongly disagree with an unsatisfactory item. Finally, adding the scores of each item together will provide the overall attitude score.

It can argue that the score "3" is not very clear, because it is known "undecided". It might correspond to no opinion or an on-the-fence opinion and therefore the central value (in this case "3") of the scale is quite unclear. For example, 50 out of 100 would be "undecided" in an overall scale ranging from "strongly for" and "strongly against" responses. However, Coolican (2004) stated that a number of advantages of using the Likert scale technique: it enables to complete and maintain respondents' direct involvement; it has been proven to have a higher degree of validity and reliability; and it is effective at measuring changes over time. Furthermore, he describes that attitude scales are highly structured measures which respondents provide the most appropriate response. Therefore, a Likert scale was used in this research work as its advantages stated by Coolican (2004).

### 3.4.2 Think-aloud Approach

The thinking aloud is an approach, which has traditionally been used as a psychological research method (Ericsson and Simon, 1984; Neilsen, 1993). However, 'usability king' Neilson Jacobson has stated this approach is being heavily used by evaluating human computer interfaces (Neilsen, 1993). Furthermore, he describes that conducting a think-aloud experiment with real participants is one of the most fundamental evaluation approaches in software applications testing. The testing process involves participants using the system or prototype to complete a predetermined set of tasks while being observed and verbalised their thoughts and comments. The observations, thoughts and comments are then analysed to address the research question.

There is a large volume of published studies that critically discuss the think-aloud approach (Dix, et al., 2004; Hertzum and Jacobsen, 2001; Neilsen, 1993; Nielsen, Clemmensen and Yssing, 2002; Nørgaard and Hornbaek, 2006; Preece, Rogers and Sharp, 2002; Ramey, et al., 2006). Ramey, et al. (2006) describes that think-aloud is the most important approach in system evaluation, because it discovers more problems than any other existing measure. Perhaps, this is the main reason that think- aloud is more popular as frequently applied techniques among the academic researchers and industrial expertise (Nielsen, Clemmensen and Yssing, 2002).

The most significant benefit of the think-aloud approach is that it enables to design the system from actual users, who are representatives of general population. Participants provide the researcher an understanding of their views of using the system based on verbalising their thoughts. This enables the researcher to recognize participants' major confusions (Neilsen, 1993). The think-aloud process is used for researchers to gain participants' behaviour through an examination. This examination analysis can reveal the causes of system problems. In addition, participants' voice and behaviour provides insight into their effective reactions such as frowns, sighs and scowls which convey dissatisfaction, frustrations or maybe happiness. Therefore, observational data gathered through a think-aloud approach is quite beneficial and powerful to design a system.

The think-aloud method is flexible so that it provides a vital benefit for system evaluators. Some researchers realise that the think-aloud approach can affect the type and amount of data collected. Therefore, they may attempt to manipulate the think-aloud session to make sure the exploration of research areas to match the objectives of the experiment (Ramey, et al., 2006). This is a significant benefit of the approach, when concerning on certain aspects of an interface such as designing a game interface for learning. In addition, this approach can help identify the suspected areas to be problematic or to uncover reasoning for problems.

Despite these advantages of the think-aloud approach there are some bottlenecks, which are worth concerning when evaluating this technique. Ramey, et al. (2006) describe the validity and reliability of the think-aloud approach has been overlooked due to its usefulness in convincing researchers that problems exist. However, it would have been important to concern the lack of soundness when investigating the effectiveness of this approach.

Dix, et al. (2004) stated that the usefulness of the think-aloud approach is influenced by the effectiveness of observation and subsequent analysis. This depends on researchers who facilitate to conduct think-aloud sessions. As suggested by Preece, Rogers and Sharp (2002), the system evaluation though users is strongly controlled by the researcher, who is in charge of the test. On the one hand, this is an advantage for practitioners to ensure goals of the systems evaluation. However, on the other hand, the think-aloud approach does not provide necessary guidelines for researchers to perform reliable analysis (Hertzum and Jacobsen, 2001). This is due to a combination of unclear goal analysis, unclear evaluation procedures and unclear problem criteria. In addition, Norgaard and Hornbaek (2006) argued that improper setting up of think-aloud sessions and incomplete analysis of results could generate research findings, which may not represent the actual user experience. These issues can enhance the possibility of inconsistency in research findings from different researchers.

The interruptive role of the researcher may cause for inconsistency in the think-aloud approach, which leads to the lack of participants' concentration (Preece, Rogers and Sharp, 2002 and Ramey, et al., 2006). Perhaps the researcher forces participants to spend more or less time on a certain aspect of system evaluation, which could mislead to represent participants' natural use of advices. Norgaard and Hornbaek (2006) stated that the researcher vary the way they conduct the think aloud sessions because they desire participants to confirm known issues which they may have foreseen or already known problems.

The laboratory environment where think aloud sessions take place is unnatural which may affect participants' behaviour. Participants interact with the system during the think aloud test while verbalising their comments in a completely unnatural environment whilst being observed. Therefore, they tempt to comment based on their experience while completing tasks in the laboratory environment. This may increase their cognitive load, which impacts their problem solving behaviour (Nielsen, Clemmensen and Yssing, 2002; Preece, Rogers and Sharp, 2002; Ramey, et al., 2006). In addition, participants are not allowed to take phone calls, check emails, talk to their friends, or occupy with any task, which they may regularly do in their normal life (Preece, Rogers and Sharp, 2002). Furthermore, observation in the laboratory environment is also an obstructive approach where participants are usually aware that they are being monitored by the researcher (Dix, et al., 2004).

The results of think aloud sessions are limited to what they truly represent in terms of the user population to test the system. Conducting think aloud experiment with a large sample size is time and resources consuming, therefore, the findings of such evaluation typically reflect on a few users (Hong and Landay, 2001). The literature revealed that ongoing disagreements around the sample population size to conduct a think aloud test. Generally between five to fifteen participants are advised. However, 'usability king' Nielson suggests that five participants are enough to find major problems (Neilsen, 1993).

In summary, Dix, et al. (2004) stressed the simplicity of the think-aloud approach. This technique is described as a "quick and dirty" approach gaining a huge number of results quickly via using a simple methodology (Preece, Rogers and Sharp, 2002; Ramey, et al., 2006). Nielson (1993) stated that the wealth of qualitative data which may collect from a relatively small number of participants taking part. Although, qualitative data is known valuable, this research study compares findings gained from such data, with findings derived from quantitative data reported in Chapter 4 of this book. In addition, the idea of strengthening findings of this book by combining finding from methods which independently collect both quantitative (questionnaire) and qualitative (think-aloud) data will be reported.

## 3.5 Data Analysis

The study reported in chapter 4, empirically investigated what key elements that should be addressed in the game design framework for computer users to thwart phishing attacks. A theoretical model therefore derived from Technology Threat Avoidance Theory (TTAT) was used in the game design framework. Because the model examined how users avoid malicious IT threats using a given safeguarding measure. A quantitative analysis, based on Likert style questionnaire, approach was adopted to evaluate the game design framework described in this study. Then the study instrumented SPSS, which is a statistical software package for data analysis was developed by IBM (Davis, Bagozzi and Warshaw, 1983). It is much easier to use and the most commonly used statistical techniques.

The statistical tests applied to the data must ensure whether or not the data can be analysed during the planning stage of the study. This will allow the hypobook to be either supported or rejected (Breakwell, Hammond and Fife-Schaw, 1995). As the first step of this study, Cronbach's alpha, which is known as a coefficient alpha was calculated to measure the reliability of the questionnaire (Pallant, 2007). In addition, the Kaiser-Meyer-Olkin (KMO) value measure was used to assess the adequacy of the sample size. Then the study employed a multiple regression analysis to test the theoretical model derived from TTAT was used in the game design framework with the following parameters: phishing attack and anti-phishing education as a malicious IT threat and safeguarding measure respectively. The multiple regression analysis is based on correlation, but it allows more sophisticated investigation of the interrelationship among a set of variables (Davis, Bagozzi and Warshaw, 1983). Furthermore, it enables to discover how well a set of variables is able to predict a particular outcome. Bryman and Cramer (2005) describes that one of the most vital explanations of the relationship between variables is the correlation. The degree of correlation between variables indicates the strength, significance and the direction (positively or negatively) of the relationship. The study reported in chapter 6 focused on evaluating the game design framework through the developed mobile game prototype using MIT App Inventor Emulator. Although, the proposed game design framework through an empirical investigation reported in chapter 4, it would have been more beneficial to determine users' impact on the framework after their engagement with the mobile game prototype. The lack of user satisfaction causes difficulties for them to interact with the system (Pinelle, Wong and Stach, 2008), which may hinder their learning take place. Therefore, the study engaged a usability study using Systems Usability Scale (SUS), which is a quantitative analysis based on Likert style questionnaire, as the first step to evaluate the user subjective satisfaction of the mobile game prototype interface. The SUS is a simple, but yet a reliable questionnaire consisting of 10 items that use to evaluate the subjective perception of an individual to interact with the software system without considering the personal sense of taste (Brooke, 1996; Finstad, 2006). Tullis and Stetson (2004) have revealed that SUS yield among the most reliable results across sample sizes compared to other questionnaires such as Questionnaire for User Interface Satisfaction (QUIS) and Computer System Usability Questionnaire (CSQU) (Finstad, 2006). Their findings concluded that, sample sizes of at least 12-14 participants are required to get reasonably reliable results. The SUS uses a five-point Likert scale with anchors for strongly agree and strongly disagree. Therefore, data gathered using SUS to evaluate the user subjective satisfaction of the mobile game prototype interface was analysed by using SPSS version 18.

Then a think-aloud study was employed along with a pre and post-test to evaluate the game design framework. This study attempted to evaluate participants' understanding and awareness of phishing attacks through the mobile game prototype. Furthermore, it emphasises what key elements that should be addressed in the game design framework. Therefore, an experimental protocol was designed including user study instructions for participants to follow up the think-aloud study. According to Norgaard and Hornbaek's (2006) study, the data analysis was conducted in two phases. As the first step, the study segmented the recordings applying keywords to each segment. Then the study attempted to analyse and try to form a coherent interpretation of segments that shared keywords as the second step.

## 3.6 Summary

This methodology chapter has discussed the general methodologies and techniques used for the work conducted of this book. First, an overview of the research question was presented. Then, the general research approach was discussed along with a justification of the selection approach. After that, a comprehensive explanation of the data collection techniques and procedures was given. Finally, an explanation of the data analysis process and the tests that have been conducted to draw conclusions from the studies reported in this book were presented.

# Chapter 4

# A Game Design Framework to Avoid "Phishing Attacks"

## 4.1 Introduction

The design of games is a double-edged sword. When its power is properly harnessed to serve good purposes, it has tremendous potential to improve human performance. However, when it is exploited for violation purposes, it can pose huge threats to individuals and society. Therefore, the design of educational games is not an easy task and there are no all-purpose solutions (Moreno-Ger, et al., 2008). The notion that game based education offers the opportunity to embed learning in a natural environment, has repeatedly emerged in the research literature (Andrews, et al., 2003; Facer, et al., 2004; Gee, 2003; Roschelle and Pea, 2002; Wegerif, 2003; Walls, 2012).

The premise behind this study is to develop a game design framework, which enhances user avoidance behaviour through motivation to protect themselves against phishing attacks. A theoretical model derived from Technology Threat Avoidance Theory (TTAT) was used to develop the game design framework, which is shown in Figure 4.1 (Liang and Xue, 2010). The TTAT describes individual IT users' behaviour of avoiding the threat of malicious information technologies such as phishing attacks (Liang and Xue, 2009). The model examines how individuals avoid malicious IT threats by using a given safeguarding measure. The safeguarding measure does not necessarily have to be an IT source such as anti-phishing tools; rather it could be behaviour such as anti-phishing education (Liang and Xue, 2010).

Consistent with TTAT (Liang and Xue, 2009), the user's IT threat avoidance behaviour is determined by avoidance motivation, which, in turn, is effected by perceived threat. Perceived threat is also influenced by the interaction of perceived severity and susceptibility. User's avoidance motivation is also determined by three constructs such as safeguard effectiveness, safeguard cost and self-efficacy.

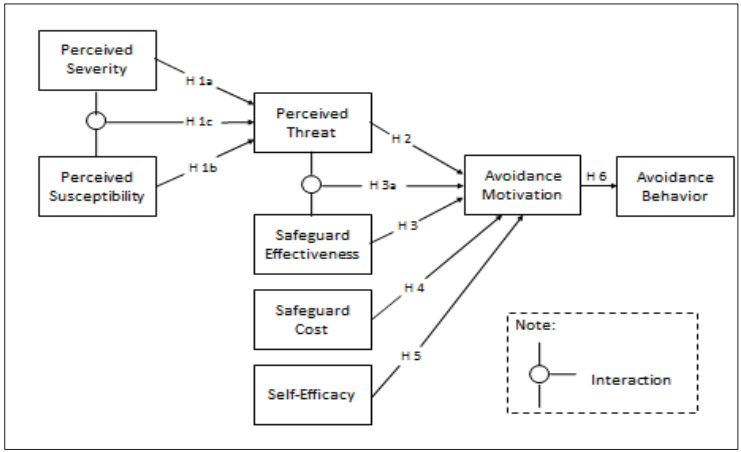

**Figure 4.1: Research model derived from TTAT (Liang and Xue, 2010)**

Safeguard effectiveness is described as the individual assessment of a safeguarding measure regarding how effectively it can be applied to avoid the malicious IT threat (Liang and Xue, 2010). For example, the individual assessment regarding how effectively anti-phishing education can be applied to avoid a phishing attack. Safeguard cost is a payback for safeguard effectiveness. This refers to the physical and cognitive efforts such as time, money, inconvenience and comprehension required using the safeguard measure (Liang and Xue, 2009). Self-efficacy is defined as individuals' confidence in taking the safeguard measure. This is an important determinant of avoidance motivation. Previous research has revealed that individuals are more motivated to perform IT security related behaviours as the level of their self- efficacy increases (Kaiser, 1974; Ng, Kankanhalli and Xu, 2009; Woon, Tan and Low, 2005). In addition, the research model posits that avoidance motivation is influenced by an interaction between perceived threat and safeguard effectiveness.

The TTAT identifies the issues that the game design framework needs to address. The proposed game design framework attempts to develop threat perceptions such that individuals will be more motivated to avoid phishing attacks and use safeguarding measures. A key aspect of this is that they realise the effectiveness of safeguarding measures, lower safeguard costs, and increase self-efficacy.

Sheng, et al. (2010) have conducted a role-play survey with 1001 online survey respondents to study who falls for phishing attacks. The study revealed that women are more susceptible than men to phishing and participants between the ages of 18 and 25 are more susceptible to phishing than other age groups. Participants were included from a diverse group of staff and student, including people who were concerned about computer security. The current study was targeted towards to examine participants' phishing threat avoidance behaviour by using anti-phishing education. Therefore, the survey was only administered to participants' ages ranged from 18 to 25 those who had not already completed the questionnaire before.

## 4.2 Aims and Objectives

This research study is the first step in the development of a game design framework to enhance user avoidance behaviour through motivation to thwart phishing attacks. The aim of this study is to investigate what are the key elements that should be addressed in the game design framework to avoid phishing attacks.

The objectives are as follows:
- To identify the key elements that should be addressed in the game design framework to avoid phishing attacks.
- To evaluate the game design framework using phishing attack (malicious IT threat) and game based anti-phishing education (safeguarding measure).
- To formulate a game design framework to thwart phishing attacks.

## 4.3 Pilot Study

A pilot study is a rehearsal, which is conducted before the main study takes place (Milne, Orbell and Sheeran, 2002; Compeau and Higgins, 1995; Sonderegger and Sauer, 2010). It helps the researcher to determine whether or not the study is appropriate in terms of reliability and validity. If any problems are encountered during the pilot study, adjustments are made before the main study. A quantitative analysis, based on Likert style questionnaire, approach was adopted to evaluate the game design framework described in this study.

### 4.3.1 Questionnaire Design

The questionnaire was constructed based on Liang and Xue's theoretical model and relevant research literature (Liang and Xue, 2009; Liang and Xue, 2010; Rosenstock, 1974; Saleeby, 2000; Smith, Milberg and Burke, 1996; Champion and Scott, 1997; Higgins, 1995; Davis, 1989; Davis, Bagozzi and Warshaw, 1983). Perceived threat was measured on the basis of substantive meaning (Rosenstock, 1974). The questionnaire items related to this aspect assessed respondents' perception of the likely harm, danger, peril or damage that a phishing attack imposes. Perceived susceptibility was developed based on health behaviour research (Saleeby, 2000); and was used to evaluate the likelihood and possibility of a phishing attack's occurrence.

TTAT speculates that computer users' well-being includes two dimensions: computer performance and information privacy. However, Liang and Xue (2009) argued that a malicious IT attack could damage both dimensions. Therefore, severity perception of computer users should relate to the two dimensions. Perceived severity was measured by the number of items based on the privacy literature in IS (Smith, Milberg and Burke, 1996) and practitioner research that report the negative impact of phishing attacks (Brody, Mulig and Kimball, 2007; Dhamija, Tygar and Hearst, 2006; Downs, Holbrook and Cranor, 2006a, 2007b; Grinter, et al., 2006; Jagatic, et al., 2007; Miller and Garfinkel, 2005; Schneier, 2000;). The items developed in their research were based on users' concerns about both the loss of personal and confidential information and degraded computer performance related to processing speed, Internet connection, and software applications.

The items of safeguard effectiveness were developed based on relevant health behaviour research (Downs, Holbrook and Cranor, 2007; Saleeby, 2000). For example, a number of items in this subscale were derived for safeguard cost based on Milne, Orbell and Sheeran and Saleeby's studies (Champion and Scott, 1997; Saleeby, 2000). Self-efficacy was measured with items developed by Compeau and Higgins (1995), making minor amendments to adapt it to the anti-phishing education context. The number of items developed for avoidance motivation was based on the behavioural intention measures from technology adoption research (Davis, 1989; Compeauand Higgins, 1995), with a focus on threat avoidance rather than IT adoption. Finally, threat avoidance was measured with three self-developed items.

Therefore, the pilot study questionnaire contained four items for perceived threat, four items for perceived severity, three items for perceived susceptibility, four items for safeguard effectiveness, three items for safeguard cost, six items for self-efficacy, three items for avoidance motivation, and three items for avoidance behaviour. In total 30 items were evaluated using a five-point scale Likert at 1= 'Strongly disagree' and 5= 'Strongly agree'. A sample set of questionnaire is shown in Table 4.1. A complete set of questionnaire is shown in Appendix 1.

**Table 4.1: A sample set of questionnaire**

| 1. Perceived Susceptibility | Strongly disagree 1 | Disagree 2 | Neutral 3 | Agree 4 | Strongly agree 5 |
|---|---|---|---|---|---|
| It is extremely likely that my computer will be infected by a phishing attack in the future | ☐ | ☐ | ☐ | ☐ | ☐ |
| My chances of getting phishing attacks are great. | ☐ | ☐ | ☐ | ☐ | ☐ |
| I feel phishing attack will not infect my computer in the future | ☐ | ☐ | ☐ | ☐ | ☐ |
| 2. Perceived Severity | Strongly disagree 1 | Disagree 2 | Neutral 3 | Agree 4 | Strongly agree 5 |
| A phishing attack would steal my personal information from my computer without my knowledge | ☐ | ☐ | ☐ | ☐ | ☐ |
| Phishing attack would invade my privacy | ☐ | ☐ | ☐ | ☐ | ☐ |
| I feel phishing attack would not steal my personal information from my computer without my knowledge | ☐ | ☐ | ☐ | ☐ | ☐ |
| I feel phishing attack would not invade my privacy | ☐ | ☐ | ☐ | ☐ | ☐ |
| 3. Perceived Threat | Strongly disagree 1 | Disagree 2 | Neutral 3 | Agree 4 | Strongly agree 5 |
| Phishing attacks pose a threat to me | | | | | |
| A phishing attack is a danger to my computer | ☐ | ☐ | ☐ | ☐ | ☐ |
| It is risky to use my computer if it being phishing attacked | ☐ | ☐ | ☐ | ☐ | ☐ |
| I feel a phishing attack will not cause any harm to my computer | ☐ | ☐ | ☐ | ☐ | ☐ |

### 4.3.2 Participants

A pilot study questionnaire survey was run with sixteen first year undergraduate students from the Department of Information Systems and Computing, Brunel University, London. Participants were invited by walking to the Data and Information practical lab session asking for their help. A summary of the demographics of the participants in the pilot study is shown in Table 4.2.

**Table 4.2: Participant demographics in the pilot study**

| Characteristics | Amount |
|---|---|
| Sample Size | 16 |
| Gender | |
|       Male | 10 |
|       Female | 6 |
| Age range (18 - 25) | 16 |
| Average hours spent per week on the Internet | |
|       0-5 | 0% |
|       6-10 | 19% |
|       11-15 | 12% |
|       16-20 | 19% |
|       20+ | 50% |

### 4.3.3 Procedure

The pilot study questionnaire survey was conducted in-person. First, participants were asked them to go through model consent form (see Appendix 1). Then the individual participants were asked whether or not they knew what the term "Phishing Attack" means. Those who gave a positive response were asked to give a short verbal description to confirm their understanding, whilst negative responders were read a brief definition of phishing attack and given a short verbal description. Then participants were asked to complete the questionnaire items, which consist of eight constructs such as perceived severity, perceived susceptibility, perceived threat, safeguard effectiveness, safeguard cost, self-efficacy, avoidance motivation and avoidance behaviour. They were already informed that they are of course welcome to clarify them any questions when difficulty of understanding. This strategy was used to determine the participants' readability and understanding of questionnaire items during the pilot study. Therefore, the corresponding adjustments were addressed before the main study.

### 4.3.4 Results

The pilot study instrumented SPSS, which is a statistical software package for data analysis; was developed by IBM (Davis, Bagozzi and Warshaw, 1983). Cronbach's alpha, which is known as a coefficient alpha was used to measure the internal consistency of the questionnaire (Pallant, 2007). Previous research has indicated that an alpha score that is greater than 7.0 indicates that there is a good level of internal scale consistency (Pallant, 2007; Cronbach, 1951; Zaharias and Poylymenakou, 2009). Therefore, Cronbach's alpha was calculated for each construct of the questionnaire and is summarised in Table 4.3.

**Table 4.3: Cronbach's alpha scores for the questionnaire constructs in the pilot study**

| Constructs | Cronbach's alpha(>0.70) |
|---|---|
| Perceived Susceptibility | 0.716 |
| Perceived Severity | 0.869 |
| Perceived Threat | 0.770 |
| Perceived Safeguard Effectiveness | 0.904 |
| Perceived Safeguard Cost | 0.938 |
| Self-Efficacy | 0.798 |
| Avoidance Motivation | 0.751 |
| Avoidance Behaviour | 0.880 |

### 4.3.5 Summary

Based on the feedback obtained from the wording of some measurement items of each construct was slightly revised. The final questionnaire contained four items for perceived threat, four items for perceived severity, three items for perceived susceptibility, four items for safeguard effectiveness, three items for safeguard cost, six items for self-efficacy, three items for avoidance motivation, and three items for avoidance behaviour. Therefore, total 30 items were used in the main study to measure 8 constructs in the research model using a five-point scale Likert at 1= 'Strongly disagree' and 5= 'Strongly agree' (see Appendix 2).

### 4.4 Main Study

The same questionnaire employed in the pilot study was used in the main study. The aim of the survey was to identify the key elements that should be addressed in the game design framework. Therefore, the 30 items were used to measure the 8 constructs of Liang and Xue's theoretical model using a five-point scale Likert at 1= 'Strongly disagree' and 5= 'Strongly agree'. Pallant (2007) argued that the sample size calculation is approximately equal to Likert scale (in this case 5) into the number of items (in this case 30) measure. Therefore, the main study survey was conducted with 151 participants.

### 4.4.1 Participants

The questionnaire was administrated to 151 participants, who were undergraduate students from Brunel University and Bedfordshire University. Participants' ages ranged from 18 to 25, with a gender split of 67 percent male and 33 percent female. They had average of 16 -20 hours per week of Internet experience (SD=1.19). Each participant took part in the survey on a fully voluntary basis. A summary of the demographics of the participants in the main study is shown in Table 4.4.

**Table 4.4: Participant demographics in the main study**

| Characteristics | Amount |
|---|---|
| Sample Size | 151 |
| Gender | |
| Male | 101 |
| Female | 50 |
| Age range (18 - 25) | 151 |
| Average hours spent per week on the Internet | |
| 0-5 | 3% |
| 6-10 | 12% |
| 11-15 | 14% |
| 16-20 | 14% |
| 20+ | 57% |

## 4.4.2 Procedure

The questionnaire was handed out to participants' in-person by the researcher (Appendix 2). First, the nature of the research was explained to each participant individually and they were given an informed consent form to read and sign (Appendix 1). They were also told that they were free to withdraw from the study at any time without having to give a reason for withdrawing. Then the individual participants were asked whether or not they knew what the term "Phishing Attack" means. Those who gave a positive response were asked to give a short verbal description to confirm their understanding, whilst negative respondents were read a brief definition of a phishing attack and also given a short verbal description. Then participants were asked to complete the questionnaire, which measured the eight constructs; perceived severity, perceived susceptibility, perceived threat, safeguard effectiveness, safeguard cost, self-efficacy, avoidance motivation and avoidance behaviour. After completing the questionnaire, participants were thanked for their valuable time and effort in taking part in the study.

### 4.4.3 Results

As in the pilot study, Cronbach's alpha was calculated for each construct to measure the internal consistency of the questionnaire items. The results of this analysis are summarised in Table 4.5. Previous research has been shown the minimum level of Cronbach's alpha is 0.7 to be internally consistent of a set of items as a group (Cronbach, 1951; Pallant, 2007; Zaharias and Poylymenakou, 2009).

**Table 4.5: Cronbach's alpha scores for the questionnaire constructs in the main study**

| Constructs | Cronbach's alpha (>0.70) |
|---|---|
| Perceived Susceptibility | 0.730 |
| Perceived Severity | 0.766 |
| Perceived Threat | 0.701 |
| Perceived Safeguard Effectiveness | 0.803 |
| Perceived Safeguard Cost | 0.805 |
| Self-Efficacy | 0.714 |
| Avoidance Motivation | 0.753 |
| Avoidance Behaviour | 0.762 |

In addition, the Kaiser-Meyer-Olkin (KMO) value measure was used to assess the adequacy of the sample and the KMO value should be greater than 0.6 for a satisfactory analysis to proceed (Cronbach and Meehl, 1955). For the sample used in this study the KMO = 0.718.

### 4.4.4 Model Testing

The study employed a multiple regression analysis to test the Liang and Xue's theoretical model using the following parameters: phishing attack and anti-phishing education as a malicious IT threat and safeguarding measure respectively. This is because to address the research question in this study, which is, what key elements should be addressed in the game design framework to enhance avoidance behaviour through motivation for computer users to thwart phishing attacks. Multiple regressions analysis is not only just one statistical technique, but also a family of techniques that can be used to explore the relationship between one continuous dependent variable and a number of independent variables or predictors (Davis, Bagozzi and Warshaw, 1983). Multiple regression analysis is based on correlation, but permits a more sophisticated investigation of the interrelationship among a set of variables. Therefore, this statistical technique makes it ideal for investigating what elements interrelated to enhance avoidance behaviour through motivation for computer users to thwart phishing attacks in the game design context. Moreover, it assists to identify the best predictor elements from the theoretical model that enhance avoidance behaviour through motivation in the game design framework. The model testing results are shown in Figure 4.2.

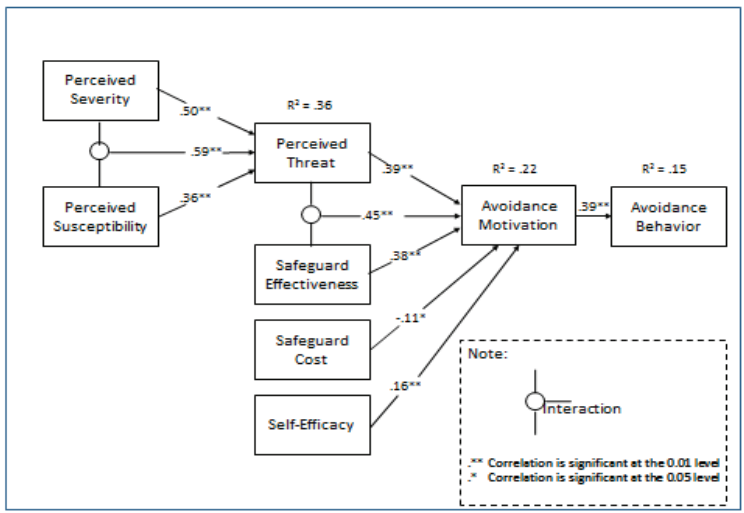

**Figure 4.2: Model testing results**

The model calculated **R square** value for perceived threat, avoidance motivation and avoidance behaviour, which was defined as how much of variance in the dependent variable is explained by its independent variables in the model (Davis, Bagozzi and Warshaw, 1983). In the results for the model in this study 36 percent of variance is explained in perceived threat, 22 percent of variance in avoidance motivation, and 15 percent of variance in avoidance behaviour. Pearson correlation analysis was then employed to describe the strength and direction of the linear relationship between two constructs. The results indicate that perceived threat is significantly determined by perceived severity ($r = .499^{**}$, and Sig. $= .000$) and perceived susceptibility ($r = .357^{**}$, and Sig. $= .000$). Avoidance motivation is significantly determined by perceived threat ($r = .386^{**}$, and Sig. $= .000$). According to Liang and Xue's and Baron and Kenny's research, these results show that the influences of perceived susceptibility and severity on avoidance motivation are fully mediated by perceived threat.

As Figure 4.2 shows, avoidance motivation is also significantly determined by safeguard effectiveness ($r = .381^{**}$, and Sig. $= .000$), self-efficacy ($r = .162^*$, Sig. $= .047$) and Safeguard cost ($r = -.112^*$, Sig. $= .037$). Finally, avoidance motivation is found to be significantly influenced by avoidance behaviour ($r = .390^{**}$, and Sig. $= .000$).

To evaluate the interaction effects of both perceived susceptibility and severity and perceived threat and safeguard effectiveness, Chin, Marcolin and Newsted's (2003) product-indicator approach was used. Interaction variables were created by cross multiplying the items of perceived susceptibility and severity and perceived threat and safeguard effectiveness (Liang and Xue, 2010). As Figure 4.2 shows, the interaction between perceived severity and susceptibility was statistically significant on perceived threat ($r = .588^{**}$, and Sig. $= .000$). Finally, the interaction between perceived threat and safeguard effectiveness was statistically significant on avoidance motivation ($r = .452^{**}$, and Sig. $= .000$).

In summary, the model testing results provided support to all of the hypotheses. Moreover, age, gender and Internet experiences were included as control variables on avoidance motivation and avoidance behaviour in the model testing. However, none of these control variables was found to have a statistically significant effect on either avoidance motivation or avoidance behaviour. This is similar to the finding of Liang and Xue's (2010) empirical study.

## 4.4.5 Game Design Framework

This study empirically investigated what key elements that should be addressed in the game design framework for computer users to avoid phishing attacks through motivation. The elements of a theoretical model derived from TTAT were used to address in the game design framework. Figure 4.2 shows the model testing results. The model accounts for 36 percent of variance in perceived threat, 22 percent of variance in avoidance motivation, and 15 percent of variance in avoidance behaviour. Perceived threat is significantly determined by perceived severity (r = .499**, and Sig. = .000), perceived susceptibility (r = .357**, and Sig. = .000) and their interaction (r = .588**, and Sig. = .000). Therefore, perceived severity and perceived susceptibility elements addressed in the game design framework for computer users to thwart phishing attacks. As Figure 4.2 shows, avoidance motivation is significantly determined by perceived threat (r = .386**, and Sig. = .000), safeguard effectiveness (r = .381**, and Sig. = .000), and safeguard cost (r = -.112*, Sig. = .037), and self-efficacy (r = .162*, Sig. = .047). However, it is interesting to note that safeguard cost negatively effects avoidance motivation though it is significantly determined by avoidance motivation. This is because the user's motivation to avoid the IT threat is expected to be reduced by the potential cost of using the safeguard measure (Liang and Xue, 2010). Therefore, perceived threat, safeguard effectiveness, safeguard cost and self-efficacy elements should be addressed in the game design framework. Finally, avoidance motivation is found significantly influence avoidance behaviour (r = .390**, and Sig. = .000).

In summary, this study results provided support to determine what elements that should be addressed in the game design framework for computer users to avoid phishing attacks through motivation. Therefore, perceived threat, safeguard effectiveness, safeguard cost, self-efficacy, perceived threat and perceived susceptibility elements were addressed in the game design framework. The game design framework is shown in Figure 4.3.

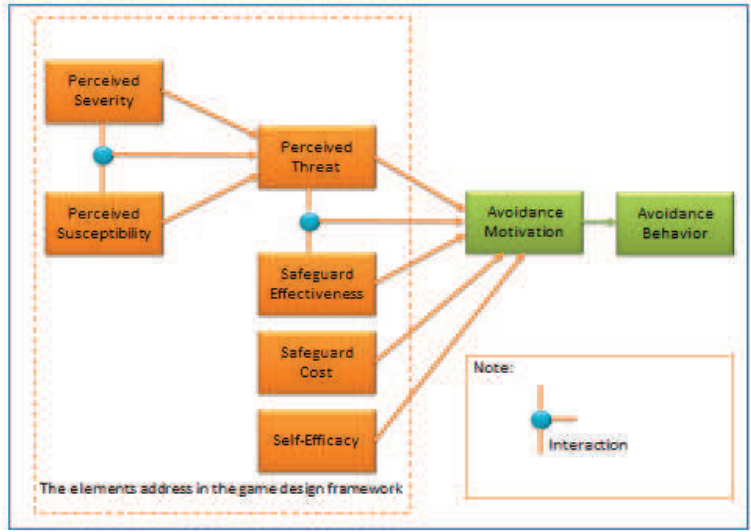

**Figure 4.3: Game Design Framework (Arachchilage and Love, 2013)**

## 4.4.6 Discussion

This study empirically investigated a game design framework for computer users to thwart phishing attacks. Therefore, phishing attack and anti-phishing education were considered as a malicious IT threat and safeguarding measure respectively in order to test a theoretical model derived from TTAT (Liang and Xue, 2010). The study paid particular attention to threat perception because it plays a vital role in influencing computer users' avoidance behaviour. Data analysis results showed in the Figure 4.2, the model is able to explain a considerable amount of variance in users' motivation to avoid IT threats (22 percent) and actual avoidance behaviour (15 percent). Therefore, this study conveys a simple, yet powerful message to motivate computer users to avoid malicious IT threats.

However, it is interesting to note that the amount of variance in users' avoidance behaviour is quite low thus it is significant (Pallant, 2007). There is a possible explanation for this result. When users decide that the IT threat can be avoided by using the safeguarding measure, they may take a problem-focused coping approach. However, when the IT threat could not be avoided completely, they may take an emotion-focused coping approach (Liang and Xue, 2010; Liang and Xue, 2009; Rhoa and Yub, 2011). Lazarus and Folkman (1984) asserted two types of coping could be performed to deal with the threat; problem-focused and emotion-focused. Problem-focused coping referred to adaptive behaviors that take a problem-solving approach. It directly deals with the malicious IT threat by taking a safeguarding measure such as updating password regularly, disabling cookies, and installing and configuring safeguarding IT. When people face the problem as a challenge, they seem to take a problem-oriented coping behavior and treat the problem as a thing that can be controlled. In contrast, emotion-focused coping, the problem identified as a threat and loss, people tend to perceive it as a thing cannot be solved by them and hence, take an emotional coping behavior. Beaudry and Pinsonneaut (2001) stated that if users perceive the malicious IT threat, they take problem-focused coping or if they believe that the threat is not avoidable, they will inactively avoid the threat by performing emotion-focused coping. Therefore, it can be argued in the current study, that users' emotion-focused coping behaviour would have caused for avoidance behaviour of phishing threat, which will account for the variance of avoidance behaviour.

Computer users have to be convinced and feel that such malicious IT threats exist in the cyberspace and are avoidable. The study found some evidence in the data analysis results that the model is able to explain a respectable amount of variance in threat perception (36 percent). This figure is little higher than Liang and Xue's (2010) empirical study, which is 33 percent. Therefore, perceived threat element is significantly important to address in the game design framework for computer users to enhance avoidance behaviour through motivation to thwart phishing attacks. Furthermore, the study demonstrates threat perception that users need to be aware of likelihood and severity of being attacked by malicious IT threat. If users actually perceive the threat, they are more motivated to avoid it. The safeguarding measure was evaluated from three aspects; taking into account safeguard effectiveness, cost related to safeguard measure and users confident of using the safeguard. If the level of effectiveness of the safeguarding measure is high then users are more motivated to avoid threats. So, the safeguard effectiveness element is important in the game design framework for computer users to thwart phishing threats. Users' high confidence in taking the safeguard measures influences their motivation to avoid threats. Therefore, self-efficacy should also be included in the game design framework for avoiding threats through motivation.

When the safeguard cost is high, users are less motivated to avoid threats. Liang and Xue (2009 and 2010) describes when time, money, inconvenience and comprehension needed to use the safeguarding measure is high, users are less motivated to avoid threats. The current study results also demonstrated safeguard cost negatively affects avoidance motivation. Therefore, safeguard cost should address in the game design framework, as a payback to safeguarding effectiveness. Liang and Xue's (2010) model testing results did not support the interaction between perceived severity and susceptibility on perceived threat. Surprisingly, this study revealed that perceived threat is significantly determined by the interaction between perceived severity and susceptibility ($r = .588**$, and Sig. = .000).

Moreover, this study emphasises that avoidance motivation is significantly determined by the interaction of perceived threat and safeguarding measure ($r =$ $.452**$, and Sig. = .000). This result contradicts with Liang and Xue's (2010) findings regarding the interaction between perceived threat and safeguard effectiveness on avoidance motivation. However, they suggest the interaction between perceived threat and safeguard effectiveness can be viewed from two perspectives. First, when the threat level is high, perceived threat can be viewed to negatively moderate the relationship between safeguard effectiveness and avoidance motivation. Second, when the level of safeguard effectiveness is high, it can be viewed to negatively moderate the relationship between perceived threat and avoidance motivation. However, this study does provide evidence to address the interaction effect of perceived threat and safeguard effectiveness in the game design framework.

## 4.5 Summary

This study attempted to develop a game design framework, which enhances computer users' avoidance behaviour through motivation to prevent themselves from phishing attacks. The study empirically investigated what elements should address in the game design framework for computer users to thwart phishing attacks. A theoretical model derived from TTAT was used develop the game design framework. To test the model, phishing attack and anti-phishing education were considered as a malicious IT threat and safeguarding measure respectively.

Finally, the current study results provided support to define what are the key elements that should be addressed in the game design framework for computer users to thwart phishing attacks. Therefore, perceived threat, safeguard effectiveness, safeguard cost, self-efficacy, perceived severity and perceived susceptibility elements should be incorporated into the game design framework for computer users to avoid phishing attacks through motivation.

The next study in Chapter 5 will attempt to design and evaluate a mobile game prototype using MIT App Inventor Emulator as a tool to educate computer users against the dangers of phishing attacks. The study will use the game design framework developed on the results from the study reported in this chapter.

# Chapter 5

# Designing a Mobile Game for Computer Users to Thwart "Phishing Attacks"

## 5.1 Introduction

The current study focuses on designing a mobile game as a tool to educate computer users creating awareness of phishing attacks. Phishing is an online identity theft, which aims to steal sensitive information such as username, password and online banking details from victims. To prevent this, phishing education needs to be considered (Downs, Holbrook and Cranor, 2007; Kumaraguru, et al., 2007 and Sheng, et al., 2007). Mobile games can facilitate to embed learning in a natural environment. Therefore, this study attempts to introduce a mobile game design prototype based on a story, which simplifies and exaggerates real life. The study derived a game design framework, which was introduced in Chapter 4 (Figure 4.2). It describes how users avoid malicious IT threats using a given safeguarding measure. Furthermore, the study findings help to define what are the elements that should be addressed in the game design framework for computer users to thwart phishing attacks. Therefore, the study findings suggests perceived threat, safeguard effectiveness, safeguard cost, self-efficacy, perceived severity and perceived susceptibility elements that should be addressed in the game design framework for computer users to avoid phishing threats through motivation.

The current study attempts to design and then develop a mobile game prototype using MIT App Inventor Emulator as an educational tool to protect computer users from phishing attacks. The MIT App Inventor emulator is a visual "blocks" programing language that was initially developed by Google and then moved to MIT centre for mobile learning for further research and developments (Wolber, 2011). To accomplish this study, the elements of the game design framework were addressed in the mobile game design prototype context. Furthermore, the mobile game design principles were used as guidelines for structuring and presenting information in the game design context. The overall mobile game design was aimed to enhance computer users' avoidance behaviour through motivation to protect themselves against phishing threats. The prototype game application was developed on MIT App Inventor Emulator that supports mobile Android platform. Finally, for testing the application through users, the mobile game prototype application was deployed on a HTC One X touch screen smart phone. The study believes that training computer users to protect themselves against phishing attacks through a mobile game prototype would be an aid to enable the cyberspace a secure environment.

## 5.2 Aims and Objectives

The aim of this research study is to design a mobile game prototype as an educational tool to teach computer users to protect themselves against phishing attacks. Therefore, it asks the following questions: The first question is how does the system developer identify which issues the game needs to be addressed? Once the developer has identified the salient issues, they are faced with second question, what principles should be used to structure this information. The elements of the game design framework proposed in Chapter 4 (Figure 4.2) were used to address those mobile game design issues and the mobile game design principles were used as guidelines for structuring and presenting information in the game design context. Then a game prototype was designed for mobile android platform using MIT App Inventor Emulator.

The objectives are as follows:

- To identify the elements that should be addressed in the mobile game design prototype for computer users to avoid themselves from phishing attacks.
- To design a storyboard for the game based on a story, which simplifies and exaggerates real life.
- To design and develop the mobile game design prototype using MIT App Inventor Emulator for computer users to thwart phishing attacks.

## 5.3 Game Design Issues

The main focus of the proposed game design is to educate computer users to thwart phishing attacks. To answer this question issues drawn from the game design framework of threat avoidance will be used to explore the principles needed for structuring the design of the game in the context of computer use. The game design theoretical framework introduced in Chapter 4 (Figure 4.2) describes individual IT users' behaviour of avoiding the threat of malicious information technologies such as phishing attacks. The framework examined how individuals avoid IT threats by using a given safeguarding measure. The safeguarding measure does not necessarily have to be an IT source such as anti-phishing tools; rather it could be behaviour (Liang and Xue, 2010).

Consistent with the game design framework introduced in Chapter 4 (Figure 4.2), users' IT threat avoidance behaviour is determined by avoidance motivation, which, in turn, is affected by perceived threat. Perceived threat is influenced by perceived severity and susceptibility as well as their interaction. The user's avoidance motivation is also determined by the three constructs such as safeguard effectiveness, safeguard cost and self-efficacy. In addition, the game design framework posits that perceived threat is influenced by an interaction between perceived severity and susceptibility. Moreover, avoidance motivation is influenced by the interaction effect of perceived threat and safeguard effectiveness.

Whilst the proposed game design framework forms the issues that the game design needs to address, it should also indicate how to structure this information and present in a game context. Therefore, the mobile game design attempts to develop threat perceptions, making individuals more motivated to avoid phishing attacks and use safeguarding measures.

## 5.4 What to teach?

Possible phishing attacks can be identified in several ways, such as carefully looking at the website address, so called Universal Resource Locator (URL), signs and content of the web page, the lock icons and jargons of the webpage, the context of email message, and the general warning messages displayed in the website (Downs, Holbrook and Cranor, 2007; Wu, Miller and Garfinkel, 2005). Previous research has identified that existing anti-phishing techniques based on URLs are not robust enough for phishing detection (Garera, et al., 2007; Iacovos and Sasse, 2012; Sheng, et al., 2007). Garera, et al. (2007) strongly argued it is often possible to differentiate phishing websites from legitimate ones by carefully looking at the URL without having any knowledge of the content of the corresponding website or sings and symbols such as "VeriSign" signs and "Padlock" icons. Therefore, in this research argues that teaching people not to fall for phishing through URLs is still infancy, but yet well-designed anti-phishing education based on URLs can contribute to stop users falling for phishing.

The objective of the anti-phishing mobile game design, reported in this book, is to teach user how to identify the phishing website addresses (URLs). Therefore the game design should develop an awareness of identifying the features of website addresses. For example, legitimate websites usually do not have numbers at the beginning of their URLs such as http://81.153.192.106/www.hsbc.co.uk.

## 5.5 Game Description

The proposed mobile game design is described in two parts; story along with mechanism and technology used to develop the game.

### 5.5.1 Story and Mechanism

Funny stories are a great tonic for maintaining users' attraction in eLearning (Taran, 2005). Storytelling techniques are used to grab attention, which also exaggerates the interesting aspect of reality. Stories can be based on personal experiences or famous tales or it could also be aimed to build a storyline that associates content units, inspire or reinforce.

The game is based on a scenario of a character of a small fish and 'his' teacher who live in a big pond. The more appropriate, realistic and content relevant the story, the better the chances that it will trigger users. The main character of the game is the small fish, who wants to eat worms to become a big fish. The game player role-plays as a small fish. However, he should be careful of phishers those who try to trick him with fake worms. This represents phishing attacks by developing threat perception. Each worm is associated with a website address, so called Unified Resource Locator (URL) which appears as a dialog box. The small fish's job is to eat all the real worms which associate legitimate website addresses and reject fake worms which associate with fake website addresses before the time is up. This attempts to develop the severity and susceptibility of the phishing threat in the game design.

The other character is the small fish's teacher, who is a mature and experienced fish in the pond. If the worm associated with the URL is suspicious and if it is difficult to identify, the small fish can go to 'his' teacher and request help. The teacher could help him by giving some tips on how to identify bad worms. For example, "website addresses associate with numbers in the front are generally scams," or "a company name followed by a hyphen in a URL is generally a scam". Whenever the small fish requests help from the teacher, the time left score will be reduced by certain amount (in this case by 100 seconds) as a payback for safeguard measure. This attempts to address the safeguard effectiveness and the cost needs to pay for the safeguard in the game design.

The game prototype design consists of total 10 URLs to randomly display worms including five good worms (associated with legitimate URLs) and five fake worms (associated with phishing URLs). If the user correctly identified all good worms while avoiding all fake worms by looking at URLs, then he will gain 10 points (in this case each attempt possible to score 1 point). If the user falsely identified good worms or fake worms, each attempt loses one life out of total lives remaining to complete the game (Figure. 5.4). If the user requested help from the big fish (in this case small fish's teacher) each attempt loses 100 seconds out of total remaining time to complete the game, which is 600 seconds (Figure. 5.5). The consequences of the player's actions are shown in Table 5.1

Table 5.1: Scoring scheme and consequences of the player's action

|  | Good worm (associated with legitimate URL) | Bad worm (associated with phishing URL) |
|---|---|---|
| Player eats | Correct, gain 10 points (each attempt = 1 point) | False negative, (each attempt loses one life out of 5 lives) |
| Player reject | False positive, (each attempt loses one life out of 5 lives) | Correct, gain 10 points (each attempt= 1 point) |
| Request help | Each attempt loses 100 out of 600 seconds | Each attempt loses 100 out of 600 seconds |

The game is designed to randomly generate worms associate with URLs. The user is presented a worm each time associated with a different URL. The URL can be either phishing or legitimate. When the user moves from the beginning to the end of the game, the complexity of URLs is dramatically increased. This helps the user to gain the conceptual knowledge of identifying URLs. Therefore, self-efficacy of preventing from phishing attacks will be addressed in the game design.

The game design is based on a story and presented to the player using digital objects. By creating attractive digital objects in the game design, not only immerse in an augmented physical environment, but also immerse into an augmented social environment. The overall game design attempts enhance individual users' avoidance behaviour through motivation to protect themselves against phishing attacks.

### 5.5.2 Technology

The storyboard technique pays attention on people who will use the solution and the value it will bring. Stories are rich, the fleshed-out descriptions of settings; people, activities, goals, motivations and values presented in a coherent, causally well connected way to form a pictorial representation (Gruen, 2000). The process of creating a story through a storyboard ensures the factors necessary to create an effective solution. Therefore, the game design was initially sketched in a storyboard using ink pen, post-it notes and papers based on the above-mentioned story and then implemented using MIT App Inventor emulator.

The content of the game including URLs and training tips was hard coded using MIT App Inventor emulator. The emulator provided a great deal of flexibility and made easy to quick update the content in developing the mobile game prototype. This is one of the main reasons to use MIT App Inventor emulator to develop a low-fidelity prototype. The current research employed the approach of the URL classification used in Sheng, et al. (2007) and Dhamija, Tygar and Hearst (2006) studies. Therefore, the game was designed total 10 URLs to randomly display including five good worms and five bad worms. The list of URLs is shown in Table 5.2.

Attractive digital objects were integrated such as sounds and graphics with the game to better engage the user within the gaming environment. For example, a sound effect to provide a feedback to the player based on his selection of either good or bad worms. In addition, a light water bubbling sound was played in the background throughout the game to feel the player that he lives in the pond.

**Table 5.2: List of URLs displayed in the game**

| Game Focus | Real or Phishing | Examples | "Tips/ Training messages" from big fish |
|---|---|---|---|
| Appropriate URL | *Real* | *http://www.nationwide.co.uk/default.htm* | "URLs with well-known domain and correctly spelled are legitimate" |
| IP address URL | *Phishing* | *http://147.46.236.55/PayPal/login.html* | "Don't trust URLs with all numbers in the front" |
| Miss spelled URL | *Phishing* | *www.paypa1.com* | "Don't trust URLs with misspelled known websites" |
| Appropriate URL | *Real* | *www.smile.co.uk/* | "URLs with well-known domain and correctly spelled are legitimate" |
| Sub domain URL | *Phishing* | *www.argos.co.uk.myshop.com* | "Don't trust URLs with large host names that contained a part of the well-known web address" |
| Similar and deceptive domains | *Phishing* | *http://www.msn-verify.com/* | "Company name followed by a hyphen usually means, it's a scam website" |
| Appropriate URL | *Real* | *http://www.halifax.co.uk/aboutonline/home.asp* | "URLs with well-known domain and correctly spelled are legitimate" |
| Similar and deceptive domains | *Phishing* | *www.ebay-security.com* | "Companies don't use security related keywords in their domains" |
| Miss spelled URL | *Phishing* | *www.online.ll0ydstsb.co.uk* | "Don't trust URLs with misspelled known websites" |
| Appropriate URL | *Real* | *https://ibank.barclays.co.uk/* | "URL with 'https://' usually a legitimate website" |

## 5.6 Game Based Learning

The game based education system in learning environment is an increasingly relevant trend (Kumaraguru, et al., 2007). Although, the motivational and immersive factors of game based learning have been researched in literature, there is still a lack of systematic design and development behind educational games (Moreno-Ger, et al., 2008). This poor design and development cause little success in game based learning (Kumaraguru, et al., 2007; Sheng, et al., 2007). The objective of the mobile game design, reported in this study, is to teach computer users to thwart phishing attacks. Therefore, the study attempts to design a mobile game prototype based on learning science principles along with mobile game design principles to ensure users learn how to avoid phishing URLs.

### 5.6.1 Learning Science Principles

The current study employed several learning science principles to the game design. Learning science literature claims that effective learning takes place if the teaching methodology is goal-oriented, contextual and interactive and challenging (Quinn, 2005). Learners have specific goals to achieve with the goal-oriented learning. They are being challenged and trained during the process of achieving the goal. Learning is more effective if related materials are presented in an attractive form within the context where learners can relate to. There is a large volume of published studies describing the effectiveness of games for interactive teaching conceptual and procedural knowledge (Gee, 2003). For example, a URL consists of a protocol part and a domain name part. Therefore, checking a URL assigned to a worm, which contained an IP address, is likely a phishing URL.

In particularly game based education, learning science has established an interactive environment that enhances the user motivation when adhere to design principles in the educational game context (Gee, 2003; Quinn, 2005; Repenning and Lewis, 2005). Therefore, this research study employed three learning science principles to design a mobile game prototype: reflection, story-based agent and conceptual-procedural knowledge (Sheng, et al., 2007).

The reflection principle is a process which learners are made to stop and think about what they are learning. In other words, the content of teaching throughout the game should reflect on learners. Educational games provide an opportunity for learners to reflect their new subject knowledge learned (Donovan, Bransford and Pellegrino, 1999). This principle is presented in the game design context, providing a real time audio feedback on each worm based on the user's decision. This helps the user to stop and think for a while and then reflect on the knowledge gained from each worm associated with either real or phishing URLs.

The story-based agent environment principle helps to guide learners throughout the learning process using agents or characters. These characters can be verbal or visual objects such as cartoon or real-life characters. The story-based agent environment principle describes how the use of story-based agents enhances the learning process (Repenning and Lewis, 2005; Sheng, et al., 2007). The current study employed a number of agents such as small fish and 'his' teacher. Moreover, the story starts with a scenario of a character of a small fish and 'his' teacher who live in a big pond. People learn from stories, because stories are easy to remember and the way organise events in a meaningful framework. The story telling technique tends to stimulate the cognitive process of the learner (Klein, 1999; Mayer, 2001). Previous research has revealed that story based agent conditions perform better learning than non-story based agent conditions (Maldonado, et al., 2005; Moreno, et al., 2001).

The conceptual-procedural knowledge principle influences one another in mutually supportive ways and builds an iterative process (Johnson and Koedinger, 2002). The current study employed conceptual-procedural knowledge principle by displaying a randomly generated worm associated with a URL each time. The URL can be either fake or real. When the user moves the game from the beginner level to the end, the complexity of different URLs is dramatically increased. This helps to develop the procedural knowledge of URLs in the game design. A URL has a format that consists of a protocol part and a domain name part. When the user moves through the game, URLs assigned to each worm are checked. In this process, the user learns the different patterns of URLs. In addition, the game design attempts to teach the user specific procedural tips such as "Don't trust URLs with all numbers in the front" or "Don't trust URLs with misspelled known websites". Therefore, this attempts to address the conceptual-procedural knowledge of URLs in the game design context.

## 5.6.2 Mobile Game Design

The above-mentioned scenario based on the theoretical framework empirically investigated in Chapter 4, should combine with a set of guidelines that focused on designing an educational mobile game (Amory and Seagram, 2003; Prensky, 2001). This set of guidelines is known as game design principles, which describe how the user interacts with the mobile game. Prensky (2001) has proposed that a mobile game can be described in terms of six structural elements. Those elements were used in the game design as guidelines for structuring and presenting information.

1. **Rules:** rules organise the game. The story developed in the mobile game design based on the game design framework describes the rules.
2. **Goals and objectives:** goals and objectives are the player struggle to achieve. The player has a goal to solve the task. This is designed in the mobile game to complete it from the beginning to the end by eating real worms associated with legitimate URLs while avoiding fake worms associated with fake URLs.
3. **Outcome and feedback:** outcome and feedback measure the progress against goals. The user obtains the real time feedback on the current status in the game. For instance, when the player taps on a real worm, which is associated with a legitimate URL, a real time audio feedback is played "wow well done". Similarly, when the player taps on a fake worm, which is associated with a fraudulent URL, also a real time audio feedback is displayed "oh try again".
4. **Conflict, competition, challenge, and opposition leading to players' excitement:** These are addressed in the mobile game design as the opportunity to gain points against the given lives of the player.
5. **Interaction:** Interaction is the social aspect in the game design. This is accomplished by providing a real time feedback, fantasy and rewards or gaining points in the game design. By creating attractive digital objects such as a big fish (in this case, the teacher), worms and pond in the game design, not only immerse the player in an augmented physical environment, but also immerse into an augmented social environment.

6. **Representation or story:** This exaggerates the interesting aspect of reality. The representation is addressed in the game design through the scenario or the story developed using digital objects such as the small fish (game player), 'his' teacher, and worms.

Furthermore, the work not only focuses on the design of a mobile game (Korhonen and Koivisto, 2006), but also it should focus on the usability of the game application (Gong and Tarasewich, 2004). The usability of the game design often attempts to recreate a typical playing environment to emulate how a player would typically play the game in the real world while allowing him to achieve the goal. For example, the player should notice when something in the game changes such as decreasing the time period to complete as they move throughout the game or move the digital objects in the game using the keys in the keypad, touch stick or finger without any difficulty.

The game scenario and design principles (a set of guidelines) were combined together to focus on designing an educational mobile game for computer users to thwart phishing attacks. Therefore, the current study attempts to design a storyboard based on the game scenario and design principles as the initial step for the game.

## 5.7 Storyboard Design

A storyboard is a short graphical representation of a narrative, which can be used for a variety of tasks (Truong, Hayes and Abowd, 2006). The exercise of creating storyboards has drawn from a long history with particular communities such as those for developing films, television program segments and cartoon animations (Gruen, 2000). Storyboarding is a common technique in Human Computer Interaction (HCI) to demonstrate the system interfaces and contexts of use (Truong, Hayes and Abowd, 2006). One of the best ways to visualise the game is "storyboarding" it to create a sequence of drawings that shows different scenes and goals (Gruen, 2000). Storyboard sketches are used in this study to do brainstorming and getting down the general flow of the game on a paper. Stories are rich; the fleshed-out descriptions of settings, people, motivation, activities, goals and values presented in a coherent and causally connected way.

As interactive computing moves off the desktop to mobile devices, storyboards should demonstrate not only the details related to the specific interface design context but also the high concepts surrounding the user's motivation and emotion during the systems use (Truong, Hayes and Abowd, 2006). This is because the lack of the user's motivation can cause less satisfaction in the system use. This situation leads the user reluctant to engage the system, which also directly causes not to produce the expected level of output by the system. Storyboards can illustrate a visualise scenario of how an application feature works in designing new technologies. The designer creates an information design scenario that specify the representation of a task's objects and actions that may help the end-user perceive, interpret and make sense of proposed functionalities. The interaction design scenario specifies how the user would interact with the system to perform new activities. Therefore, in the current study, the process of creating a story can influence to keep the game player's attention to the factors necessary to make an effective solution.

The story-based design methods can be beneficial in two specific ways (Rosson and Carroll, 2001). First, use case descriptions are important in understanding how technology reshapes user activities. Second, the story can be created before the system is built and its impacts are figured out. Thus, it has been shown that designers often use storyboards to visualise the depiction of information.

There is a number of existing research utilised storyboarding tools such as Silk (Landay and Myers, 1996), DENIM (Lin, et al., 2000) and DEMAIS (Bailey, Konstan and Carlis, 2001) to create interaction scenarios to convey how the end-user interacts with the envisioned system. However, designers brainstorm individually their design alternatives while capturing their ideas in quick sketches still using ink pen or pencil and papers (Truong, Hayes and Abowd, 2006). Because this is the easiest and quickest way to visualise the idea instead of using drawing tools such as Microsoft

PowerPoint and Microsoft Paint, or much more advanced graphics tool such as Adobe Photoshop and Adobe Illustrator. In addition, paper prototypes are low-fidelity prototypes more license to suggest changes. A prototype design can be either hi- fidelity or low-fidelity prototype. There are some important trades off here in developing high-fidelity prototypes and low-fidelity prototypes. The hi-fidelity prototype does look and feel like a more real to the user. Consequently, users may be more reluctant to critique the interface. Therefore, a quick and dirty paper prototype is more powerful to design storyboards (Sefelin, Tscheligi and Giller, 2003). A list of requirements for the game design sketch is shown in Table 5.3. Finally, the storyboard was sketched using ink pen, post-it notes and papers based on the above story and game design principles, which is shown in Figure 5.1 (Truong, Hayes and Abowd, 2006).

**Table 5.3: A list of requirements for the game design sketch**

| The elements of the game design framework | Questionnaire survey results | Game design sketch |
|---|---|---|
| **1. Perceived Susceptibility**<br><br>Perceived susceptibility is described as an individual's subjective probability that a malicious IT attack will negatively affect him or her (Liang and Xue, 2009; 2010; Workman, et al., 2008; Ng, et al., 2009). | Perceived susceptibility significantly influences on perceived threat ($r = .357**$, and Sig. $= .000$). | Each worm displayed in the game is associated with a website address (URL) which appears as a dialog box. The small fish's (the game player) job is to eat all the real worms which associate legitimate website addresses while rejecting fake worms which associate with fake website addresses before the time is up. This attempts to develop the susceptibility of the phishing threat in the game design. |

| 2. Perceived Severity | | |
|---|---|---|
| Perceived severity is described as the extent to which an individual perceives that negative consequences caused by a malicious IT attack will be severe (Liang and Xue, 2009; 2010; Woon, et al., 2005; Workman, et al., 2008). | Perceived severity significantly influences on perceived threat (r = .499**, and Sig. = .000) | If the small fish (the game player) falsely identified good worms or fake worms, each attempt loses one life out of total lives remaining to complete the game. This attempts to develop the severity of the phishing threat in the game design. |
| **3. Perceived Threat** | | |
| Perceived threat is described as the extent to which an individual perceives the malicious IT threat as dangerous or harmful (Liang and Xue, 2009; 2010). | Perceived threat (R square value = 22 percent) is significantly determined by perceived severity (r = .499**, and Sig. = .000), perceived susceptibility (r = .357**, and Sig. = .000) and their interaction effect (r = .588**, and Sig. = .000) Perceived threat significantly influences on Avoidance motivation (r = .386**, and Sig. = .000). | The main character of the game is the small fish (the game player), who wants to eat worms to become a big fish. However, he should be careful of phishers those who try to trick him with fake worms assigned with phishing URLs. This represents phishing attacks by developing threat perception in the game design. |
| **4. Perceived Safeguard Effectiveness**<br><br>Safeguard effectiveness is described as the individual assessment of a safeguarding measure regarding how effectively it can be applied to avoid the malicious IT threat (Liang and Xue, 2009; 2010; Anderson and Agarwal, 2006; Ng, et al., 2009; Woon, et al., 2005). | Perceived safeguard effectiveness significantly influences on avoidance motivation (r = .381**, and Sig = .000). | The other character is the small fish's teacher, who is a matured and experienced fish in the pond. If the worm associated with the URL is suspicious and if it is difficult to identify, the small fish (the game player) can go to 'his' teacher and request help. The teacher could help him by giving some tips on how to identify bad worms. This attempts to address safeguard effectiveness in the game design. |
| **5. Perceived Safeguard Cost**<br><br>Safeguard cost is a payback for safeguard effectiveness. This refers to the physical and cognitive efforts such as time, money, inconvenience and comprehension required using the safeguard measure (Liang and Xue, 2009; 2010; Woon, et al., 2005). | Perceived safeguard cost significantly influences, but negatively on avoidance motivation (r = -.112*, Sig = .037). | Whenever the small fish (the game player) requests help from the teacher, the time left score will be reduced by a certain amount (in this case by 100 seconds) as a payback for safeguard measure. This attempts to address the cost needs to pay for the safeguard in the game design. |

| 6. Self-Efficacy | Self-efficacy significantly influences | The game is designed to randomly generate worms |
|---|---|---|
| Self-efficacy is described as individuals' | on avoidance motivation (r = .162*, | associate with URLs. The small fish (the game |
| confidence in taking the safeguard measure | Sig. = .047). | player) is presented a worm each time associated |
| (Liang and Xue, 2009; 2010; Kaiser, 1974; Ng, | | with a different URL. The URL can be either |
| Kankanhalli and Xu, 2009; Woon, Tan and Low, | | phishing or legitimate. When the user moves from |
| 2005; Ng, et al., 2009; 2005; Workman, et al., | | the beginning to the end of the game, the complexity |
| 2008). | | of URLs is dramatically increased. This helps the |
| | | user to gain the conceptual knowledge of identifying |
| | | URLs. Therefore, self-efficacy of preventing from |
| | | phishing attacks will be addressed in the game |
| | | design. |

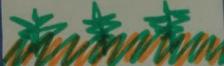
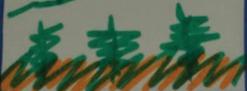
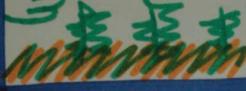

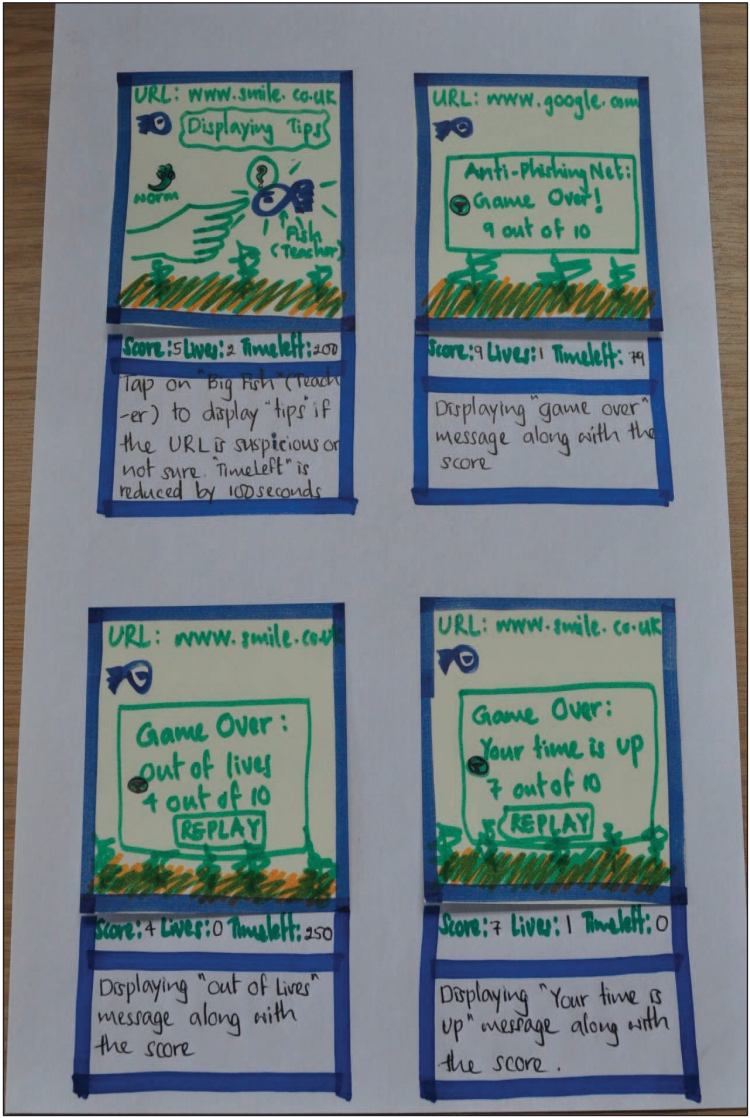

**Figure 5.1: The Storyboard of the anti-phishing mobile game design (continued)**

## 5.8 Mobile Game Prototype

This research study implemented a mobile game prototype for computer users to thwart phishing attacks based on the above mentioned storyboard design. To explore the viability of using a game to prevent from phishing attacks, a working prototype model was developed for a mobile telephone using MIT App Inventor Emulator (Fig. 2). Google app inventor is an application originally introduced by Google and now maintained by MIT (Massachusetts Institute of Technology), which is also a high level visual programming tool (Wolber, 2011). It uses a visual "blocks" language for creating mobile applications, which is easy and less time consuming to develop prototypes compare to the other traditional programming languages (MIT App Inventor, 2012). Therefore, this study employed MIT App Inventor Emulator as a visual block programming language to create a mobile game prototype model, which is shown in Figure 5.2, 5.3, 5.4 and 5.5.

The advantage of creating a workable mock-up enables designers to visually convey the envisioned interaction for a specific class of applications (Truong, Hayes and Abowd, 2006). For example, a mobile game mock-up can act as the low-fidelity prototype versions of the application, thus designers can use to study the usability of the overall application.

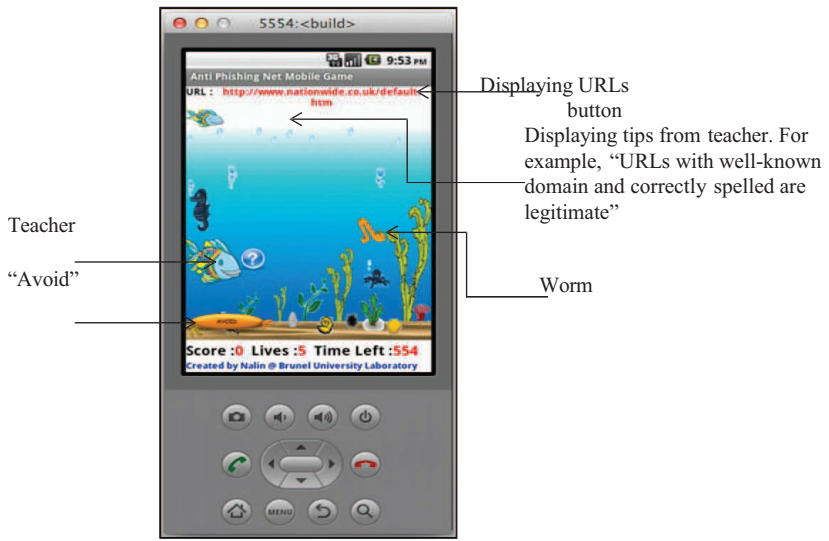

Displaying URLs button

Displaying tips from teacher. For example, "URLs with well-known domain and correctly spelled are legitimate"

Teacher

"Avoid"

Worm

**Figure 5.2: Main Menu of the game prototype design displayed on the MIT App Inventor Emulator. (Appendix 3)**

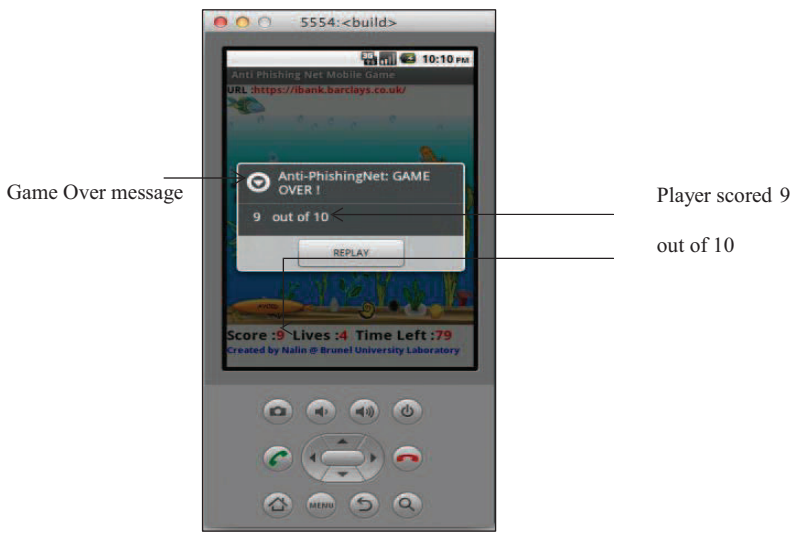

Game Over message

Player scored 9

out of 10

**Figure 5.3 Game Over message on the MIT App Inventor Emulator. (Appendix 3)**

Out of Lives message —— 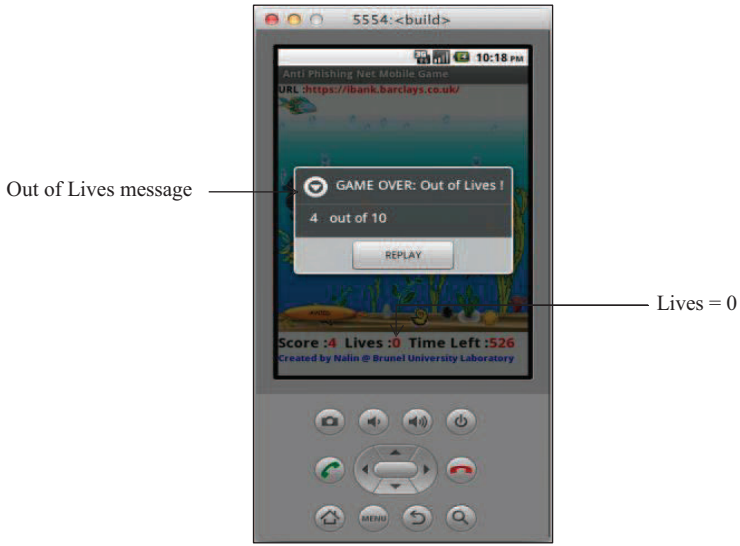

—— Lives = 0

**Figure 5.4: Out of Lives message on the MIT App Inventor Emulator.**
**(Appendix 3)**

Your time is up message —— 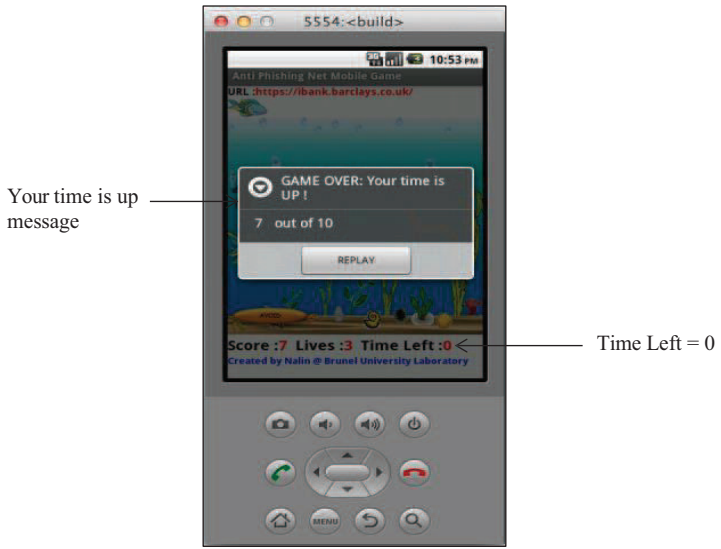

—— Time Left = 0

**Figure 5.5: Your time is up message on the MIT App Inventor Emulator**
**(Appendix 3)**

The player is given instructions before starting the game. Then the main menu of the mobile game mock-up is appeared along with underwater background sound effects (Figure 5.2). A light water bubbling sound is played in the background throughout the game to make the user feelthat he is in the pond. A URL is displayed with each worm and the worms are randomly generated.

If the worm associated with the URL is legitimate, then the user is expected to tap on the worm in order to increase the score. However, if the user fails to identify the legitimate URL, then the remaining lives will be reduced by one point. On the other hand, if the worm associated with the URL is phishing, then the user is also expected to tap on "AVOID" button to reject the URL in order to increase the score. If the user fails to do this, then the remaining lives will be reduced by one point. If the worm associated with the URL is suspicious and if it is difficult to identify, the user can tap on big fish (in this case, teacher fish) to request help. Then some relevant tips will be displayed just below the URL.For example, "website addresses associate with numbers in the front are generally scams. Whenever the user taps on the big fish asking help, the time left is reduced by 100 points (in this case 100 seconds).

Finally, the user will gain 10 points if all the given URLs were correcly identifed within 5 lives and 600 seconds to complete the game.

## 5.9 Summary

The current research study focuses on designing a mobile game as a tool to educate computer users by creating awareness of phishing attacks. It addressed two questions: The first question is how does the systems developer identify which issues the game needs to be addressed? Then they are faced with the second question, what principal should guide them to structure this information? The study employed a theoretical game design framework introduced in Chapter 4 (Figure 4.2) in order to address those game designing issues and mobile game design principles along with learning science principles were used as a set of guidelines for structuring and presenting information in the game design context. The objective of the anti-phishing mobile game design prototype was to teach the user how to identify phishing website addresses (URLs), which is one of the many ways to identify a phishing attack. The overall game design was targeted to enhance avoidance behaviour through motivation to protect computer users from phishing attacks. This study believes by educating computer users against malicious IT threats would make a considerable contribution to enable the cyberspace a secure environment.

As previously elaborated, this chapter has attempted to design and then develop a mobile game prototype for computer users to thwart phishing attacks. The next chapter will discuss how the theoretical game design framework introduced in Chapter 4 (Figure 4.2) was evaluated through the developed mobile game prototype.

# Chapter 6

# Evaluation of Game Design Framework Using a Mobile Game Prototype

## 6.1 Introduction

This study focuses on evaluating the game design framework introduced in Chapter 4 through the mobile game prototype developed using MIT App Inventor Emulator. Even though, the game design framework was introduced using an empirical investigation, it would have been further useful to determine users' impact on the framework after playing the mobile game prototype. Sefelin, Tscheligi and Giller (2003) reported based on an empirical investigation that low-fidelity prototypes help users to understand and criticise the overall system and also provide suggestions for its improvement. From a user perspective, it would be much easier to get the user's reliable and accurate feedback of the game design framework after the engagement with the game prototype. This is due to the facts that they can see how the mobile game prototype works and understand the practical implication of it. Therefore, the mobile game prototype application was deployed on a HTC One X touch screen smart phone to conduct the current experiment on users.

In recent years, there has been an increasingly amount of literature on usability and user experience in the gaming context (Bernhaupt, et al., 2008; Nacke, 2009). While a variety of definitions of the term usability have been suggested, ISO 9241-11 usability standard defines usability as "the extent to which a product can be used by specified users to achieve specified goals with effectiveness, efficiency and satisfaction in a specified context of use"(ISONORM 9241 Part 1-17, 2008; Korhonen, Elina and Koivisto, 2006; Nacke, 2009). Effectiveness is the ability of users to complete tasks using the system and the quality of the output of those tasks. Efficiency is the level of resources consumed in order to perform those tasks. Satisfaction is the user's subjective reaction of performing tasks in the system use. In general, it is extremely difficult to define the usability of a system without exactly specifying who the intended end-users of the system are, what the performing tasks are and what the characteristics of the physical, social and organisational environments are, in which the system will be used (Brooke, 1996).

The potential demand of mobile games is attracting the interest of a lot of different sectors and industries, including the education industry though game based education. Game based education provides joy and fun when it provides sufficient challenges for the player (Korhonen, Elina and Koivisto, 2006). However, mobile games still face some obstacles such as limited memory, limited processor power and limited screen size (Hashim, Ahmad and Rohiza, 2010; Tolentino, et al., 2011). In addition, educational mobile games are yet in their infancy due to the lack of potential education and entertainment; they are also often qualified as boring and without challenges (Van Eck, 2006).

Thinking beyond issues related to fun and challenges, usability of mobile game interfaces is another issue that affects user satisfaction (Tolentino, et al., 2011). For example, the lack of user satisfaction can be caused by difficulty in understanding commands, little intuitiveness, inappropriate response time, excessive or insufficient amount of information, and faulty navigations through the mobile game interface (Pinelle, Wong and Stach, 2008). Tolentino, et al. (2011) stated that there is a positive correlation between in-game experience and user satisfaction. Korhonen, Elina and Koivisto (2006) claimed that a good gaming experience has high requirements of the user interface design. Furthermore, they state that the user interface should be designed to be convenient, reliable and usable so that the player can enjoy the game and have fun rather than struggle with the interface.

In addition, it is worthwhile to state that the game design interface itself has a responsibility to attract players and retain with the game without frustrating them and leading them to quit playing the game. Higher education institutions constantly claim that poorly designed game based education can cause the lack of students' satisfaction, which causes obstacles to effective education (Munoz-Organero, et al., 2010; Irvine and Thompson, 2003).

Difficulties in interaction with the mobile game interface provide obstacles for the player to progress and navigate through the game. When the player cannot precede the game against the offered challenges, it is more likely that he is unable to absorb the information to learn. Consequently, learning through the game does not take place. Korhonen, Elina and Koivisto (2006) stated that the general usability of mobile games is a significantly important factor in terms of measuring how the player interacts with the game. If players are not interested in the user interface, they hence do not want to struggle with it. In other words, the user interface of the mobile game should facilitate them to continue the game play without having a specific attention to it. Therefore, to identify the general usability aspect of mobile game prototype application developed in Chapter 5 is an important goal. This focuses on the player's success in dealing with the interface. A set of questionnaire items was used to evaluate the general aspects of interface usability in the mobile game prototype.

The study reported in Chapter 4 empirically investigated as the first attempt what key elements that should be addressed in the game design framework for computer users to avoid phishing attacks through motivation. This research study is the second attempt to empirically evaluate the game design framework introduced in Chapter 4 (Arachchilage, Love and Beznosov, 2016). The game design framework describes what elements influence to enhance the user avoidance behaviour through motivation to thwart phishing attacks.

## 6.2 Aims and Objectives

The aim of this research study is to evaluate the game design framework introduced in Chapter 4 through the mobile game prototype. The current study also answers the question: can a mobile game be designed that protects users against phishing attacks? The objectives are as follows:

- To identify the elements that should be addressed in the game design framework to avoid phishing attacks.

- To explore which elements impact on the game design framework after the user's engagement with the mobile game prototype.
- To formulate a game design framework to thwart phishing attacks.

## 6.3 Pilot Study

The current pilot study employed a usability study of the game prototype as the first step to assess the subjective satisfaction of the mobile game prototype interface. This is based on the notion that mobile game based learning takes place if participants are satisfied with the overall game prototype. Then followed by a think aloud experiment as the second step of the research along with a pre and post-test to assess the game design framework introduced in Chapter 4 through the mobile game prototype.

Think-aloud testing is a widely employed prototype testing method in the area of Human-Computer Interaction (Nørgaard and Hornaek, 2006). Previous research has reported that it is a widely used as instrument to study cognitive processes, such as problem solving, human-computer interaction, reading and writing (Rooden, 1998). Participants in a think-aloud study are asked to carry out a task or series of tasks, while verbalising their thoughts. The experimenter records all verbalisations and writes them down in a verbal report. Finally, the recorded and written data is analysed in accordance to the research question(s). The current pilot study employed a quantitative data analysis approach to evaluate the usability of the game prototype whilst a qualitative data analysis approach is used for the think-aloud study. The main purpose of a think-aloud study along with a pre- and post-test is to evaluate the game design framework after the usability study of the mobile game prototype.

The pilot study was conducted along with a pre- and post-test before the main study took place. This was launched to assess the game design framework introduced in Chapter 4 through the mobile game prototype deployed on a HTC One X touch screen smart phone. If any problems were encountered during the pilot study, corresponding adjustments were addressed in the main study.

### 6.3.1 Data Collection Techniques

The pilot study employed both qualitative and quantitative data collection approaches. A quantitative data collection technique was employed to collect data about the usability of the mobile game prototype. Therefore, a set of questionnaire items was used to evaluate the general aspects of interface usability of the mobile game prototype. In addition, a think-aloud study was employed to collect data about the user impact on the game design framework introduced in Chapter 4 through the mobile game prototype. A qualitative data analysis approach was employed to evaluate the game design framework. Therefore, an experimental protocol was designed to conduct the think-aloud study along with the pre- and post-test.

### 6.3.2 Questionnaire Design

A variety of questionnaires has been designed and reported in the literature to measure the general usability of interactive systems, including Questionnaire for User Interaction Satisfaction (QUIS) (Chin, Diehl and Norman, 1988), System Usability Scale (SUS) (Brooke, 1996), Computer System Usability Questionnaire (CSQU) (Lewis, 1995) and Microsoft Production Reaction Cards (Benedek and Miner, 2002). In the industry, some usability laboratories have been using their own questionnaire to assess subjective reactions of participants in a usability test (Tullis and Stetson, 2004). The current research study employed a tool called the SUS, which is used to measure users' subjective satisfaction of mobile game interface usability.

John Brooke (1996) initially developed the SUS while he was working in the Digital Equipment Corporation. The SUS is a commonly known, freely distributed, simple and reliable questionnaire consisting of 10 items that aim to evaluate the subjective perception of the individual interacting with software without considering personal taste (Brooke, 1996; Finstad, 2006). Tullis and Stetson (2004), as well as Finstad (2006) have revealed that SUS provides superior assessments compared to other questionnaires such as QUIS and CSQU. Their results concluded that, for the conditions of the study, sample sizes of at least 12-14 participants are needed to get reasonably reliable results. The SUS uses a five-point Likert scale with anchors for strongly agree and strongly disagree. However, Finstad (2006) revealed that a significant amount of non-native English speakers failed to understand the word "cumbersome" in Item 8 of the SUS, that is, "I found the system to be very cumbersome to use" (Finstad, 2006; Lewis and Sauro, 2009). Since the current usability study represents participants from a multinational background, the word "cumbersome" was replaced with "awkward" as Finstad (2006) stated in his study. Finally, a set of questionnaire items was reproduced from Brooke's SUS to measure the users' subjective satisfaction of the mobile game prototype interface as shown in Table 6.1.

**Table 6.1: A set of SUS questionnaire items**

| No | Statement | Strongly Disagree 1 | Disagree 2 | Neutral 3 | Agree 4 | Strongly Agree 5 |
|---|---|---|---|---|---|---|
| 1 | I think that I would like to use this mobile game frequently | ☐ | ☐ | ☐ | ☐ | ☐ |
| 2 | I found the mobile game unnecessarily complex | ☐ | ☐ | ☐ | ☐ | ☐ |
| 3 | I thought the mobile game was easy to use | ☐ | ☐ | ☐ | ☐ | ☐ |
| 4 | I think that I would need the support of a technical person to be able to use this mobile game | ☐ | ☐ | ☐ | ☐ | ☐ |
| 5 | I found the various functions in this mobile game were well integrated | ☐ | ☐ | ☐ | ☐ | ☐ |
| 6 | I thought there was too much inconsistency in this mobile game | ☐ | ☐ | ☐ | ☐ | ☐ |
| 7 | I would imagine that most people would learn to use this mobile game very quickly | ☐ | ☐ | ☐ | ☐ | ☐ |
| 8 | I found the mobile game very awkward to use | ☐ | ☐ | ☐ | ☐ | ☐ |

| 9 | I felt very confident using the mobile game | | | | | |
|---|---|---|---|---|---|---|
| 10 | I needed to learn a lot of things before I could get going with this mobile game | | | | | |

### 6.3.3 Experimental Protocol Design

An experimental protocol was designed to carry out a think-aloud study along with a pre- and post-test to evaluate the game design framework. The experimental protocol included user instructions that helped participants to carry out the think-aloud study, which is presented in Figure 6.1.

**Experimental Protocol: Think-aloud User Study Instructions**

This experiment – "User Study" contributes to evaluate a game design framework for personal computer users to thwart phishing attacks. There are some instructions you need to follow in the order given below.

**Session 1: Phishing test (Pre-test)**

- **Task Description:** You are given 10 suspect "phishing" websites and are required to identify legitimate websites from phishing websites. Phishing is an online identity theft, which aims to steal sensitive information such as username, password and online banking details from victims. For example, the attacker creates a fraudulent website which has the look-and-feel of the legitimate website. Users are invited to fraudulent websites by sending emails and steal their money. The test is PC based, with the overall score being displayed at the end of the test.

- **Task:** To take the test, please type the link below into the web address (URL) of your browser and follow the instructions given to you.
  http://people.brunel.ac.uk/~cspgnag/index.php

**Session 2: Playing the mobile game**

- **Task Description:** The mobile game prototype is designed and implemented to teach home computer users to prevent themselves from phishing attacks.

- **Task:** You are given fifteen minutes to complete the mobile game training activity. The experimenter will help you with the game, which runs on Samsung Galaxy Mini handset. As you do so, you are asked to express your opinions and comments to the mobile game in the specific context of knowledge, understanding and awareness of phishing attacks. After the mobile game training activity, you are asked to complete a survey to evaluate your subjective satisfaction of mobile game prototype interface.

**Session 3: Phishing test (Post-test)**

- **Task Description:** You are given more 10 suspect "phishing" websites and are required to identify legitimate websites from phishing websites. The test is PC based, with the overall score being displayed at the end of the test.

- **Task:** To take the test, please type the link below into the web address (URL) of your browser and follow the instructions given to you.
  http://people.brunel.ac.uk/~cspgnag/index1.php

**Figure 6.1: Experimental protocol: Think-aloud user study instructions**

## 6.3.4 Participants

The questionnaire pilot study was conducted with eight participants' in-person to observe a wide array of questions relating to measure the user's subjective satisfaction of the mobile game prototype interface. The results indicated a significantly higher usability score; therefore an initial hypobook was formed and subsequently used to inform the main think aloud study. Otherwise, some work is needed in order to refine and redesign the mobile game prototype until results indicate reasonable higher usability scores.

The above mentioned usability study was followed by a think-aloud study with the same participants. A think-aloud pilot study along with a pre- and post-test was run with eight participants' in-person to observe their understanding, knowledge and awareness of phishing attacks through a mobile game prototype, so that an initial hypobook could be formed and subsequently used to inform the main study.

Virzi (1992) revealed that 90 percent of problems through the prototype testing were identified in a study with less than ten participants. However, usability expert Jakob Nielsen (2000) encouraged researchers to use only five participants in a prototype testing during the pilot study. He argued that using more than ten participants may have a very low ratio of problems discovered to resources consumed.

Sheng, et al.'s (2010) study showed participants between the ages of 18 to 25 are more susceptible for phishing attacks than other age groups. Therefore, undergraduate students' ages ranged from 18 to 25 of the department of Computer Science and Technology at the University of Bedfordshire, UK, who are computer users, were recruited while attending their Applied Programming practical lab sessions. For the purpose of this study, each individual participant was taken to a vacant room after finishing their practical lab session. Before launching the pilot study, the permission was taken from the field chair and head of department. A summary of demographics in the pilot study is shown in Table 6.2.

**Table 6.2: Participant demographics in the pilot study**

| Characteristics | Amount |
|---|---|
| Sample Size | 8 |
| Gender | |
|       Male | 6 |
|       Female | 2 |
| Age (18 - 25) | 8 |
| Experience using mobile device | |
|       Mobile phone | 2 |
|       Smart phone | 6 |
| Average hours per week on the Internet | |
|       0-5 | 0 |
|       6-10 | 2 |
|       11-15 | 3 |
|       16-20 | 0 |
|       20+ | 3 |

## 6.3.5 Procedure

The pre- and post-tests were based on an Apple MacBook where the participants will receive their score at the end of each test. First and foremost, each individual participant was explained the nature of the think-aloud experimental study and asked to sign a consent form (see Appendix 1). They were also informed that the experiment is about testing their understanding of phishing threat awareness through the mobile game prototype. Then the individual participants were asked whether they knew what the term *'phishing attack'* means. Those who gave a positive response were asked to give a short verbal description to confirm their understanding, whilst negative responders were read a brief definition of phishing attack and gave a short verbal description. To begin the experiment, participants were asked to follow think-aloud user study instructions given in the experimental protocol (Figure 6.1). They were also informed that they are welcome to clarify anything related to the experiment. In the pre-test, participants were presented with ten websites and asked to differentiate phishing websites from legitimate ones. After evaluating 10 websites, participants were given fifteen minutes to complete a mobile game based training activity on a HTC One X touch screen smart phone. Initially, the game was designed with 10 suspect URLs where the participant's responsibility is to identify legitimate URLs from phishing ones. Brooks (1996) stated that the SUS is generally used after the respondent has had an opportunity to use the system being evaluated, however before any debriefing or discussion takes place. Therefore, after engaging 15 minutes with the mobile game activity, participants were asked to fill in a survey (SUS questionnaire items in Table 6.1), which was used to measure participants' subjective satisfaction of mobile game prototype interface. The survey was followed by a post-test where participants were shown ten more websites to evaluate. The score was recorded during the pre- and post-tests to observe how participants' understanding and awareness of phishing threat developed through the mobile game based learning. More than half of the websites were phishing websites based on popular brands, whilst the rest were legitimate websites from popular brands. For the purpose of this test, recently being attacked phishing websites were captured from PhishTank.com (PhishTank, 2011) from November 1 to November 28, 2011. All phishing website URLs were selected within 7 hours of being reported. During the mobile game based training activity and post-test a think-aloud study was employed where participants talked about their opinions and experience of phishing threat awareness and understanding through the mobile game prototype.

## 6.3.6 Results

The initial focus of this study was to evaluate the general usability of a mobile game prototype. Therefore, the usability study employed SPSS, which is a statistical software package for its data analysis.

Cronbach's alpha was used to measure the internal consistency of how closely related a set of items are as a group (Pallant, 2007). Previous research has revealed that the given alpha greater than 0.70 is adequate with the number of items in the scale (Cronbach, 1951; Pallant, 2007; Zaharias and Poylymenakou, 2009). Therefore,

Cronbach's alpha was calculated (Cronbach's alpha = 0.878) for a set of SUS questionnaire items to measure the internal consistency among them.

In the current experiment, participants talked about their experience and opinions of mobile game prototype to thwart phishing threats. The results were encouraging; however it highlighted some areas where the mobile game prototype needed improvements.

This research study revealed that the mobile game was somewhat effective at teaching participants to look at the URL in their browser's address bar when assessing a website whether or not legitimate. Participants scored 49 percent in the pre-test and 78 percent in the post-test of identifying phishing or legitimate websites after playing the mobile game. There is a considerable improvement of participants' results of 29 percent in the post-test during the pilot study. Therefore, the study findings provide evidence that the mobile game prototype can teach personal computer users to thwart phishing attacks. All the participants responded that their decisions were based on looking at the address bar when evaluating the second set often websites in the post-test. Furthermore, they stated that they made only very few attempts to look at the address bar when evaluating the first ten websites in the pre-test. There, many participants used incorrect strategies to determine the website legitimacy. For example, one of the common strategies consisted of checking whether or not the website was designed professionally. However, this may not be a useful strategy as many phishing websites are exact replica of legitimate websites. The attacker can easily mimic any professional website from the source code of the particular page provided by the browser. Moreover, they highlighted that the mobile game was somewhat effective at teaching different patterns of URLs to differentiate phishing URLs from legitimate ones.

The study findings showed that the participants learned most of the URL-related concepts that tried to teach using the mobile game prototype. For example most participants seemed to understand that URLs beginning with numbers are usually a sign of scam. However, the study found that some participants applied the lessons learned from the mobile game incorrectly. For example, some participants misapplied the rule about URLs that have all numbers in the front are usually scams through

*www2.fdic.gov* as a phishing website because the URL contained the number 2 after the "*www*". This is because the mobile game prototype was not included the URL contained the numbers after the *www* such as *www2* or *www4*. However, participants were tested on their ability to identify the above type of URLs in the post-test. Therefore, the URL '*www2.fdic.gov*' included in the mobile game prototype before launching the main study.

### 6.3.7 Summary

The current pilot study attempted to evaluate the participants' subjective satisfaction of the mobile game prototype interface. Therefore, the study employed SUS developed by John Brook with 10 items using a five-point Likert scale at 1= 'Strongly disagree' and 5= 'Strongly agree'. The study recruited eight participants, ages ranged from 18 to 25 from the department of Computer Science and Technology at the University of Bedfordshire, UK. The participants filled in the questionnaire after engaging 15 minutes with the mobile game prototype deployed on a HTC One X touch screen smart phone. Based on the feedback of the survey study, Cronbach's alpha was calculated to measure the internal consistency of a set of SUS questionnaire items. This value (Cronbach's alpha = 0.878) was significantly higher than Nunnally recommended a minimum level of Cronbach's alpha, which is 0.70 (Zaharias and Poylymenakou, 2009). Therefore, the current usability survey followed by a think-aloud study since participants were satisfied with mobile game prototype.

The think-aloud pilot study along with a pre- and post-test employed to evaluate the game design framework introduced in Chapter 4. A pilot test of the mobile game prototype was conducted through a user study; the same eight participants who used in the usability study were tested on their ability to identify phishing websites from a set of legitimate and phishing websites before (pre-test) and after (post-test) playing the mobile game. In addition, they talked about their experience and opinions of the mobile game prototype to thwart phishing threats. The pilot study results were encouraging and the study found that the mobile game prototype could teach personal computer users to thwart phishing attacks. However, it highlighted some areas where the mobile game prototype needed some improvements. Those highlighted issues were addressed in the mobile game prototype before conducting the main study.

## 6.4 Main Study

The same usability study and think-aloud experiment used in the pilot study were employed in the main study. Therefore, the main study used a set of SUS questionnaire items developed in the pilot study (Table 6.1). A quantitative data analysis approach using a survey was occupied to assess the subjective satisfaction of the mobile game prototype interface through the participants. Then the survey followed by a think-aloud study, along with pre- and post-tests as used in the pilot study.

The think-aloud experiment was employed to evaluate the game design framework introduced in Chapter 4. The think-aloud study attempted to investigate how the user impacts on the elements of the proposed mobile game design framework to make phishing awareness after their engagement with the mobile game prototype. Therefore, the study employed the experimental protocol developed in the pilot study (Figure 6.1). The researcher recorded all the verbalisations made during the think-aloud study. A qualitative data analysis approach was then occupied to evaluate the game design framework through the mobile game prototype.

## 6.4.1 Participants

The think-aloud study along with a pre- and post-test was run with 20 participants to observe their understanding, knowledge and awareness of phishing attacks through the mobile game prototype. Participants were recruited from the Brunel University, London, UK, who are computer users. They were invited to participate by sending email messages and posting a message on Facebook asking for their help. They were invited to the St. John's building computer laboratory at Brunel University. Before conducting the think-aloud study, the permission was taken from the ethical committee at Brunel University, as well as from the participants themselves.

Participants' ages ranged from 18 to 25, with a gender split of 65 percent male and 35 percent female. The majority has more than 20 hours per week of Internet experience. All of them had the experience of Internet shopping at least a single time. Furthermore, all of them had experience in using a smart phone for more than a year. Each participant took part in the think-aloud study on a fully voluntary basis. A summary of demographics in the main study is shown in Table 6.3.

**Table 6.3: Participants demographics in the main study**

| Characteristics | Amount |
|---|---|
| Sample Size | 20 |
| Gender | |
| Male | 13 |
| Female | 7 |
| Age (18 - 25) | 20 |
| Education | |
| Less than high school diploma | 0 |
| High school diploma | 4 |
| Undergraduate degree | 16 |
| Postgraduate degree | 0 |
| Average hours per week on the Internet | |
| 0-5 | 0 |
| 6-10 | 0 |
| 11-15 | 0 |
| 16-20 | 0 |
| 20+ | 20 |

## 6.4.2 Procedure

The data collection was made using the think-aloud study along with a pre and post-test. The think-aloud study was conducted in-person with each participant approximately one hour. First and foremost, each individual participant was explained the nature of the think-aloud experimental study and asked to sign a consent form (see Appendix 1). They were informed that the think-aloud experiment is about testing the participant's understanding of phishing threat awareness through the mobile game prototype. They were also told that they are free to withdraw from the experiment study at any time and without having to give a reason for withdrawing. To begin the experiment, participants were asked to follow think-aloud user study instructions given in the experimental protocol that is shown in Figure 6.1. The pre and post-tests were based on an Apple MacBook Pro computer where the participants receive their score at the end of each test.

In the pre-test, participants were presented with the same ten websites as those participants were in the pilot study and asked them to identify phishing websites from legitimate ones. A list of ten website addresses were used in pre-test is shown in Table 6.4. After evaluating the first 10 websites, participants were given fifteen minutes to complete a mobile game based training activity on a HTC One X touch screen smart phone. After engaging 15 minutes with the mobile game prototype, participants were asked to fill in a survey. The SUS questionnaire items of the survey used to measure the users' subjective satisfaction of the mobile game prototype interface that is shown in Table 6.1. Then participants were shown the same ten more websites in the post-test as those participants were in the pilot study. A list of ten website addresses used in post-test is shown in Table 6.5. The score was recorded during the pre- and post-tests, to observe participants' understanding and awareness of phishing threat through the given mobile game prototype. More than half of the websites were phishing websites based on popular brands, whilst the rest is legitimate websites from popular brands. For the purpose of this experiment, recently being attacked phishing websites were taken from PhishTank.com (PhishTank, 2011) from November 1 to November 28, 2011. All phishing website URLs were captured within seven hours of being reported. Participants talked about their opinions and experience of phishing threat awareness through the mobile game prototype during the study. Moreover, they talked about their opinions in terms of motivation, behaviour, threat perception, threat severity and susceptibility, cost, knowledge and effectiveness of mobile game prototype to protect themselves from phishing threats.

**Table 6.4: List of ten website addresses used in pre-test**

| Real or Phishing | Website Name | Website address |
|---|---|---|
| Phishing | Santander | 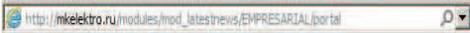 |
| Phishing | PayPal | 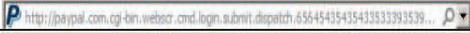 |
| Real | HSBC | 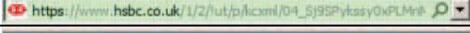 |
| Phishing | Halifax | 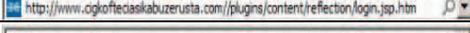 |
| Real | eBay | 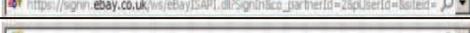 |
| Phishing | Western Union | 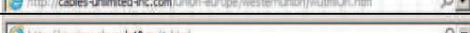 |
| Phishing | eBay | 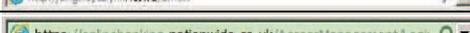 |
| Real | Nationwide | 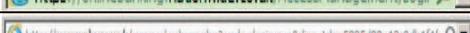 |
| Phishing | Money:hq | 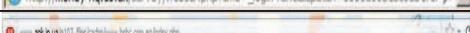 |
| Phishing | HSBC | 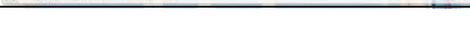 |

**Table 6.5: List of ten website addresses used in post-test**

| Real or Phishing | Website Name | Website address |
|---|---|---|
| Phishing | PayPal | 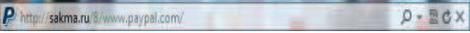 |
| Phishing | HABBO | 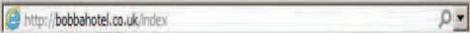 |
| Real | FDIC | 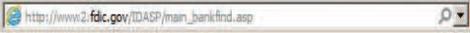 |
| Phishing | Littlewoods | 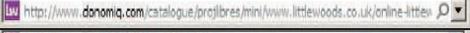 |
| Real | Lloyds TSB | 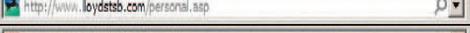 |
| Phishing | Facebook | 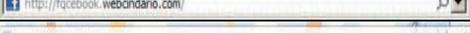 |
| Phishing | Santander | 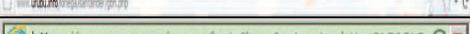 |
| Real | UPS | 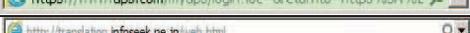 |
| Phishing | eBay | 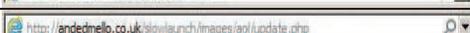 |
| Phishing | AOL | 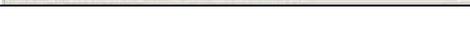 |

### 6.4.3 Data Collection Instruments

The data collection was done through observations, field notes, audio recording and fill in a survey from 20 think-aloud sessions, and that is, the period of time from the participant arrival to the test and until the participant leaves. Each participant spent approximately one hour with the experiment.

The data collection of the main study was occupied in two phases. First, data were collected by asking participants to fill in the survey (Table 6.1) after engaging 15 minutes with the mobile game prototype. This technique was employed to assess the users' subjective satisfaction of mobile game prototype interface. A quantitative data analysis approach was employed to analyse the data collected from the survey. Second, the data collected from the think-aloud method and participants' score of pre and post-tests. A qualitative data analysis approach employed to analyse the data collected from the think-aloud study.

### 6.4.4 Results Analysis

The findings of this research study aims to determine whether or not the participant avoidance behaviour enhances through motivation to protect themselves from phishing attacks through the mobile game prototype. The think-aloud study results discussed how the mobile game prototype impacts on game design framework elements introduced in Chapter 4 such as perceived threat, safeguard effectiveness, safeguard cost, self-efficacy, perceived severity and perceived susceptibility. In addition, pre- and post-tests were used to determine whether or not anti-phishing education takes place after the game play activity. Moreover, the study findings contributed to the existing literature concerning the view that the mobile game based education approach for preventing malicious IT threats (such as phishing attacks) is effective.

## 6.4.5 SUS Study Results

The purpose of this study was to evaluate the general usability of the mobile game prototype. Therefore, the usability study employed SPSS, a statistical software package for data analysis, developed by IBM (Finstad, 2006).

The SUS survey study showed the initial pool of items to 20 regular computer users to elicit feedback through a questionnaire survey after engaging 15 minutes with the mobile game prototype. Based on the feedback of the survey study, initially, Cronbach's alpha was calculated to measure the internal consistency of a set of items as a group (Cronbach's alpha = 0.901). Nunnally has suggested that the minimum level of Cronbach's alpha is 0.70. Therefore, the study continued to assess the subjective user's satisfaction of mobile game prototype interface.

The current usability study, which aimed to evaluate the mobile game prototype, employed the SUS scoring approach introduced by Brooke (1996). The SUS produces a single number representing a composite measure of the general usability of the mobile game prototype application. To obtain the SUS score of the mobile game prototype, initially the sum of score contributions from each item was calculated. Each item's score contribution ranges from 0 to 4. For items 1, 3, 5, 7 and 9 the score contribution is the scale position minus 1. For items 2, 4, 6, 8 and 10, the score contribution is 5 minus the scale position. Finally, multiply the sum of the scores by 2.5 to obtain the overall value of mobile game prototype usability. The SUS scores range from 0 to 100. Therefore, to accomplish this study, the user satisfaction with the mobile game prototype application deployed on a HTC One X touch screen smart phone was measured using the SUS. The scores are summarized in Table 6.6.

In general, the participants' subjective satisfaction of the mobile game prototype application was significantly high with 84 percent (83.62 out of 100) (Brooke, 1996). Participants also noted that they found the mobile game prototype application to be very usable and felt that they could learn its usage very quickly. This is mainly because the game player had to follow three functionalities, which are easy to remember when interacting with the mobile game. First, tap on the worm to make it eat if the worm was associated with a legitimate website address. Second, tap on the "avoid" button if the worm was associated with a fake website address. Third, tap on the "big fish" image icon (the small fish's teacher) to request help if the worm associated with the website address is suspicious and difficult to identity. Therefore, the minimum number of functionalities that could easily be remembered to complete the game helped to enhance the participant subjective satisfaction of mobile game interface usability. Moreover, participants demonstrated they had higher confidence after using the mobile game prototype. Participants were also able to learn the game quickly and they stated they are quite interested in using the mobile game frequently.

**Table 6.6: Summary of System Usability Scale of the mobile game prototype application**

| No | Statement | Average Score | Standard Deviation |
|----|-----------|---------------|--------------------|
| 1 | I think that I would like to use this mobile game frequently | 3.95 | 0.759 |
| 2 | I found the mobile game unnecessarily complex | 1.50 | 0.607 |
| 3 | I thought the mobile game was easy to use | 4.55 | 0.510 |
| 4 | I think that I would need the support of a technical person to be able to use this mobile game | 1.70 | 0.865 |
| 5 | I found the various functions in this mobile game were well integrated | 4.20 | 0.696 |
| 6 | I thought there was too much inconsistency in this mobile game | 1.65 | 0.671 |
| 7 | I would imagine that most people would learn to use the mobile game very quickly | 4.45 | 0.686 |
| 8 | I found the mobile game very awkward to use | 1.60 | 0.503 |
| 9 | I felt very confident using the mobile game | 4.35 | 0.587 |
| 10 | I needed to learn a lot of things before I could get going with this mobile game | 1.60 | 0.754 |
| | **Average Overall Satisfaction Score (Ranges from 0-100)** | **83.62** | |

**Total Score = 33.45**
**SUS Score = 33.45\*2.5 = <u>83.62</u>**

In summary, the survey result analysis showed that the participants' subjective satisfaction of the mobile game prototype was significantly high. Therefore, and since participants showed higher degree of satisfaction using the mobile game prototype, the study continued with a think-aloud result analysis. The think-aloud study results showed the impact of mobile game prototype on the elements of game design framework introduced in Chapter 4. Furthermore, pre- and post-tests demonstrated whether or not the anti-phishing education took place through the mobile game prototype.

### 6.4.6 Think-aloud Study Results

The data analysis of the think-aloud study was conducted in two phases, which were based on Norgaard and Hornbaek's (2006) study. First, the study segmented the recordings through the application of keywords to each segment. The purpose of having a think-aloud study was to evaluate the game design framework through the mobile game prototype. The keywords were taken from the elements in the game design framework introduced in Chapter 4 (Fig 4.3). Therefore, the audio recordings were mainly segmented into eight keywords: avoidance behaviour, avoidance motivation, perceived threat, safeguard effectiveness, safeguard cost, perceived severity, perceived susceptibility and self-efficacy. Second, the study attempted to analyse and try to form a coherent interpretation of segments that shared keywords. Therefore, the study findings were organised in eight areas. Table 6.7 summarises these areas and main findings within each of them.

**Table 6.7: Overview of results**

**N refers to the number of sessions in which a finding was made (out of 14 sessions in total)**

| Area of attention | Main findings | N | Example of quotes |
|---|---|---|---|
| **Avoidance behaviour** | I play the mobile game to avoid phishing attacks<br><br>OR<br><br>Updating knowledge through the mobile game is very useful to avoid phishing attacks | 20 | "The game was useful and I liked to play the game to avoid phishing"<br><br>"I like to play the game to learn about phishing rather than reading books, articles or papers"<br><br>"You can see my avoidance behaviour has increased by looking at the scores of pre and post-test" |
| **Avoidance motivation** | I'm interested in playing the mobile game to avoid phishing attacks<br><br>OR<br><br>I feel gaining mobile game based education to avoid phishing attacks is somewhat useful | 20 | "Wow! This game is useful and a fantastic idea"<br><br>"This game is really interesting. I think I love to play it again and again"<br><br>"Wow! This game is great! Can I download it from the Internet?"<br><br>"Yes, this game is important, because at the end of the day it's our money, we do not want to lose it"<br><br>"I like to play the game a little longer"<br><br>"I would recommend this game to my family, peers and friends" |
| **Perceived threat** | Phishing attacks pose a threat to my computer<br>OR<br>A phishing attack is a danger to my computer | 20 | "Yes, this is a simple game anybody can play"<br>"I will be the first one who will buy this game, tell me when everything is done"<br><br>"Yes, I feel phishing is a huge threat, because at the end of the day attackers steal our money"<br>"I feel phishing threat is harmful to my computer"<br>"Yes, I feel phishing threat is dangerous to my computer" |
| **Perceived severity** | A phishing attack would steal my personal information from my computer without my knowledge<br>OR<br>Phishing attack would invade my privacy | 20 | 'Phishing attacks would steal my banking details"<br>'Phishing attacks would steal my username, passwords, credit or debit card details"<br>'Phishing attackers can use my personal details for crimes" |
| **Perceived susceptibility** | It is extremely likely that my computer will be infected by a phishing attack in the future<br>OR<br>My chances of getting phishing attacks are great. | 20 | "Yes, I now feel that my computer also may be infected by a phishing attack in the future"<br>"It is very easily that I can fall for phishing"<br>"Hmm, there is a high probability that my computer also will be infected by a phishing attack" |
| **Safeguard effectiveness** | The mobile game based education would be useful for detecting phishing attack<br><br>OR<br><br>The mobile game based education increased my knowledge about phishing attacks | 20 | "Yes, the game really helps me to identify good websites from bad ones"<br><br>"This mobile game is useful and fantastic for learning"<br><br>"This mobile game teaches how to avoid phishing threat"<br><br>"Actually, I liked the way how the game teaches"<br><br>"Yes, I learnt a lot about detecting phishing attacks through the game"<br><br>"I don't think that I would not be able to get this much of knowledge by reading a book or article about phishing" |
| **Safeguard cost** | It will take less time to gain phishing education through the mobile game<br><br>OR<br><br>It will less cost money to gain anti-phishing education through the mobile game, if downloaded from the Internet | 18 | "The game is simple"<br><br>"The game does not take long time to play"<br><br>"I like to download the game for free"<br><br>"I don't mind to pay for this game" |
| **Self-efficacy** | I gained knowledge about phishing attacks through the mobile game<br><br>OR<br><br>I feel gaining anti-phishing knowledge | 20 | "Yes, this game improved my knowledge about phishing"<br><br>"In the past, I really didn't check the URLs, now I know it is very important"<br><br>"Yes, this game taught me how to identify phishing URLs" |

| | |
|---|---|
| through the mobile game does really helped me for detecting phishing attacks | "I didn't know the meaning of 'https://' is secure version of http before playing the game."<br><br>"Yes, I think that the game was really good in teaching different patterns of URLs"<br><br>"I did not know by looking at the URL I can decide the legitimacy of the website before playing the game"<br><br>"Now I know how to identify the difference between the good and bad website" |

### 6.4.7 Summary of Results

The current study empirically evaluated the game design framework introduced in Chapter 4 (Figure. 4.3) through the mobile game prototype developed using MIT App Inventor Emulator. Therefore, a think-aloud study was conducted along with the pre- and post-test to assess the framework. The study employed 20 participants with each participant participating for an approximately one-hour session. In the pre-test, participants scored 55 percent and after the game play activity in the post-test their score was 84 percent. There is a significant improvement of 29 percent of the participants' phishing avoidance behaviour compared to the pre-test. This is a significant improvement of overall participants' phishing avoidance behaviour through the mobile game. 18 participants scored above 80 percent whilst five participants scored full marks (100 percent) in the post-test. However, all participants scored above 50 percent in their post-test. Therefore, study results concluded that the mobile game prototype is somewhat effective at teaching participants to look at URLs in their browser's address bar when assessing the website legitimacy.

In addition, all participants talked about their opinions about phishing threat awareness after using the mobile game in the think-aloud study. All of them believed that the mobile game was somewhat effective to enhance their avoidance behaviour through motivation to protect themselves against phishing attacks. Furthermore, they talked about their opinion on how the mobile game prototype impacts on game design framework elements introduced in Chapter 4. The game design framework elements are perceived threat, safeguard effectiveness, safeguard cost, self-efficacy, perceived severity and perceived susceptibility. The study revealed that perceived threat, safeguard effectiveness, perceived susceptibility, perceived severity and self-efficacy positively impacted while safeguard cost negatively impacted on avoidance behaviour through motivation to protect themselves against phishing threats. Therefore, the current study findings provided evidence of addressing the above elements in the game design framework for personal computer users to thwart phishing threat.

## 6.4.8 Discussion

The main study employed 20 regular computer users ages ranged from 18-25 year in order to elicit audio feedback of phishing threat awareness through the mobile game prototype. All participants were personal computer users and had some experience of Internet shopping. They also had some experience of using a smart phone for at least one year. The data were collected in 20 think-aloud study sessions along with pre- and post-tests where each participant spent approximately one hour in each session. The study results found that the mobile game prototype teaches personal computer users to protect themselves from phishing attacks. The participants scored 55 percent in the pre-test and 84 percent in the post-test after playing the mobile game in the main study. The overall score had been increased by 29 percent in the post-test in the main study. Five participants scored full marks (100 percent) during their post-tests, while eighteen participants scored above 80 percent. It is interesting to note that participants' avoidance behaviour has increased in the post-test, after playing the game. Furthermore, twelve participants stated: "*You can see my phishing knowledge has increased by looking at the scores of pre and post-tests*". These results back up the findings of Liang and Xue's (2010) theoretical model. The model explained a considerable amount of variance in users' avoidance behaviour (21 percent). Their findings showed actual avoidance behaviour is significant to avoid spyware attacks using given anti-spyware software as a safeguarding measure. In addition, the game design framework introduced through an empirical investigation in Chapter 4, showed that users' avoidance behaviour is significantly important to combat phishing threat using a given game based anti-phishing education (safeguarding measure). The game design framework explained a considerable amount of variance in participants' avoidance behaviour, which is 15 percent. Therefore, the current study results described that participants had a great impact on the avoidance behaviour element of the game design framework after playing the mobile game prototype.

All participants talked about their opinions and experience of phishing threats awareness through the mobile game in the think-aloud study. All of them were convinced that the mobile game is somewhat effective to enhance avoidance behaviour through motivation to protect them from phishing threat. Their common argument was that books, papers, articles and lecture notes are still in infancy boring. Those materials cannot provide fun with immediate feedback whereas this type of mobile game based education can actually provide both. This would have motivated them to play the game to learn about phishing threats. One participant responded that: "*I will now go and read more about phishing threats*". This statement describes how much they were motivated to learn about phishing threats. Another participant also stated: "*I will recommend this game to my family, peers and friends*". This shows that they are really motivated to share their experience with others. All participants said that this mobile game is really interesting and that they would love to play it again and again. Therefore, the current study conveys a simple, yet powerful message that the mobile game prototype enhances the motivation of personal computer users to avoid phishing attacks. This backs up the findings of Liang and Xue's (2010) theoretical model. The model explained a considerable amount of variance in users' avoidance motivation (56 percent). Their findings showed that users' IT threat avoidance behaviour is determined by avoidance motivation. In addition, the game design framework introduced in Chapter 4, showed that users' avoidance motivation is significantly important to combat phishing threat and also avoidance behaviour is determined by avoidance motivation. Therefore, the current study results showed that the mobile game prototype has a significant impact on the avoidance motivation element of the game design framework.

All twenty participants said that they felt that such phishing threats exist in the cyberspace after playing the game and they believed that an attack might occur at any time to their personal computer systems. The threat perception enhanced their motivation to avoid phishing threats. One participant stated: "*I feel phishing is a huge threat after playing the game, because at the end of the day attackers may steal our sensitive information such as username, password and credit/debit card information if we are unaware of phishing threats*". Furthermore, the participant mentioned that the risk of being phished is relatively high due to the pervasiveness of Internet technology. Participants rated the danger to be very high, in case that a real phishing attack occurs. A few participants discussed the fact that attackers could not only disclose their sensitive information but also use that information for crimes which is even more dangerous. One participant stated: "*I think phishing attacks not only steal my money, but attackers can also use my personal information for crimes*". Therefore, it seemed like severity and susceptibility of phishing attacks have developed through the mobile game prototype where participants perceived phishing as a dangerous threat. This findings support the findings of Liang and Xue's (2010) theoretical model. They argued that computer users have to be convinced and feel that such malicious IT threats exist in the cyberspace and are avoidable. Users' failure to feel the threat perception causes them to not act to avoid it. Their data analysis results found that the model is able to explain a respectable amount of variance in threat perception (33 percent). However, this figure is slightly lower than the finding of the game design framework introduced in Chapter 4, which is 36 percent. The current study findings demonstrated the threat perception that users need to be aware of the likelihood and severity of being attacked by phishing threat. One participant stated: "*Now I must be really careful of when I do online transactions*". The study revealed that users perceived that such a threat is existent in the cyberspace: they also sensed that there is likelihood and that it could be severe if the threat actually occurred. The findings demonstrated that the mobile game prototype had a significant impact on perceived severity and susceptibility elements of the game design framework. Therefore, the perceived threat element is significantly important to address in the game design framework for computer users to enhance avoidance behaviour through motivation to thwart phishing attacks

All twenty participants believed that the mobile game prototype is an effective safeguarding measure to thwart phishing threat. One participant stated: "*Now only I realised the worth of looking at the URL to identify good website from bad ones. I never knew that the attacker can mimic URLs to launch a phishing attack before playing the game.*" All participants mentioned that the mobile game prototype is an effective approach that motivated them to learn about phishing threat. Therefore, they believed the mobile game is an effective way of educating people to combat phishing. Furthermore, they described that mobile game prototype enhanced their avoidance motivation to identify good websites from bad ones, helping to avoid phishing threat. Their main argument was reading books or articles to learn phishing threat is neither practical nor interesting, but rather time consuming and boring. One participant stated: "*Reading books or articles take much longer yet boring where this mobile game delivers the same knowledge effectively and easy to learn with fun and less time*". Moreover, participants stressed that the mobile game was somewhat effective in gaining knowledge with fun. When they were stuck in the middle of playing the game they could always ask for help. One of the most important things was that participants got immediate feedback during the game. One participant stated: "*The game is very effective in identifying good URLs from bad ones. If I'm not sure the legitimacy of URL, then I can always ask 'BIG' fish in the game.*" The finding of the current study showed that the mobile game is an effective safeguarding measure to thwart phishing attacks. One participant stated: "*I like to play this game not because I have fun with the game, but I will learn something*". It can be argued that if the mobile game prototype is effective, then it influences to enhance participants' avoidance motivation to thwart phishing attacks. Therefore, the current findings supported the findings of Liang and Xue's (2010) theoretical model. They empirically investigated that safeguard effectiveness can motivate users to avoid malicious IT threats. In addition, the game design framework introduced in Chapter 4, empirically proved that the safeguard effectiveness element should be addressed in the game design framework for personal computer users to thwart phishing threat. Therefore, the current study provided evidence that avoidance motivation is significantly determined by safeguard effectiveness in the game design framework.

All participants in the think-aloud study stated that the game is simple and it does not take too long to play. They mentioned that they would only like to download the game online if it was freely available. One participant stated: "*If the game is too expensive and takes too long to play, I don't buy it then.*" However, a couple of participants showed their interest of purchasing the mobile game online. One participant stated: "*I like to buy the game online. I'm a bit more suspicious if a useful game is online for free download. Because it can be a virus attack sometimes.*" The other participant said: "*I do not hesitate to pay for useful things. I really liked the game. Therefore, I would pay for it*". Participants' motivation to play the game is determined by the cost that they have to pay to download the game and the time it takes to play the game (Liang and Xue, 2010). When the game is freely available online and does not take too long to play, participants showed their motivation to play the game to avoid phishing threat. It explained that avoidance motivation is determined by safeguard cost that describes time and money efforts for playing the mobile game. Therefore, the safeguard cost element should be addressed in the game design framework for personal computer users to thwart phishing threat. Furthermore, the current study findings backed up the findings of the game design framework introduced in Chapter 4. It empirically investigated that safeguard cost had a negative impact on users' avoidance behaviour through motivation to protect themselves from phishing attacks. In addition, the current findings supported the findings of Liang and Xue's (2010) theoretical model. Their findings also revealed safeguard cost negatively impacted on users' avoidance motivation thwart phishing attacks. Therefore, the current study provides evidence to address the safeguard cost in the game design framework.

All participants believed that the mobile game prototype was somewhat effective in teaching how to identify good URLs from bad ones. The evidence was obvious from the scores of pre- and post-tests. In the pre-test, participants scored 55 percent while in the post-test, after playing the game, they scored 84 percent. The score has been increased by 29 percent in the post-test. The score of post-tests supported opinions and statements observed in the think-aloud study. Therefore, the findings demonstrated that learning has taken place through the mobile game prototype, which reflected on the score of participants' post-test. A few participants responded that before playing the game they had not known https:// is for indicating a secure version of "http", which allows secure online transactions such as transferring money via online banking. For example, one participant stated: "*I learnt a lot of things about phishing which I had not known 30 minutes ago. For example, I hadn't known 's' of https stands for 'security'"*. Almost all participants responded that before playing the game they did not know the different patterns of URLs which attackers use to trick people to fall for phishing. For example, the URLs beginning with numbers such as http://147.46.236.55/PayPal/login.html are usually scams. Furthermore, they said that they did not know "www2" or "www3" is a legitimate format of URLs.All participants concluded saying, "*I learnt a lot of new things about phishing which I had not known through the mobile game"*. All participants believed that the mobile game prototype was somewhat effective in teaching participants to identify good URLs from bad ones. The study argued that effective learning about phishing threat through the mobile game motivated participants to avoid phishing attacks. The current study findings supported the findings of game design framework introduced in Chapter 4. It empirically investigated self-efficacy positively impacts on users' motivation to thwart phishing attacks. The current findings also backed up the finding of Liang and Xue's (2010) theoretical model. Their findings also revealed that self-efficacy that reflected on users' knowledge about phishing attacks, positively impacted on users' avoidance motivation to thwart phishing attacks. Therefore, current study findings provided evidence to address the self-efficacy in the game design framework.

Many authors argued that the current security education on phishing threats still provides little protection to end-users (Herley, 2009; Iacovos and Sasse, 2012; Kumaraguru, et al., 2009; Sheng, et al., 2007). However, a considerable amount of the literature claimed that well-designed end-user security education could be a recommended approach to combating phishing attacks (Allen, 2006; Hiner, 2002; Iacovos and Sasse, 2012; Kumaraguru, et al., 2008; Schechter, 2007; Sheng, et al., 2007; Timko, 2008). Sheng, et al. (2007) conducted a study about anti-phishing education comparing the effectiveness of a game with existing online training materials and a tutorial they created based on the game. They found that participants who played the game performed better at identifying phishing websites than participants who completed the two other types of training. This is accordance with the current study, which found that some evidence of well-designed mobile game based education could protect personal computer users from phishing attacks. In addition, this study attempted to assess the game designed framework introduced in Chapter 4 (Figure. 4.3) through the mobile game prototype.

However, only five participants were able to successfully differentiate legitimate websites from phishing websites by looking at URLs in the post-study. The current study identified possible explanations for these results. The mobile game functioned properly, but was still a low-fidelity prototype. Previous research has claim the necessity of high-fidelity prototypes for ubiquitous computing applications such as mobile applications, because low-fidelity approaches are insufficient to precisely capture interactivity (Lim, et al., 2006; Liu and Khooshabeh, 2003). It would have been useful to develop a high-fidelity mobile game prototype with some attractive graphics with visual objects and then test on users. In addition, limited display size of the mobile phone might have caused a problem for participants especially those who have visual problems.

Previous research revealed that mobile ubiquitous applications are still in their infancy due to the limitations of processor power, memory and specifically screen size (Hashim, Ahmad and Rohiza, 2010; Tolentino, et al., 2011). Other than this, participants would have played the game much longer than they were given. One participant stated: "*If I played the game a bit longer, I would have scored more marks in the post-test*". In this case, the participants had enough time to familiar with the gamming environment. Furthermore, it would also have been useful if the amount of URLs included in the game were increased. Then participants would have had the opportunity to test their phishing knowledge on various types of URLs. They may also have experienced different combination of URLs patterns that would help them to develop self-confidence in their knowledge about phishing threat.

In summary, this study found that the mobile game prototype enhanced the user avoidance behaviour through increasing the motivation to protect themselves from phishing attacks after the game play. Furthermore, the results provided support to assess the game design framework introduced in Chapter 4 (Figure. 4.3). Therefore, the study highlighted that perceived threat, safeguard effectiveness, safeguard cost, self-efficacy, perceived threat and perceived susceptibility elements have a significant impact on avoidance behaviour through motivation to thwart phishing attacks as addressed in the game design framework.

## 6.5 Summary

A game design framework has been introduced in Chapter 4 (Figure. 4.3) through an empirical investigation. The study findings showed the game design framework enhances computer users' avoidance behaviour through motivation to prevent themselves from phishing attacks. The current study started as a second attempt to evaluate the game design framework through a mobile game prototype developed in Chapter 5. Even though the game design framework was empirically evaluated in the Chapter 4, it would have been useful to stress the participants' impact on the framework after their engagement with the game play activity. The study employed a usability study using SUS as the first step to assess the subjective satisfaction of mobile game prototype interface. Then, a think-aloud study was employed along with a pre- and post-test in order to evaluate the game design framework. An experimental protocol was designed, which included user study instructions for participants to undertake the think-aloud study.

Finally, the current study found the results were encouraging. In the pre-test, participants scored 55 percent whilst scoring 84 percent in their post-test. The study findings showed that learning has taken place through the mobile game prototype. Participants' avoidance behaviour has also been increased by 29 percent, which is reasonably high. Therefore, the mobile game prototype was able to teach participants to protect themselves from phishing attacks. Furthermore, the study found that the mobile game prototype enhanced user avoidance behaviour through motivating to protect themselves from phishing attacks after the game play event. Finally, the study highlighted perceived threat, safeguard effectiveness, safeguard cost, self-efficacy, perceived threat and perceived susceptibility elements have a significant impact on avoidance behaviour through motivation to thwart phishing attacks as addressed in the game design framework in Chapter 4 (Figure. 4.3).

Chapter 7 will focus a discussion investigating the effectiveness of the developed mobile game prototype compared to traditional online learning to thwart phishing threats. Moreover, it will discuss the implications of the research work reported in this book.

# Chapter 7

# Discussion of Mobile Game Prototype and Public Anti-phishing Educational Website

## 7.1 Introduction

The study reported in Chapter 6 showed that the developed mobile game prototype is somewhat effective in teaching computer users on how to identify phishing URLs. The overall mobile game prototype was designed to enhance the user's avoidance behaviour through motivation to protect themselves against phishing attacks. A discussion reported in this chapter examines the effectiveness of the developed mobile game prototype compared to traditional online learning to thwart phishing threats. Furthermore, it discusses the implications of the research work reported in this book. Therefore, an evaluation study is conducted to investigate the effectiveness of the developed mobile game prototype compared to traditional online learning to thwart phishing threats. To accomplish this study, a website developed by Anti-Phishing Work Group (APWG) for the purpose of public anti-phishing education initiative was employed as a traditional web based learning source. A think-aloud experiment along with a pre- and post-test was conducted through a user study. Total 40 participants were asked to follow think-aloud user study instructions given in an experimental protocol. Participants were assigned into two groups: 20 participants with a group those who played the mobile game (as the study reported in Chapter 6) and the other 20 participants with another group those who read the APWG public education initiative website. Finally, the study used Paired-samples t-test to compare the means scores for the participants' pre- and post-tests while analysing the data gathered from the think-aloud study. The study findings revealed that the participants who played the mobile game were better able to identify fraudulent websites compared to the participants who read the website without any training.

## 7.2. Mobile Game Prototype

The study reported in Chapter 5, developed a mobile game prototype as an educational tool to teach computer users how to protect themselves against phishing threats. The study asked the following questions: The first question is how does one identify which issues the game needs to be addressed? Once the salient issues were identified, the second question is what principles should guide the structure of this information. The elements from the game design framework introduced in Chapter 4 (Figure 4.3) were used to address those mobile game design issues and the mobile game design principles were used as a set of guidelines for structuring and presenting information in the mobile game design context. The objective of the anti-phishing mobile game prototype was to teach the user how to identify phishing URLs. The overall mobile game design was focused to enhance avoidance behaviour through motivation of computer users to thwart phishing threats.

The study reported in Chapter 6 included total 20 participants, the age range between 18 and 25, from a diverse group of staff and students at Brunel University and the University of Bedfordshire, including people who were concerned about computer security. The pre- and post-tests were based on Apple MacBook Pro where the participants received their score at the end of each test. Total 20 participants were asked to follow think-aloud user study instructions given in an experimental protocol. They were also informed that they are welcome to clarify anything related to the experiment. In the pre-test, participants were presented with ten websites (shown in Table 6.4) and asked to differentiate phishing websites from legitimate ones. After evaluating 10 websites, participants were given fifteen minutes to complete a mobile game based training activity on a HTC One X touch screen smart phone. Initially, the game was designed with 10 suspect URLs (shown in Table 5.2) where the participant's responsibility is to identify legitimate URLs from phishing ones.

Then the study employed System Usability Scale (SUS) to measure the users' subjective satisfaction of the mobile game interface usability since the work reported in Chapter 6 developed the mobile game prototype. Therefore, after engaging 15 minutes with the mobile game activity, 20 participants were asked to fill in a survey as shown in Table 6.6. Then all 20 participants followed a post-test where participants were shown ten more websites to evaluate (shown in Table 6.5). The score was recorded during the pre- and post-tests to observe how participants' understanding and awareness of phishing threats developed through the mobile game based training activity.

During the mobile game based training activity, a think-aloud study was employed where the participants were asked to talk about how they identified phishing URLs from legitimate ones. Additionally, they talked about their opinions and experience of phishing threat awareness and understanding through the mobile game prototype.

## 7.3 Public Anti-phishing Educational Website

The Anti-Phishing Work Group (APWG) was established in 2003 as an industry association focused on amalgamating the global response to cyber-crime (Anti- Phishing Work Group, 2003). The organisation provides a forum for public such as responders and managers of cyber-crime to discuss phishing and other cyber-crime issues, to consider potential technology solutions, to access data logistics resources for cyber-security applications and for cybercrime forensics, to cultivate the university research community dedicated to cyber-crime and to advise government, industry, law enforcement and treaty organisations on the nature of cybercrime.

The public phishing education section of the Anti-Phishing Work Group website is developed for learning more about phishing education. For example, what is phishing threat, how it could be severe, what is the usefulness of having a safeguarding measure, where to report a suspected phishing email or website and phishing education to thwart phishing attacks. Therefore, the public Anti-phishing education initiative section of the APWG website was used as a traditional online learning source in this research study, which is shown in Figure 7.1 and 7.2.

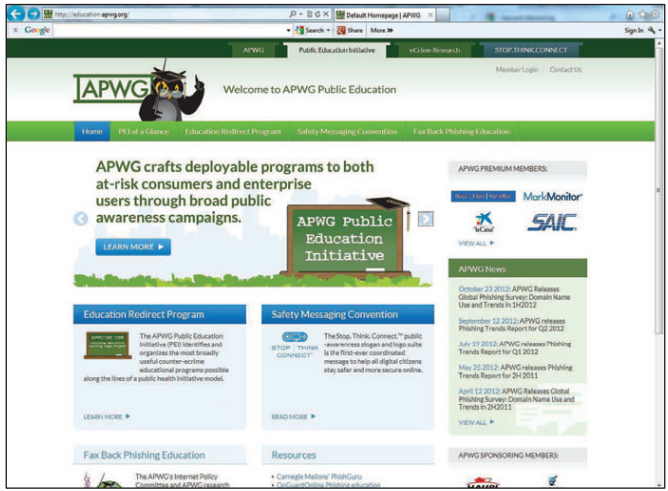

**Figure 7.1: Welcome page of APWG Public Education (APWG, 2003)**

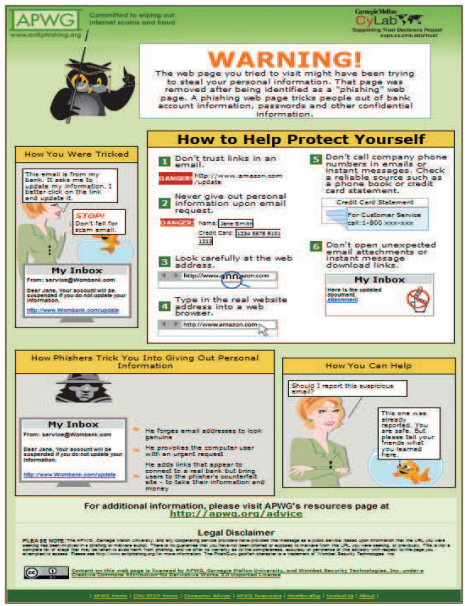

**Figure 7.2: Educational redirect program section of APWG Public Education (APWG, 2003)**

## 7.3.1 Participants

Sheng, et al. (2010) conducted a role-play survey with 1001 online survey respondents to study who falls for phishing attacks. Their study showed participants between the ages of 18 to 25 are more susceptible for phishing attacks than other age groups. The current study included 20 participants from a diverse group of staff and students at Brunel University and the University of Bedfordshire, including people who were concerned about computer security. Participants' ages ranged from 18 to 25, with a gender split of 67 percent male and 33 percent female. They had average of more than 20 hours per week of Internet experience. Each participant took part in the think-aloud study on a fully voluntary basis. A summary of the demographics of the participants in the think-aloud study is shown in Table 7.1.

**Table 7.1: Participant demographics**

| Characteristics | Amount (Mobile game prototype) | Amount (APWG educational website) |
|---|---|---|
| Sample Size | 20 | 20 |
| Gender | | |
| Male | 13 | 13 |
| Female | 7 | 7 |
| Age (18 - 25) | 20 | 20 |
| Experience using mobile device | | |
| Mobile phone | 0 | 0 |
| Smart phone | 20 | 20 |
| Average hours per week on the Internet | | |
| 0-5 | 0 | 0 |
| 6-10 | 0 | 0 |
| 11-15 | 0 | 0 |
| 16-20 | 0 | 0 |
| 20+ | 20 | 20 |

### 7.3.2 Procedure

The pre- and post-tests were based on Apple MacBook Pro where the participants received their score at the end of each test. First and foremost, each individual participant was explained the nature of the think-aloud experimental study and asked to sign a consent form. They were also informed that the experiment is about testing their understanding of phishing threat awareness using APWG public education initiative website. Then the individual participants were asked whether they knew what the term 'phishing attack' means. Those who gave a positive response were asked to give a short verbal description to confirm their understanding, whilst negative responders were read a brief definition of phishing attack and gave a short verbal description. To begin the experiment, total 20 participants were asked to follow think-aloud user study instructions given in an experimental protocol. They were also informed that they are welcome to clarify anything related to the experiment.

In the pre-test, participants were presented with ten websites (shown in Table 6.4) and asked to differentiate phishing websites from legitimate ones. After evaluating first 10 websites, participants were asked to walkthrough the APWG public education initiative website for 15 minutes. Then all 20 participants followed a post-test where participants were shown ten more websites to evaluate (shown in Table 6.5). The score was recorded during the pre- and post-tests to observe how participants' understanding and awareness of phishing threats developed through the APWG public education initiative website.

During the mobile game based training and public education initiative website reading activities, a think-aloud study was employed where participants talked about their opinions and experience of phishing threat awareness and understanding through either mobile game prototype or APWG public education initiative website.

## 7.4 Results

In the think-aloud experiment, participants talked about their experience and opinions of either mobile game prototype or APWG public education initiative website to thwart phishing threats. The results were encouraging; however it highlighted some areas where the APWG public education initiative website needed improvements.

The study reported in Chapter 6, measured the participants' subjective satisfaction of the mobile game prototype using the SUS scoring approach introduced by Brooke (1996). The score was significantly high with 84 percent (83.62 out of 100) (Brooke, 1996). Then the current research study employed Paired-samples t-test to compare the means scores for the participants' pre- and post-tests (Pallant, 2007). Participants who played the mobile game, scored 55 percent in the pre-test and 84 percent in the post- test of identifying phishing or legitimate websites after playing the mobile game prototype (Table 7.2 and Table 7.3). There was a statistically significant increase in the post-test of participants who played the mobile game ((Pre-test: M= 55.00, SD=17.911 and Post-test: M=84.00, SD=13.139), t(19)= -7.97, p<0.005 (two-tailed)) compared to those who read APWG public education initiative website (Pre-test: M= 60.00, SD= 17.770 and Post-test: M= 62.50, SD=25.930, t(19)= -0.036, p>0.005

(two-tailed)).

It has been seen that there is a considerable improvement of participants' results of 29 percent in the post-test after their engagement with the mobile game prototype (p<0.005 (two-tailed)). This is a significant improvement of overall participants' phishing avoidance behaviour through the mobile game prototype. 18 participants scored above 80 percent whilst five participants scored full marks (100 percent) in the post-test. However, all participants scored above 50 percent in their post-test. The overall participants' score is shown in Figure 7.3. Additionally, the individual participant's score during their engagement with the mobile game prototype is shown in Figure 7.4.

**Table 7.2: Paired Samples t-test Statistics**

| Paired Samples Statistics | | Mean | N | Std. Deviation | Std. Error Mean |
|---|---|---|---|---|---|
| Pair 1 | MobilePreTest | 55.00 | 20 | 17.911 | 4.005 |
| | MobilePostTest | 84.00 | 20 | 13.139 | 2.938 |
| Pair 2 | WebPreTest | 60.00 | 20 | 17.770 | 3.974 |
| | WebPostTest | 62.50 | 20 | 25.930 | 5.798 |

**Table 7.3: Paired Samples Test**

| Paired Samples Test | | Paired Differences | | | | | t | df | Sig. (2-tailed) |
|---|---|---|---|---|---|---|---|---|---|
| | | Mean | Std. Deviation | Std. Error Mean | 95% Confidence Interval of the Difference | | | | |
| | | | | | Lower | Upper | | | |
| Pair 1 | MobilePreT - MobilePostT | -28.500 | 15.985 | 3.574 | -35.981 | -21.019 | -7.973 | 19 | .000 |
| Pair 2 | WebPreT - WebPostT | -2.500 | 31.267 | 6.992 | -17.133 | 12.133 | -.358 | 19 | .725 |

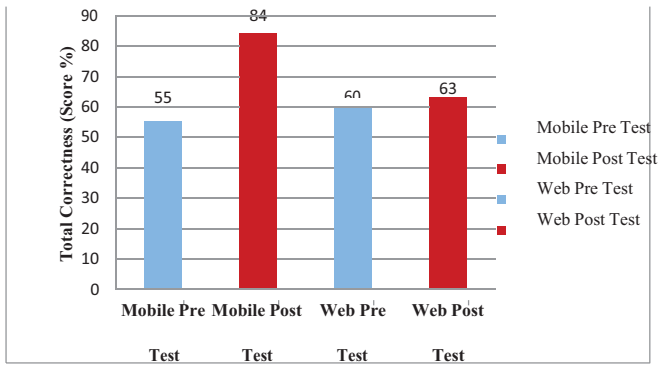

**Figure 7.3: Total correctness for the test groups. N=20 in all conditions. The game condition shows the greatest improvement (by 29%).**

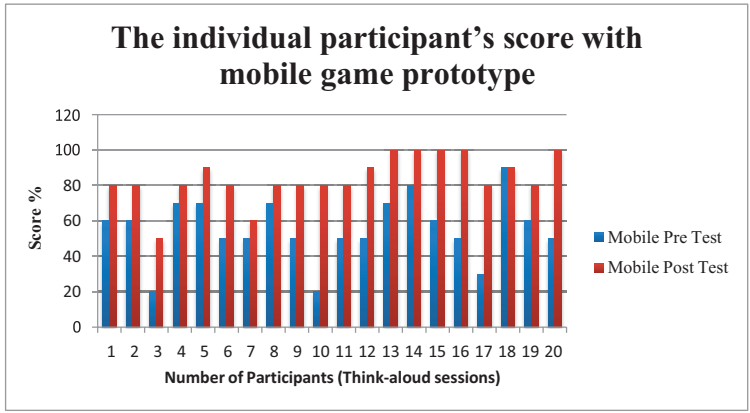

**Figure 7.4: The individual participant's score during their engagement with the mobile game prototype**

All participants responded that their decision was based on looking at the address bar (URL) when evaluating the second set of ten websites in the post-test. Furthermore, they stated that they made only very few attempts to look at the address bar when evaluating the first ten websites in the pre-test. There, many participants used incorrect strategies to determine the website legitimacy. For example, one of the common strategies consisted of checking whether or not the website was designed professionally. However, this may not be a useful strategy as many phishing websites are an exact replica of legitimate websites. The attacker can easily mimic any professional website from the source code of the particular page provided by the browser. Moreover, all participants highlighted that the mobile game was somewhat effective at teaching different patterns of URLs to differentiate phishing URLs from legitimate ones.

Participants those who read the APWG public education initiative website, scored 60 percent in the pre-test and 63 percent in the post-test of identifying phishing or legitimate websites after reading the APWG public education initiative website (Table 7.2). There is a slightly little improvement of participants' results of 3 percent in the post-test during the think-aloud study. 7 participants scored below 50 percent whilst five participants scored above 80 percent in the post-test. Only 12 participants scored above 50 percent in the post-test. None of the participants scored full marks (100 percent) in their post-test. The overall participants' score is shown in Figure 7.3. Additionally, the individual participant's score during their engagement with the APWG public education initiative website is shown in Figure 7.5. It can therefore be argued that participants were not much aware of how to avoid phishing threats by just reading the APWG public education initiative website. The current research study identified possible explanations to these results. All participants who read the website during the think-aloud study stated that they are bored of reading the website, since it contains a lot of text. Their argument was reading a website to gain phishing knowledge does consume a lot of time though it is useful in the long run. Only a few participants responded that their decision was based on looking at the address bar when evaluating the second set of ten websites in the post-test.

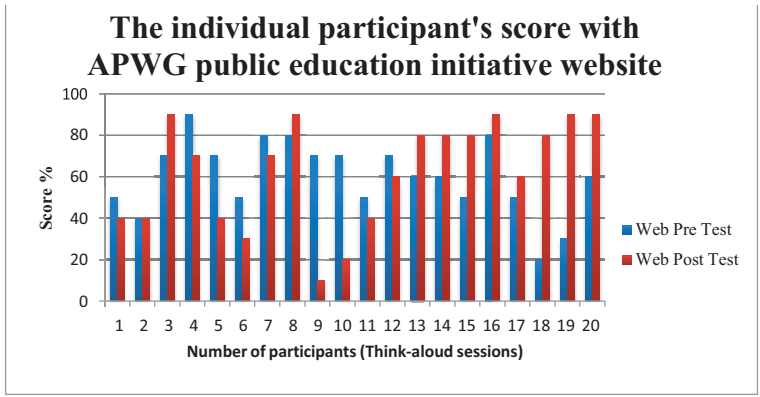

**Figure 7.5: The individual participant's score during their engagement with the APWG public education initiative website**

Furthermore, they described that the website does tell us to look at the URL; however, it does not tell us how to identify good URLs from bad ones more specifically. They also believed that the website does not provide a way of testing their knowledge. For example, it would be useful to give the user a test after reading the website for self-assessment their knowledge gained against phishing attacks. This helps to obtain an immediate feedback on what they have learnt. Additionally, video tutorials can be embedded into the website where the user can better engage with learning process. Therefore, the current study concluded that the mobile game prototype is somewhat effective in teaching computer users to thwart phishing attacks than traditional online learning.

## 7.5 Research Implications

The research work reported in this book focuses on how one can educate computer users against phishing threats, because phishing attacks continuously jeopardise innocent computer users (Iacovos and Sasse, 2012; Kumaraguru and Cranor, 2009; Purkait, 2012). Usability experts claim that end-user education and training does not work to combat against phishing (Sheng, et al., 2007; Evers, 2007). It can therefore be argued that existing anti-phishing security education such as through web-based training, contextual training and embedded training materials has not been appropriately designed for users. Yet theory-based empirical research that enhances computer users' phishing awareness through their voluntary IT threat avoidance behaviour is lacking (Iacovos and Sasse, 2012; Purkait, 2012). The current research study findings revealed that the participants who played the mobile game were better able to identify fraudulent websites compared to the participants who read the APWG public education initiative website without any training. This is because the developed mobile game prototype was based on the game design framework introduced in Chapter 4. The framework formulated through an empirical investigation to enhance computer users' phishing threat avoidance behaviour though their motivation. The study incorporated perceived threat, safeguard effectiveness, safeguard cost, self- efficacy, perceived severity and perceived susceptibility elements into the game design framework for computer users to thwart phishing attacks. As a result, the current study found that the mobile game is somewhat effective in enhancing the computer user's phishing awareness. Therefore, it is worth understanding how the human aspect of avoiding malicious IT threats can influence in designing security education. For example, when designing security education though web-based training materials. Furthermore, it can argue this proposed game design framework not only can protect computer users from phishing attacks but also other cyber-attacks such as botnets, zombies, virus, trojans, spyware, spam, malware, etc.

The research reported in this book can inform IT security education programs in several aspects. First, it recommends the worth of security awareness, education and training programs. Individuals are more motivated to avoid malicious IT threats and use of safeguarding measures if these security educational programs help them to develop threat perceptions, comprehend the effectiveness of safeguarding measures, lower safeguard costs and increase their self-efficacy. Second, this research suggests that security awareness and training programs should stress both the likelihood of malicious IT threats and the severity of losses caused by the threats. The game design framework introduced in Chapter 4 suggests that emotion-focused coping behaviour is a possible culprit that confuses users' motivation systems at high threat levels. Therefore, public security education initiatives should draw attention to possible emotion-focused coping approaches and help computer users to understand how they unintentionally engage in emotion-focused coping behaviour. Furthermore, individuals should be trained how to stop emotion-focused coping behaviour when they are dealing with malicious IT threats. Such training and education would help computer users to emphasis on the problem-focused coping approach, thus reducing the negative effects of high threat perceptions.

Most organisations do conduct security education, training and awareness programs to regulate employees' malicious IT threat avoidance behaviour and adhere to best security practices against malicious IT threats (Gordon, et al., 2006). However, users at home in a non-work setting are free from those regulations and acting on their own and are easily prey to malicious IT threats. In addition, organisations may consist of either a centralised or decentralised IT security environment (Warkentin and Johnston, 2006). In the centralized organisational IT security environment, security is managed at the enterprise level and employees have no choice to opt out. Those organisations with centralised IT security environment may have sufficient IT infrastructure and strict IT security policies and procedures developed by professional security expertise. In contrast, in the decentralized organisational IT security environment, employees involve in voluntary protective actions such as configuring and enabling their personal firewall/proxy and updating their own antivirus and/or anti-phishing software. These employees in the decentralised IT security environment are highly likely to engage in unsafe computer behaviours and become the weakest link in their organisational security system. Therefore, security education, training and awareness are needed to help these uses to combat against malicious IT threats. The mobile game prototype design introduced in Chapter 5 as an educational tool focused to teach computer users how to protect themselves against phishing attacks. This is because mobile games can facilitate to embed learning in a natural environment. Therefore, a game based learning approach can contribute to the effectiveness of such IT security programs.

Some organisations such as banks provide help and guidance on their official website to protect their customers against phishing scams. In addition, they provide free anti-virus software such as Kaspersky Internet Security for their customers. However, the studies reported in Chapter 6 and 7 revealed that reading articles, books, lecture notes or webpages to gain anti-phishing knowledge is still in infancy boring. On the other hand, how many of computer users are competent enough to properly install, configure and use anti-virus software on their personal computer system is questionable today. Most of them believe that by only installing an anti-virus application does entirely protect their computer system from cyber-attacks. Unfortunately, this is not true. However, organisations like banks should be aware the human element is the weakest link in personal computer use and need to better educate them to combat malicious IT threats. This can be achieved by allowing their customers to free download security educational games (e.g. the game prototype developed in Chapter 5) for their computers and mobile devices.

## 7.6 Summary

The current study attempted to evaluate the effectiveness of a mobile game compared to a traditional website in order to protect computer users against phishing attacks. Therefore, a mobile game prototype was developed based on the game design framework introduced in Chapter 4 (Figure 4.3) that aimed to enhance avoidance behaviour through motivation to protect computer users against phishing threats. The APWG public education initiative website was used as a traditional web based learning source. The experiment was conducted through a user study. A think-aloud study was employed along with a pre- and post-test with total 40 participants, where 20 participants were asked to play the mobile game prototype (as reported in Chapter

6) and the other 20 participants were asked to read the website. The study findings revealed that the participants, who played the mobile game, were better able to identify fraudulent websites than the participants, who read the website. Therefore, the study believes that teaching computer users how to prevent from phishing threats using a mobile game, would contribute to enable the cyberspace a secure environment.

However, only five participants who played the game were able to successfully differentiate legitimate websites from phishing websites by looking at URLs in the post-study. The current study identified possible explanations for these results. The mobile game functioned properly, but was still a prototype. It is useful to develop a proper mobile game rather than a prototype with some attractive graphics with visual objects, including more complex URLs and then test on a different sample size to confirm these findings. In addition, limited display size of the mobile phone might have caused a problem for participants especially those who have visual problems. Additionally, it can argue that if participants were allowed to play the game much longer they would have scored more in the post-test. Future research can be conducted with a larger sample size allowing participants to play the game much longer using a high-fidelity mobile game rather than a prototype.

Chapter 8 will focus a discussion based on the findings of Chapter 4, 5, 6 and 7 with focus on the relevant literature. It will highlight some limitations and areas of improvements. Furthermore, it will open more branches for future research based on these studies.

# Chapter 8

# Discussion and Conclusion

## 8.1 Introduction

This book addressed the security awareness issues concerning computer use with specific attention to phishing threats. Phishing is a type of semantic attack that is particularly dangerous in computer use (Downs, Holbrook and Cranor, 2007; Schneier, 2000). It is well known as an online identity theft that attempts to steal sensitive information such as username, password and online banking details from its victims. The attacker generally synbookes the social engineering element as well as the technical element to launch an attack (Jagatic, et al., 2007). It could be easy to hinder the technical element by developing necessary software tools such as anti-phishing software. Though, it is hard to stop the social engineering element because most computer users tend to disclose their sensitive information through social networking websites. This could be through Internet enabled services such as Facebook, Twitter, Hi5, Orkut, Skype and even more professional social networking websites like LinkedIn. Furthermore, users can have a lack of security awareness and struggle to make sensitive trust decisions during online activities, such as online bill payments or banking transactions. These are one of the major reasons that computer users are susceptible to phishing attacks (Kumaraguru, et al., 2007). Therefore, the computer user's contribution is vital to make the cyberspace a safer place for everyone.

The purpose of this book was to investigate how to create awareness on computer users to protect themselves from phishing attacks. The motivation behind this research was that phishing attacks get more sophisticated day by day as and when attackers learn new techniques and change their strategies accordingly. As a result, innocent people disclose their confidential information such as usernames, passwords and credit or debit card information to hackers via the Internet. As reported in the Anti- Phishing Working Group (APWG) in 2011, approximately one in every 300 emails circulating over the web is considered to contain elements pointing to phishing (APWG, 2012). Furthermore, they claimed that compared with the total numbers of phishing attacks recorded in 2010, phishing numbers have increased significantly through 2011. The BBC has also constantly reported increasing phishing scams over the years. One of the most recent scams targeted LinkedIn users through email scams after hackers leaked more than six million user passwords online (BBC News, 2012).

The previous chapters reported in this book have discussed the underlining research domain, highlighted the importance of anti-phishing education to thwart phishing threats and investigated ways to improve the situation. The final chapter of this book will discuss and summarise the findings of the research and report limitations, which still leave room for improvement and further exploration.

## 8.2 Summary of the Studies' Findings

Two studies and one theoretical design were incorporated into the research work reported in this book as presented below.

## 8.2.1 Developing a Game Design Framework to Thwart Phishing Attacks

The research work reported in Chapter 4 empirically investigated the key elements that should be addressed in the game design framework for computer users to thwart phishing attacks. The elements of a theoretical model derived from TTAT, were used in the game design framework. The main findings of this study revealed that perceived threat, safeguard effectiveness, safeguard cost, self-efficacy, perceived severity and perceived susceptibility elements should be incorporated into the game design framework for computer users to avoid phishing attacks through motivation.

The study was particularly concerned about threat perception because it plays an important role in influencing the computer user's avoidance behaviour. The data analysis results in Chapter 4 showed that the model explained users' motivation to avoid IT threats influences their actual avoidance behaviour. Therefore, it conveyed a powerful message to motivate computer users to avoid malicious IT threats. A similar conclusion has been made by Liang and Xue (2010). Furthermore, when users decided that the IT threat can be avoided by using a given safeguarding measure, they may take a problem-focused coping approach. However, when the IT threat could not be avoided completely, they may take an emotion-focused coping (Liang and Xue, 2010; Rhoa and Yub, 2011). A similar conclusion has been made by Beaudry and Pinsonneaut (2001). They stated that if users perceive the malicious IT threat, they take a problem-focused coping approach; if they believe that the threat is not avoidable, they will inactively avoid the threat by performing a emotion-focused coping approach.

The study revealed that the perceived threat element is significantly important to be addressed in the game design framework for computer users to enhance avoidance behaviour through motivation to thwart phishing attacks. This is because computer users need to be convinced and feel that such malicious IT threats exist in the cyberspace and are avoidable. Users' failure to perceive the threat causes passive and inactive behaviour. Furthermore, the study described threat perception that users need to be aware of likelihood and severity of being attacked by malicious IT threat. The study described that users feel that such a threat exists in the cyberspace; they acknowledged its likelihood and also the severeness if the particular threat actually occurred. If users actually perceive the threat, they are more motivated to avoid it. Therefore, the study concluded that perceived susceptibility and severity elements should be addressed in the game design framework for computer users to truly perceive the threat.

The study found that the safeguard effectiveness element should be addressed in the game design framework for computer users to thwart phishing threats. The safeguarding measure was evaluated from three aspects; taking into account safeguard effectiveness, costs related to the safeguard measure and users confidence of using the safeguard to avoid malicious IT threats. When the level of effectiveness of the safeguarding measure is high then users are more motivated to avoid threats. Furthermore, users' high confidence in taking the safeguard measure influenced to enhance the motivation of avoiding threats. Therefore, the study revealed that self- efficacy should also be addressed in the game design framework for avoiding threats through motivation.

The current study revealed that safeguard costs negatively affect the avoidance motivation. When the safeguard cost is high, then users are less motivated to avoid threats. A similar conclusion has been made by Liang and Xue (2010); they described that when time, money, inconvenience and comprehension are required to use the safeguarding measure is high, then users are less motivated to avoid threats (Liang and Xue, 2009a; 2010b). Therefore, safeguard costs should be addressed in the game design framework as a payback to safeguarding effectiveness. However, Liang and Xue's (2010) study did not support the interaction between perceived severity and susceptibility on perceived threat. Unexpectedly, this study found that perceived threat is significantly determined by the interaction between perceived severity and susceptibility.

Moreover, the current study did provide evidence to address the interaction effect of perceived threat and safeguard effectiveness in the game design framework. The study revealed that avoidance motivation is significantly determined by the interaction of perceived threat and safeguard effectiveness. This provided contradicting results to Liang and Xue's (2010) findings. However, they elaborated that the interaction between perceived threat and safeguard effectiveness can be viewed from two perspectives. First, when the threat level is high, perceived threat can be viewed to negatively moderate the relationship between safeguard effectiveness and avoidance motivation. Second, when the level of safeguard effectiveness is high, it can be viewed to negatively moderate the relationship between perceived threat and avoidance motivation.

## 8.2.2 Developing a Mobile Game for Computer Users to Thwart Phishing Attacks

The aim of this theoretical study reported in Chapter 5 focused on designing a mobile game as a tool to educate computer users by creating awareness of phishing attacks. It addressed two issues: first, which key elements should be addressed in the mobile game? Second, which principal should be used to address these elements? The elements derived from the game design framework introduced in Chapter 4 were used to address in the mobile game context. Mobile game design principles were used as a set of guidelines for structuring and presenting those elements in the mobile game design context. The main objective of the anti-phishing mobile game design prototype was to teach users how to identify phishing website addresses (URLs) which is one of many ways to identify a phishing attack. The overall game design was targeted to enhance avoidance behaviour through motivation to protect computer users from phishing attacks.

This study has attempted to develop a mobile game prototype based on the design using MIT App Inventor. Finally, the developed mobile game prototype application was deployed on a HTC One X touch screen smart phone.

## 8.2.3 Evaluating the Game Design Framework through the Mobile Game

The final research study reported in Chapter 6 focused on evaluating the game design framework (presented in Chapter 4) through the mobile game prototype developed using MIT App Inventor Emulator (presented in Chapter 5). The objective of this study was to determine users' impact on the framework after playing the mobile game prototype, which will help to emphasis the game design framework to avoid phishing attacks. In addition, pre- and post-tests were employed in this study to determine whether or not users' phishing awareness has been improved through the mobile game prototype. Finally, this finding elaborated whether or not the user avoidance behaviour enhances through motivation to protect themselves from phishing attacks through the mobile game prototype.

All participants in the study were personal computer users and had some experience of Internet shopping. They also had some experience of using a smart phone at least one year. The study results revealed that the mobile game prototype enhanced their phishing awareness. Participants scored 55 percent in the pre-test and 84 percent in the post-test, after playing the mobile game. The overall score had been increased by 29 percent in the post-test, which is considered as a significant improvement of participants' phishing awareness. These findings illustrated that the mobile game prototype could enhance participants' avoidance behaviour through motivation to protect themselves from phishing attacks. In addition, all participants believed that the mobile game prototype is somewhat effective in enhancing their avoidance behaviour through motivation to thwart phishing attacks. Therefore, the study concluded that the game had a significant impact on avoidance behaviour through avoidance motivation to thwart phishing attacks as addressed in the game design framework in Chapter 4. These results back up the finding of Liang and Xue's (2010) theoretical model. They argued and empirically revealed that users with a stronger avoidance motivation are more likely to engage in the avoidance behaviour of using the safeguard.

All participants said that they perceived that such a threat could exist in the cyberspace and threat perception enhanced their motivation to avoid it while playing the game. They believed that the risk of being phished is relatively high due to the pervasiveness of Internet technology. However, if the actual attack occurred, they believed that the danger caused by the attack would be high. Therefore, it elaborated that the game attempted to develop both severity and susceptibility of phishing attacks where participants perceived phishing as a threat which danger to their computer use. These findings concluded that the game had a significant impact on both severity and susceptibility elements as addressed in the game design framework in Chapter 4. A similar conclusion can be found in Liang and Xue's (2010) theoretical model. They concluded that computer users have to be convinced and feel that such malicious IT threats exist in the cyberspace and are avoidable. Users' failure to feel the threat perception causes them will not act to avoid it.

The study found that the mobile game prototype is an effective approach that motivated participants to learn how to avoid phishing. Their main argument was reading books or articles to learn phishing threat are neither practical nor interesting, but it is rather time consuming and boring. However, the notion of a game teaching people about phishing is quite interesting. Moreover, it enables to gain knowledge with fun. The current study found that the mobile game prototype is usable. It can therefore be argued that if the game is neither usable nor effective, the overall score of the post-test would not be increased by a 29 percentage. Therefore, the study concluded that the game is an effective safeguard measure which motivated participants to avoid phishing threat as addressed in the game design framework in Chapter 4. A similar conclusion can be found in previous studies (Anderson and Agarwal, 2006; Ng, Kankanhalli and Xu, 2009; Woon, Tan and Low, 2005; Liang and Xue, 2010).

All participants showed interest in playing the game if it is simple and not taking too long to play. Furthermore, most of them would have liked to download the game for free. The study found that participants' motivation to play the game is determined by the cost that they have to pay for downloading and the time taken to play it. Therefore, the study concluded that the cost of the game influences participants' motivation to avoid phishing threat as addressed in game design framework in Chapter 4. A similar conclusion can be found in Liang and Xue's (2010) theoretical model, where they stated that safeguard cost negatively impact on users' avoidance motivation to thwart phishing attacks.

All participants believed that the mobile game prototype is somewhat effective in teaching them to identify good URLs from bad ones. The study argued that participants' effective learning through the game would enhance their behaviour to avoid phishing threats. On the other hand this was proved by the increase of participants' overall score by 29 percent in the post-test. Therefore, the study concluded that the game developed participants' self-efficacy, which has a correlation to knowledge to combat phishing threat. This result backs up the finding of Liang and Xue's (2010) theoretical model, which has stated that self-efficacy that reflected on users' knowledge about phishing attacks, positively impacts on users' avoidance motivation to thwart phishing attacks.

In summary, this study found that the mobile game is somewhat effective in improving the user phishing avoidance behaviour after their engagement with the mobile game. In addition, the study findings supported to evaluate the game design framework through the mobile game prototype. Therefore, the study highlighted perceived threat, safeguard effectiveness, safeguard cost, self-efficacy, perceived threat and perceived susceptibility elements have a significant impact on avoidance behaviour through motivation to thwart phishing attacks as addressed in the game design framework.

## 8.3 Originality of the Research

There are studies in the literature investigating how to educate people to protect themselves from malicious IT attacks such as phishing threats. This book contributed to develop a game design framework for computer users to protect themselves from phishing attacks. Therefore, the originality of the first study, in Chapter 4, is the idea of introducing a game design framework to avoid phishing attacks through an empirical investigation. The elements of a theoretical model derived from TTAT, were used to address the game design framework. The literature did not report a study that gave the opportunity to the researcher to introduce a game design framework to avoid phishing attacks based on TTAT.

Likewise, the work discussed in Chapter 5 of this book focuses on designing a mobile game for computer users to thwart phishing attacks. The elements derived from the game design framework introduced in Chapter 4 were used to address the mobile game while using mobile game design principles as a set of guidelines for structuring and presenting those elements in the mobile game design context. Therefore, this mobile game design is original and has not been published in any previous research studies. The strength of this study is the approach of the design that was based on the game design framework introduced in Chapter 4.

Similarly, the work reported in Chapter 6 focuses on evaluating the game design framework (presented in Chapter 4) through the mobile game prototype developed using MIT App Inventor Emulator (presented in Chapter 5). The strength of this work is the evaluation process that was carried out after designing the game framework. Therefore, the game design framework is original and has not been published in any previous studies.

One of the distinguishing characteristics reported in the research work of this book from the literature work is the methodological approach. This research depended on the quantitative and qualitative data that have been gathered from participants. The unique feature is that the qualitative data of this study were gathered from a think- aloud study along with pre and post-test in a laboratory based experiment. These data allowed the researcher to determine whether or not participants' phishing awareness has been improved through the mobile game prototype. Moreover, the data collected from quantitative approach gave a possibility to support the findings that were concluded from quantitative work.

## 8.4 Limitations and Future Work

In the first instance the research work of this book has several limitations. The study reported in Chapter 4 of this book focused on developing a game design framework for computer users to thwart phishing attacks. First, the study recruited university students' ages ranged from 18 to 25 years as a convenience sample. However, university students do not represent the population of general computer users. They are younger, may have more computer skills and are probably more knowledgeable about malicious IT threats than average users. Therefore, researchers should be cautious when generalising these findings to other computer users. Nevertheless, this study is more focused on developing a game design framework to avoid phishing attacks rather than on generalisability. Therefore, it is justifiable to use a subset of the computer user population as the sample size. In addition, it is practically infeasible to collect a random sample from the entire population of computer users. Future research is required to confirm these research findings using different samples.

Second, the study selected phishing attacks as malicious IT and anti-phishing education as the safeguarding measure in order to empirically investigate the game design framework. However, this does not necessarily mean that the source of a malicious IT threat and the safeguarding measure must be an IT and behaviour respectively. The source of a malicious IT threat could be rather a person (e.g., a hacker or an attacker) or an event (e.g., denial of service) and the safeguarding measure could be an IT (e.g., anti-phishing tool or anti-virus application) or inaction (e.g., stop downloading freebies). Therefore, future research can be conducted with different malicious IT threat sources and safeguarding measures to observe whether the findings of this study will remain as same.

The theoretical designing approach reported in Chapter 5 focused on designing a mobile game as a tool for computer users to protect themselves from phishing attacks. The objective of the developed anti-phishing mobile game was to teach user how to identify phishing website addresses which is one of many ways of identifying a phishing attack. Furthermore, future research can be conducted to design and develop a mobile game to teach other areas such as signs and content of the web page, the lock icons and jargons of the webpage, the context of the email message and the general warning messages displayed in the website.

The study reported in Chapter 6 of this book focuses on evaluating the game design framework through the mobile game prototype developed using MIT App Inventor Emulator. Although the mobile game functioned properly, it would have been useful to design and develop a high-fidelity mobile game prototype with some attractive graphics and visual objects to better interact with the user. Perhaps this may enhance their satisfaction of interacting with the game which presumably helps to improve the phishing awareness through the game. Therefore, a future research would benefit from evaluating the game design framework through a high-fidelity mobile game prototype. In addition, a different combination of phishing and legitimate websites can be used in pre- and post-tests to make sure the mobile game prototype creates awareness to avoid phishing threats. Furthermore, as future research, the same study reported in Chapter 6 can be conducted through a simulated phishing attack in a laboratory-based environment. This helps to observe how people are susceptible for real life phishing attacks before and after their engagement with the mobile game prototype.

The study reported in Chapter 7 examined the effectiveness of the developed mobile game prototype compared to traditional online learning to thwart phishing threats. The study findings showed that the participants who played the mobile game were better able to identify fraudulent websites compared to the participants who read the website without any training. Future research can be conducted to investigate a set of design guideline to better design traditional online learning systems to thwart phishing threats. This can be based on the elements of the game design framework that enhanced computer users' avoidance behaviour thorough motivation to protect themselves against phishing attacks.

## 8.5 Summary

This chapter discussed the main findings of the research work and a detailed discussion of the main finding has been presented. Furthermore, the contribution of the research work and its originality has been illustrated. Finally, limitations and areas for future work were discussed to open up more opportunities for forthcoming research.